\documentclass[aps,amsmath,amssymb, notitlepage,nofootinbib,10pt,twocolumn]{revtex4-2}

\usepackage{color}
\usepackage[colorlinks=true,linkcolor=blue,urlcolor=blue,citecolor=blue]{hyperref}
\usepackage{color}
\usepackage{bbold}
\usepackage{bm}

\usepackage{times}

\usepackage{ulem}

\usepackage{amsmath}
\usepackage{amssymb}
\usepackage{graphicx}
\usepackage{array}

\usepackage{tikz}

\usetikzlibrary{calc,fadings,decorations.pathmorphing,decorations.pathreplacing,shapes,shapes.geometric,shapes.multipart,arrows,shapes.misc,intersections,positioning,patterns}

\usepackage{xargs}

\renewcommand{\em}{\it}

\newcommand{\Eq}[1]{Eq.~\ref{#1}}
\newcommand{\Eqs}[1]{Eqs.~(\ref{#1})}
\newcommand{\eq}[1]{(\ref{#1})}
\newcommand{\rme}{\mathrm{e}}   
\newcommand{\rmd}{\mathrm{d}} 
\newcommand{\cd}{\!\cdot\!}

\newcommand{\bit}{\begin{itemize}}
\newcommand{\eit}{\end{itemize}}

\newcommand{\half}{\frac{1}{2}} 
\newcommand{\f}{\frac}
\renewcommand{\>}{\right\rangle}
\newcommand{\<}{\left\langle}
\newcommand{\ba}{\begin{align}}
\newcommand{\ea}{\end{align}}
\newcommand{\be}{\begin{equation}}
\newcommand{\ee}{\end{equation}}
\newcommand{\bea}{\begin{eqnarray}}
\newcommand{\eea}{\end{eqnarray}}
\newcommand{\nn}{\nonumber}
\newcommand{\bi}{\begin{itemize}}
\newcommand{\ei}{\end{itemize}}
\newcommand{\lf}{\left(}
\newcommand{\ri}{\right)}
\newcommand{\dd}{\mathrm{d}}

\newcommand{\Tr}{\operatorname{Tr}}

\newcommand{\black}{\color{black}}

\newcommand{\ca}{\mathcal}

\def\a{\alpha}

\def\+{\dagger}
\def\x{\bar x}

\def\det#1{\left|\begin{matrix}#1
\end{matrix}\right|}

\def\a{\alpha}

\renewcommand{\det}{\operatorname{det}}

\begin{document}

\newcommand{\bra}[1]{\< #1 \right|}
\newcommand{\ket}[1]{\left| #1 \>}

\newcommand{\brackets}[1]{\left< #1 \right>}

\newcommand{\bbra}[1]{\<\< #1 \right| \right|}
\newcommand{\kket}[1]{\left|\left| #1 \>\>}

   \newcommand{\bbraket}[2]{\<\< #1 || #2 \>\>}

\title{Renormalization group for measurement and entanglement phase transitions}
\author{Adam Nahum and Kay J\"org Wiese}
\affiliation{CNRS-Laboratoire de Physique de l'Ecole Normale Sup\'erieure, PSL, ENS, Sorbonne Universit\'e, Universit\'e Paris Cit\'e, 24 rue Lhomond, 75005 Paris, France}
\date{\today}

\begin{abstract}
We analyze the renormalization-group (RG) flows of
two effective Lagrangians, 
one for measurement induced transitions of monitored quantum systems 
and one for  entanglement transitions in random tensor networks.
These Lagrangians, 
previously proposed on  grounds of replica symmetry, 
are derived in a controlled regime for an 
illustrative family of tensor networks. They have  different forms in the two cases, and involve   distinct replica limits.
The perturbative RG is controlled by working close to a critical dimensionality, ${d_c=6}$ for measurements and ${d_c=10}$ for random tensors, where interactions become marginal.
The resulting RG flows are surprising in several ways. 
They indicate that in high dimensions $d>d_c$ there  are at least two  (stable) universality classes for each kind of transition, separated by a nontrivial tricritical point. In each case one of the two stable fixed points is Gaussian, while the other is nonperturbative.
In lower dimensions, $d<d_c$, the flow always runs to the nonperturbative regime.
This picture  clarifies the ``mean-field theory'' of these problems,  including the phase diagram of all-to-all quantum circuits.
It  suggests a way of reconciling exact results on tree tensor networks with field theory.
Most surprisingly, the perturbation theory for the random tensor network 
(which also applies to a version of the measurement transition with ``forced'' measurements) 
formally possesses a dimensional reduction property 
analogous to that of the random-field Ising model. 
When only the leading interactions are retained, perturbative calculations in $d$ dimensions reduce to those in a simple scalar field theory in ${d-4}$ dimensions. We show that this holds to all orders by writing the action in a superspace formulation.
\end{abstract} 
\maketitle

\noindent

\newcommand{\triagdiag}{{\parbox{0.8cm}{{\begin{tikzpicture}[scale=1]
\coordinate (x1) at (0,0) ;
\coordinate (x2) at  (0.7,0) ;
\coordinate (x3) at  (0.35,0.5) ;
\fill (x1) circle (1.5pt);
\fill (x2) circle (1.5pt);
\fill (x3) circle (1.5pt);
\draw [black] (x1) -- (x2) -- (x3) -- (x1);
\end{tikzpicture}}}}}

\newcommand{\smalltriagdiag}{{\parbox{0.4cm}{{\begin{tikzpicture}[scale=0.5]
\coordinate (x1) at (0,0) ;
\coordinate (x2) at  (0.7,0) ;
\coordinate (x3) at  (0.35,0.5) ;
\fill (x1) circle (1.5pt);
\fill (x2) circle (1.5pt);
\fill (x3) circle (1.5pt);
\draw [black] (x1) -- (x2) -- (x3) -- (x1);
\end{tikzpicture}}}}}

\newcommand{\propdiag}{{{\parbox{1.cm}{{\begin{tikzpicture}[scale=1]
\coordinate (x1) at (0,0) ;
\coordinate (x2) at  (0.6,0) ;
\coordinate (x1p) at  (-.2,0) ;
\coordinate (x2p) at  (0.8,0) ;
\fill (x1) circle (1.5pt);
\fill (x2) circle (1.5pt);
\draw (.3,0) circle (3mm);
\draw [black] (x1) -- (x1p);
\draw [black] (x2) -- (x2p);
\end{tikzpicture}}}}}}

\newcommand{\propdiagBlue}
{ {\parbox{1.cm}{{\begin{tikzpicture}[scale=1]
\coordinate (x1) at (0,0) ;
\coordinate (x2) at  (0.6,0) ;
\coordinate (x1p) at  (-.2,0) ;
\coordinate (x2p) at  (0.8,0) ;
\fill (x1) circle (1.5pt);
\fill (x2) circle (1.5pt);
\draw [blue] (.3,0) circle (3mm);
\draw [blue] (x1) -- (x1p);
\draw [blue] (x2) -- (x2p);
\end{tikzpicture}}}} }

\newcommand{\propdiagAmp}{{{\parbox{0.7cm}{{\begin{tikzpicture}[scale=1]
\coordinate (x1) at (0,0) ;
\coordinate (x2) at  (0.6,0) ;
\fill (x1) circle (1.5pt);
\fill (x2) circle (1.5pt);
\draw (.3,0) circle (3mm);
\end{tikzpicture}}}}}}

\newcommand{\smallpropdiag}{{{\parbox{.4cm}{{\begin{tikzpicture}[scale=0.5]
\coordinate (x1) at (0,0) ;
\coordinate (x2) at  (0.6,0) ;
\fill (x1) circle (1.5pt);
\fill (x2) circle (1.5pt);
\draw (.3,0) circle (3mm);
\end{tikzpicture}}}}}}

\newcommand{\vertex}{{\parbox{1.45cm}{{\begin{tikzpicture}[scale=1]
\coordinate (x1) at (0,0) ;
\coordinate (x1p) at (0.0,0.5) ;
\coordinate (x1pL) at (-0.02,0.5) ;
\coordinate (x1pp) at  (-.35,-.35) ;
\coordinate (x1ppL) at  (-.4,-.5) ;
\coordinate (x1ppp) at  (.35,-.35) ;
\coordinate (x1pppL) at  (.35,-.5) ;
\fill (x1) circle (1.5pt);
\draw [black] (x1pp) -- (x1) -- (x1p) ;
\draw [black] (x1) -- (x1ppp) ;
\node  at (x1pL)  {$\scriptstyle a~b$} ;
\node  at (x1ppL)  {$\scriptstyle a'b'$} ;
\node  at (x1pppL)  {$\scriptstyle a''b''$} ;
\end{tikzpicture}}}}}

\newcommand{\Ctriangle}{C_\text{triag}}
\newcommand{\Cbubble}{C_\text{bub}}

\section{Introduction}
\label{sec:introduction}

In this paper we explore renormalization group flows, close to a critical dimensionality $d_c$, for two kinds of phase transitions: 
measurement-induced phase transitions in monitored quantum systems \cite{skinner2019measurement,li2018quantum,potter2022entanglement,fisher2022random}, 
and entanglement transitions in networks of random tensors \cite{hayden2016holographic,vasseur2019entanglement}.
We find that both kinds of flow have an unexpected structure, 
with the upper critical dimensionality playing a different role to that in familiar ordering transitions.  As a result, these flows do not give perturbative access to critical exponents in low dimensions. 
However they shed light on the structure of the phase diagram in high dimensions
and in mean-field-like limits.
Surprisingly, one of the flows  also  yields a new example of  perturbative dimensional reduction \cite{ParisiSourlas1979,ParisiSourlas1981}, 
a phenomenon in which a field theory's RG flow maps onto that of a simpler theory in fewer dimensions, as a result of an underlying supersymmetric structure.

Measurement-induced phase transitions (MPTs)
occur in quantum systems whose evolution is constantly   interrupted by local measurements.
Depending on the strength of monitoring, 
quantum correlations in the evolving state may either persist or be suppressed at late times \cite{skinner2019measurement,li2018quantum,unitary2019chan,
li2019measurement,
gullans2020dynamical,
jian2020measurement,
bao2020theory,
nahum2021measurement,
szyniszewski2019entanglement,
choi2020quantum,
zabalo2020critical,
noel2022measurement,
koh2022experimental,
gullans2020scalable,
zabalo2022operator,
turkeshi2020measurement,
vijay2020measurement,
tang2020measurement,
nahum2020entanglement,
li2021conformal,
lavasani2021measurement,
sang2021measurement,
jian2021measurement,
li2021statistical,
czischek2021simulating,
li2021entanglement,
li2021statistical2,
alberton2021entanglement,
turkeshi2021measurement,
ippoliti2021entanglement,
napp2022efficient,
feng2022measurement,
sierant2022universal},
affecting, for example, whether the evolution of a quantum trajectory may be efficiently simulated.
Since the  outcomes of quantum measurements are generically random, 
dynamics of this kind have formal similarities to   disordered systems in  statistical mechanics. These similarities are particularly manifest 
if the dynamics is formulated in discrete spacetime, i.e.\ as a quantum circuit. 
However, the ``disorder'' ---  the random measurement outcomes --- 
is not drawn in advance, but generated by the system via Born's~rule.

A random tensor network (RTN) that is made up of uncorrelated local tensors, 
with ``physical'' bond indices at the boundary of the network, may either be in  an entangled or a disentanged phase \cite{hayden2016holographic, vasseur2019entanglement, nahum2021measurement,
sun2022entanglement,
li2021statistical,
medina2021entanglement,
lopez2020mean,yang2022entanglement}.
The transition between the two may be driven by varying  an effective bond dimension for the network or a parameter in the probability distribution for the tensors, or even the statistics of the network structure. 
There is a close analogy with the measurement transition,
with the tensor network  (for an appropriate geometry) playing the role of the nonunitary time-evolution operator  in the measurement problem. 
In the latter case, this evolution operator can be a circuit made up of  unitaries and projection operators/Kraus operators.
These structural similarities mean that many techniques can be applied to both problems, 
including the replica trick that is ubiquitous in disordered systems \cite{vasseur2019entanglement,jian2020measurement,bao2020theory,zhou2019emergent}. 
However, there is also a structural distinction between the two problems, 
arising from the fact that the  ensemble of random evolution operators associated with a true measurement process is generated by Born's rule.
This difference has an effect on the replica lattice model \cite{jian2020measurement, nahum2021measurement}.
(The effective replica model for a dynamical process is also highly constrained in the unitary limit, i.e.\ in the absence of measurements.)

Two basic questions about these types of phase transitions are: 
(1) Do there exist simple mean-field theories that capture their critical properties in high dimensions, 
or in models with all-to-all connectivity?
(2) Are there simple Landau-Ginsburg-Wilson-like field theories that  
capture the effect of fluctuations close to a  critical dimension?
Here we address these questions using replica Landau-Ginsburg theories that were proposed in Ref.~\cite{nahum2021measurement} on grounds of symmetry, and which are explicitly derived here for a simple class of tensor networks.

The replica trick is useful   because it allows  random models to be exchanged  (at least formally) for  effective nonrandom ones.  
The replica approach in simple tensor networks \cite{vasseur2019entanglement} or circuits 
\cite{jian2020measurement,bao2020theory,zhou2019emergent} 
built with 
random Gaussian tensors or
Haar-random unitaries leads to effective lattice ``magnets'' whose  ``spins'' $\sigma$ take values in the permutation group $S_N$ (for related models without the replica limit see \cite{hayden2016holographic,nahum2018operator,von2018operator}).
The physical meaning of these degrees of freedom will be discussed below.
Here $N$ is a replica number: 
loosely speaking, it arises from the need to average arbitrary tensor powers $\check \rho^{\otimes N}$ of an appropriately-defined density matrix $\check\rho$.
In the random tensor network, physical averages of entanglement entropies or correlation functions 
may be obtained by taking the replica limit  ${N\to 0}$. In the measurement problem averages must be taken with weights from  Born's rule, and since this probability is itself expressed in terms of the density matrix, the required replica limit is  ${N\to 1}$ instead of $N\to0$ (see Ref.~\cite{potter2022entanglement} for a review).

These lattice models give a useful picture deep inside the entangled phases, where
the effective spin $\sigma$ is long-range ordered,
and entanglement entropies 
translate to  free energy costs of domain walls that are forced into the ordered state by boundary conditions. What is less apparent is how to ``coarse-grain'' these lattice models to obtain a continuum field theory that is useful close to the critical point. 
This is because the lattice model must be formulated for arbitrary $N$ in order to apply the replica trick, and the complexity of the  degrees of freedom and their interactions increases with $N$. See \cite{potter2022entanglement,fisher2022random,nahum2021measurement,vasseur2019entanglement} for further discussions: it should be noted that there are   limits where the effective model becomes solvable \cite{jian2020measurement}.

Because of the difficulty in coarse-graining these effective models, Ref.~\cite{nahum2021measurement} instead  used a formulation in terms of the ``overlap'' between distinct Feynman trajectories of a monitored system,  or between distinct configurations of bond indices in a random tensor network.
Consider, for example, quantum circuit evolution  for a spin-1/2 system with  projective measurements. This may be thought of as a discrete path integral:
the sum over bond indices needed to ``contract'' the quantum circuit (which we denote by $K$) is 
a sum over spacetime configurations $S(x,t)$ in the computational basis.
The basic object in the replica approach 
is a tensor-product evolution operator
of the schematic form
${K^{\otimes N}\otimes (K^*)^{\otimes N}}$ 
(where $K^*$ is complex-conjugated in the computational basis): this may be thought of as a multi-layer circuit with $2N$ layers. As a result the ``path integral'' involves $2N$ spacetime configurations, one for each layer.
We denote the associated configurations by $S^a(x,t)$ for the $K$ layers, and  
$\bar S^a(x,t)$
for the $K^*$ layers, with ${a=1,\ldots, N}$ in each case. 

By considering simple models one may motivate using an ${N\times N}$ ``overlap'' matrix as the order parameter for coarse-graining. Microscopically,
\be
Y_{ab}(x,t) \sim S_a(x,t) \bar S_b(x,t).
\ee
This is in the spirit of the Edwards-Anderson order parameter for a spin glass \cite{edwards1975theory}:
 $Y_{ab}(x,t)$  measures the local similarity between two configurations, one associated with a $K$ layer and one with a $K^*$ layer.
Heuristically, phase cancellation favors the local pairing of layers: when we average over measurements and/or random gates, the configurations which are most robust to phase cancellation are those in which the random phases of the path integral amplitudes cancel in pairs between $K$ and $K^*$ layers.

The ordered states mentioned above, labelled by permutations ${\sigma\in S_N}$, correspond to configurations in which the trajectory $S^a(x,t)$ in a given ``forward'' layer $a$ is locally similar to the trajectory $\bar S^{\sigma(a)}(x,t)$ in the partnered ``backward'' layer $\sigma(a)$. 
This leads to $Y^{ab}$ being largest, for given $a$, when $b=\sigma(a)$.
We can think that, in an strongly ordered state, the expectation value $\<Y_{ab}\>$ is linearly related to the matrix representing the permutation $\sigma$, though there are subtleties in this identification which we will discuss.

Similar considerations apply    to the tensor network, with a pairing field $Y_{ab}(x)$  indexed by a  coordinate $x$ that (microscopically) runs over the bonds of the network.
In either case, the replica structure leads to a global symmetry group
\be
\label{eq:symmgp}
G_N \equiv \lf S_N\times S_N \ri \rtimes \mathbb{Z}_2
\ee
for this field, with separate permutation actions on the row and column indices of $Y$ (together with symmetry under transposition \cite{zhou2020entanglement,bao2021symmetry}).
These symmetry considerations are similar for the MPT and for RTNs, but as discussed above the required replica limits are different. Surprisingly, this leads to different structures for  minimal Landau-Ginsburg-Wilson like Lagrangians in the two cases.

The ${N\to 1}$ limit, putatively relevant to the MPT, is simpler, because 
--- in analogy to standard ordering transitions ---
we can consistently isolate a piece of $Y_{ab}$ that transforms irreducibly under the global $G_N$ symmetry.
We denote the resulting field by $X_{\alpha\beta}$, since below it will be convenient to distinguish it from $Y_{ab}$. Irreducibility is ensured by the row and column-sum constraints ${\sum_{\alpha} X_{\alpha\beta}=0}$ and 
${\sum_{\beta} X_{\alpha\beta}=0}$.
(We use ${\alpha,\beta=1,\ldots, N}$ for indices  on fields that obey such row/column sum constraints, and ${a,b=1,\ldots,N}$ for fields that do not.)
The conjectured Lagrangian, for a spatially local system, is then
\be
\label{eq:defineMPTLagrangianintro}
\mathcal{L}[X]=   \sum_{\alpha,\beta=1}^N \left( \f1{2}(\nabla X_{\alpha\beta})^2 + \frac{m^2}{2}   X_{\alpha\beta}^2 + \frac g{3!} X_{\alpha\beta}^3 \right).
\ee
Both space and time derivatives have been included in the first term, and a nonuniversal velocity has been set to one.
We must study this theory in the limit ${N\to 1}$ where the matrix $X$ becomes trivial.
Our   focus in this paper will be on the critical theory, in which the renormalized mass is set to zero by tuning a microscopic parameter such as the rate of measurements.

For a random tensor network with a Euclidean geometry, 
the conjectured Lagrangian may be written in the form 
\bea
\label{eq:defineRTNLagrangianintro}
\mathcal{L} [Y] &=&  \sum_{a,b=1}^{N}  \left(  \half \left[ \nabla Y_{ab} \right]^2 
 + \frac{m^2}2 Y_{ab}^2 
+ \frac{g}{3!} Y_{ab}^3 \right)  \nn\\ 
&&  + \frac{\sigma}2\sum_{a,b,c=1}^{N}
\lf  Y_{ab} Y_{ac} + Y_{ab} Y_{cb} \ri,
\eea
with the limit ${N\to 0}$ for the size of the matrix.
Here no constraint is imposed on the row or column sums of $Y_{ab}$.
The reason for this is discussed in the following sections: it is closely analogous to phenomena in the random-field Ising model \cite{CardyBook}.
The Lagrangian in \Eq{eq:defineRTNLagrangianintro}   also applies to the measurement problem if the Born-rule probabilities are ``overridden'' by complete postselection of measurement outcomes (giving a ``forced measurement'' phase transition).

These Lagrangians were conjectured largely on symmetry grounds \cite{nahum2021measurement}.
One of our tasks is to show that they can be derived in a quantitative fashion
for a concrete ensemble of random tensor networks --- also interpretable as nonunitary circuits generalizing the ``kicked Ising model'' \cite{akila2016particle,GopalakrishnanLamacraft2019,BertiniKosProsen2018} --- by examining the two relevant limits, ${N\to 0}$ and ${N\to 1}$.
(A caveat in the latter case is that our microscopic ensemble of nonunitary circuits does not correspond to pure Born-rule dynamics,
as it does not form a  Kraus ensemble \cite{nielsen2002quantum}:  however it shares the same replica symmetry as a true measurement problem, so is an instructive toy model.)
Our main aim in this paper is to analyze the renormalization-group properties of 
Eqs.~\ref{eq:defineMPTLagrangianintro} and~\ref{eq:defineRTNLagrangianintro}, 
by working close to the critical dimensionalities 
(six and ten respectively) where the cubic interactions in these theories become marginal.

At first sight the field theories above led to an apparent paradox when compared with other results. These Lagrangians suggest that in  high dimensions, or in models with ``all-to-all'' connectivity, we should find entanglement transitions  governed by the mean-field limits of Eqs.~\ref{eq:defineMPTLagrangianintro},~\ref{eq:defineRTNLagrangianintro} \cite{nahum2021measurement}.
On the other hand, there is another setting where entanglement and measurement transitions become exactly solvable, which is on a tree tensor network \cite{nahum2021measurement,feng2022measurement} (closely related to circuit dynamics with nonlocal gates \cite{lopez2020mean,gullans2020dynamical,vijay2020measurement}). 
The analysis of Haar-random tree ensembles 
turns out to give results quite different to the above mean-field limits, with for example an essential singularity in the order parameter close to the critical point. 
It has so far been unclear how to relate these very different solvable regimes, which  might naively have been expected to be similar. 
(On the other hand, examining deterministic tree tensor networks showed that even on trees there is more than one kind of transition \cite{deterministictreeinprep}.)

\begin{figure}
\includegraphics[width=6.5cm]{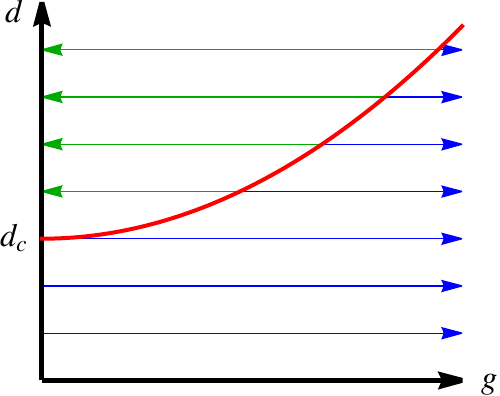}
\caption{The RG flow for \Eq{eq:MPTbetafunction}  describing the massless ${N\to 1}$ theory. The red line is an unstable (tricritical) fixed point that separates two stable universality classes for the transition: a Gaussian universality class and a strong-coupling universality class.}
\label{fig:MPTflows}
\end{figure}

The picture we find here is that, above the critical dimensions, there are at least two stable universality classes for the phase transition in ``generic'' models, i.e.\ in models lacking any special symmetry.
At weak coupling there is a stable mean-field like universality class,
while at strong coupling something else happens that is beyond the domain of simple perturbation theory.
These two stable universality classes are separated by a third universality class which is governed by a nontrivial and   perturbatively accessible unstable fixed point. 
This qualitative flow structure is shown for the theory of \Eq{eq:defineMPTLagrangianintro} in Fig.~\ref{fig:MPTflows}.
For the RTN case there is a similar flow topology for an appropriate combination of the two couplings $g$, $\sigma$ in~\Eq{eq:defineRTNLagrangianintro}.

Making a precise connection between trees and high-dimensional systems is beyond the scope of this paper, but this division into weak- and strong-coupling universality classes suggests a natural resolution of the above ``paradox'' (i.e.\ the difference between Haar-random trees and weak-coupling mean field):  
it is tempting to identify the critical behavior found previously on the tree with a version of the strong-coupling phase implied by the above RG flows.
We note that the explicit tensor network models studied here are in a different limit from  models with Gaussian random tensors or Haar-random gates studied previously: 
It is possible that the latter models are typically in the strong-coupling phase, as seems to be the case for certain   random-graph like models~\cite{nahum2021measurement}.

The ${\epsilon}$ expansion associated to 
the flow topology in Fig.~\ref{fig:MPTflows} is not  for an IR-attractive fixed point in dimensions smaller than $d_{\rm c}$, but rather for a  repulsive fixed point that appears above  the critical dimension $d_{\rm c}$, and describes   the transition between weak and strong coupling regimes.
So unfortunately we cannot use it to predict exponents in, say, {1+1} dimensions.
However, the RG behavior is  interesting in its own right. 
Surprisingly, the RTN theory (\ref{eq:defineRTNLagrangianintro}) turns out to have a hidden supersymmetric structure, which implies that the perturbative beta function in $d$ dimensions reduces to that of a simple scalar theory (in fact, the simplest possible interacting scalar theory, with a cubic interaction) in ${d-4}$ dimensions.
If we neglect terms that are subleading near the critical dimension ${d_c=10}$, the beta function for \Eq{eq:defineRTNLagrangianintro} is reduced to that of 
a superspace action of the form 
\bea\label{L-Susy-intro}
\ca S[\Upsilon] 
&=&
\int \dd^d x \int _\theta
\frac 12 \Upsilon \lf - \nabla^2_{\rm S} \ri   \Upsilon + \frac{m^2}2  \Upsilon^2 + \frac{g\sigma}{6} \Upsilon^3,
\eea
where $\Upsilon$ is a superfield that depends on both commuting and anticommuting ``spatial'' coordinates ($x$ and $\theta$ respectively) and $\nabla_S^2$ is the superspace Laplacian \cite{Wegner2016}  generalized to two anticommuting pairs of variables. 
In turn, the flow of the coupling in this theory maps onto that of a simple scalar $\Phi^3$ theory.

The original example of dimensional reduction is the random-field Ising model, where it was observed that various perturbative series
in $d$ dimensions reduced to those in the \textit{clean} Ising model in ${d-2}$ dimensions \cite{grinstein1976ferromagnetic,AharonyImryMa1976,Young1977}.
This was understood to be due to the existence of a supersymmetric formulation \cite{ParisiSourlas1979}.
In the random-field Ising model dimensional reduction holds near the upper critical dimension of six, but must fail in sufficiently low dimensions \cite{Imbrie1984,Fisher1985b,Cardy1985,BrezinDominicis1998,BrezinDeDominicis2001,Feldman2002,TissierTarjus2012b,BaczykTarjusTissierBalog2014,FytasMartin-MayorParisiPiccoSourlas2019,KavirajRychkovTrevisani2021,KavirajRychkovTrevisani2022,KavirajRychkovTrevisani2019,Rychkov2022}: this is now believed to be due to  
operators that are irrelevant near six dimensions becoming relevant somewhere around five dimensions
\cite{TissierTarjus2012b,KavirajRychkovTrevisani2022,Rychkov2022,Wiese2021}.
There is another well-known example in which dimensional reduction works not only close to the upper critical dimension, but all the way down to two dimensions:
this is the field theory describing the  statistics of random branched polymers or lattice animals \cite{ParisiSourlas1981,HsuNadlerGrassberger2005,LutherMertens2011,Cardy1985,Cardy2001,BrydgesImbrie2003,Cardy2003,
KavirajTrevisani2022}.
(In this case there is   a rigorous microscopic  demonstration of dimensional reduction in a particular model which does not use field theory \cite{BrydgesImbrie2003,Cardy2003}.)

How useful the supersymmetric formulation of the RTN field theory is remains to be seen. Here we confine ourselves to perturbative statements: in the future it will be interesting to investigate whether any non-perturbative information can be obtained.

Since the reader may query the relevance of entanglement transitions in very high dimensions, let us summarize some key outcomes.
Perhaps the most basic is that we explicitly demonstrate the relevance of the above field theories to a non-fine-tuned 
class of random tensor networks.
(In order to obtain these field theories in a controlled way we use the trick \cite{wegner1979disordered} of introducing a large number of flavors for the degrees of freedom, but in essence we consider simple tensor networks whose contraction gives rise to an effective Ising model with complex couplings.)
These derivations imply the existence of weak-coupling mean-field regimes not only in high enough dimensions but also --- more realistically --- in models with sufficiently mean-field-like connectivity. We will discuss the resulting mean-field phenomenology in more detail separately~\cite{NahumWieseToBePublished}.

The derivation   shows that the field theories serve as   effective descriptions even in low dimensions: while the   RG flow is not accessible perturbatively, other continuum techniques might be applicable. 

Another outcome is a clear demonstration that the two classes of problems we discuss have a quite different universal behavior. 
Universal differences between models involving measurements on the one hand, 
and models involving forced measurements and random tensors on the other hand, have   been  demonstrated analytically for dynamics with a tree-like structure \cite{feng2022measurement} and for free-fermion systems \cite{fava2023nonlinear,jian2023measurement}
(which show their own, distinct, kinds of measurement-induced behavior \cite{cao2019entanglement,nahum2020entanglement,alberton2021entanglement,jian2022criticality,turkeshi2021measurementinduced,buchold2021effective,sang2021measurement,kells2021topological,turkeshi2022enhanced,turkeshi2022entanglement,muller2022measurement,turkeshi2022entanglementand,lucas2022generalized,coppola2022growth,gal2022volume,merritt2022entanglement,chen2020emergent,bao2021symmetry}).

In the next sections we  plunge into the RG flows, first for the simpler case of  models with the replica symmetry of the measurement transition (${N\to 1}$) in Sec.~\ref{sec:Nequals1RG}.

We then continue to random tensor networks/forced measurements for which the relevant limit ${N\to 0}$  is more involved:
in Sec.~\ref{s:FTforNt00} 
we consider a direct perturbation expansion. The aim of the following Sec.~\ref{s:Supersymmetry and dimensional reduction} is  to give a supersymmetric formulation showing that --- perturbative --- dimensional reduction holds to all orders. 

Readers who wish to know what kind of concrete lattice models can be used to obtain these theories may wish to skip ahead to section \ref{sec:derivefieldtheory}, where 
a   concrete derivation of the field theories is given.

\section{RG for the MPT Lagrangian \protect{($N=1$)}}
\label{sec:Nequals1RG}

The putative Lagrangian for the MPT is 
\be\label{eq:defineMPTLagrangian}
\mathcal{L}(X_{\alpha\beta})=   \sum_{\alpha,\beta=1}^N \f1{2}(\nabla X_{\alpha\beta})^2 + \frac{m^2}{2}   X_{\alpha\beta}^2 + \frac g{3!} X_{\alpha\beta}^3, 
\ee
where the ${N\times N}$ matrix $X$ encodes the pairing between Feynman trajectories (Sec.~\ref{sec:introduction}),
and the $N$-dependence is non-trivial because of constraints on the fields:
\be
\sum_{\alpha=1}^N X_{\alpha \beta} =  \sum_{\beta=1}^N X_{\alpha\beta}=0 .
\ee
Space and time coordinates are grouped together (although the physical dynamics is in real time, the spacetime signature  is Euclidean).
Symmetry arguments for this form can be found in Ref.~\cite{nahum2021measurement}, and a derivation in a class of nonunitary circuits will be given in Sec.~\ref{sec:derivefieldtheory} (for which the bare strength of the interactions may be made arbitrarily small). Only the interaction that is most relevant in high dimensions is retained. The microscopic derivation shows that the coupling constant $g$ for this interaction is real and negative.
The replica limit is ${N\to 1}$, where the field and the free energy become trivial.

At the mean-field level, this action has a disordered phase for positive mass-squared (representing the strong-monitoring, or disentangled phase), where  ${\<X\>=0}$, and an ordered (weak-monitoring/disentangled) phase for negative mass-squared, 
in which the left/right permutational symmetry in \Eq{eq:symmgp} is broken down to a diagonal subgroup: for example one of the equivalent ordered states is ${\<X_{\alpha\beta}\> = (2 m^2/g) \lf N \delta_{\alpha\beta}-1\ri}$.
Our focus is on the transition between these two phases, where the renormalized mass vanishes.

Before discussing RG for this theory it is useful to recall some simpler theories.
The structure of  Feynman diagams for the Lagrangian (\ref{eq:defineMPTLagrangian}) is 
closely related to that for the analogous theory for a field with only a single index:
\bea\label{eq:definePottsLagrangian}
\mathcal{L}(\Phi_\alpha) &=&  \sum_{\alpha=1}^N \f1{2}(\nabla \Phi_{\alpha})^2 + \frac{m^2}{2}   \Phi_{\alpha}^2 + \frac g{3!} \Phi_{\alpha}^3, 
 \eea
again with the constraint ${\sum_{\alpha=1}^N \Phi_\alpha=0}$.
This is the standard Lagrangian for the $N$-state Potts model \cite{ZiaWallace1975,Amit1976}, and specifically for the ${N\rightarrow 1}$ limit of Potts which describes percolation.\footnote{In order to describe the phase transition of the Potts model with $N>1$ we would need to add higher-order terms to stabilize the potential.}
The $S_N$ symmetry of the Potts model acts by permuting the $N$ components of the order parameter $\Phi_\alpha$, and the constraint ${\sum_\alpha \Phi_\alpha=0}$ isolates an irreducible representation of this symmetry.

An even simpler theory is a  scalar field $\Phi$ (without any index) with a cubic interaction,
\be\label{eq:yangleeLagrangian}
\mathcal{L}(\Phi) = \f{1}{2} (\nabla \Phi)^2 + \f{m^2}{2} \Phi^2 +  \f{g}{3!} \Phi^3.
\ee
This is the Yang-Lee theory: in this case, unlike the cases of interest to us, the coupling $g$ is usually taken to be imaginary.

For any cubic theory  (with a single cubic coupling), 
a one-loop calculation in $6-\epsilon$ dimensions gives  \cite{Amit1976,Zinn-Justin2021} 
\be
\f{\dd g }{\dd \ell } = \f{\epsilon}{2} g - \lf \f{1}{4} C_\smallpropdiag - C_\smalltriagdiag \ri g^3,
\ee
(after a standard rescaling of the coupling constant\footnote{In discussing the perturbative RG we absorb a factor via $g \mapsto g (4\pi)^{+d/4}$ (also in Sec.~\ref{s:FTforNt00}).})
with $\ell$ the RG time, i.e.\ the logarithm of a lengthscale.
Here $C_\smallpropdiag$ and $C_\smalltriagdiag$ are group-theoretic factors
associated with the
one-loop diagrams that renormalize the propagator and the three-point vertex, respectively, i.e.\ the diagrams of the schematic form
\ba
&
\begin{tikzpicture}[scale=1]
\coordinate (x1) at (0,0) ;
\coordinate (x2) at  (0.6,0) ;
\coordinate (x1p) at  (-.2,0) ;
\coordinate (x2p) at  (0.8,0) ;
\fill (x1) circle (1.5pt);
\fill (x2) circle (1.5pt);
\draw (.3,0) circle (3mm);
\draw [black] (x1) -- (x1p);
\draw [black] (x2) -- (x2p);
\end{tikzpicture}~
,
&
&
\begin{tikzpicture}[scale=1]
\coordinate (x1) at (0,0) ;
\coordinate (x2) at  (0.7,0) ;
\coordinate (x3) at  (0.35,0.5) ;
\coordinate (x1p) at (-.2,-0.2) ;
\coordinate (x2p) at  (0.9,-.2) ;
\coordinate (x3p) at  (0.35,0.8) ;
\fill (x1) circle (1.5pt);
\fill (x2) circle (1.5pt);
\fill (x3) circle (1.5pt);
\draw [black] (x1) -- (x2) -- (x3) -- (x1);
\draw [black] (x1) -- (x1p) ;
\draw [black] (x2) -- (x2p) ;
\draw [black] (x3) -- (x3p) ;
\end{tikzpicture}~.
\end{align}
These combinatorial factors are defined such that, for the simplest case of a single scalar field without any index (Yang-Lee), we have  ${C_\smallpropdiag=C_\smalltriagdiag=1}$.
The Yang-Lee case therefore gives 
\ba
\label{beta-YL}
\f{\dd g }{\dd \ell } & = 
 \f{{\epsilon }}{2} g + \f{3}{4} g^3
 & 
 &\text{for Yang-Lee}.
\end{align}
This theory has a fixed point for \textit{imaginary} $g$ in $6-\epsilon$ dimensions, which is relevant to the Ising model in an imaginary magnetic field. 

For $N$-state Potts, 
\ba\label{eq:CvalsforPotts}
C_\smallpropdiag^{\text{Potts}} & = 1-\frac2N,
&  C_\smalltriagdiag^{\text{Potts}} &  = 1-\frac3N.
\end{align}
These nontrivial combinatorial factors arise from the constraint ${\sum_\alpha \Phi_\alpha=0}$, which leads to a non-diagonal propagator 
${\<\Phi_\alpha\Phi_{\alpha'}\> \propto (\delta_{\alpha{\alpha'}} -\frac1N)}$.
The calculation of these constants is reviewed in App.~\ref{s:Potts-comb}.
With the values in Eq.~(\ref{eq:CvalsforPotts}), the cubic term in the Potts $\beta$- function is negative in the limit $N\rightarrow 1$:
\ba
 \f{\dd g }{\dd \ell } &  = 
 \f{\epsilon}{2} g - \f{7}{4} g^3
 & 
 &\text{for percolation}.
\end{align}
This is consistent with the existence of a nontrivial fixed point in $6-\epsilon$ dimensions for percolation (which is described by the theory with real $g$).

Now consider the matrix Lagrangian (\ref{eq:defineMPTLagrangian}). 
Propagators now carry both ``row''   and ``column'' indices, 
arising from the row and column indices of the field $X_{\alpha\beta}$.
The evaluation of any given diagram involves contractions of these indices.
Crucially, the contractions of the row and column indices are independent, 
and each gives the same group theoretic factor as in the Potts case.
To see this, let us represent the propagator as a double line,
\be\label{eq:matrixtheorypropagator}
\left< X_{\alpha \beta}X_{\alpha'\beta'}\right> = 
{\fboxsep0mm{\parbox{1.8cm}{{\begin{tikzpicture}[scale=1]
\coordinate (x1) at (0,0.02) ;
\coordinate (x2) at  (0.8,0.02) ;
\coordinate (x1p) at  (0,0.25) ;
\coordinate (x2p) at  (0.8,0.25) ;
\node [left] at (x1)  {$\alpha$} ;
\node [right] at (x2)  {$\alpha'$} ;
\node [left] at (x1p)  {$\beta$} ;
\node [right] at (x2p)  {$\beta'$} ;
\draw [blue] (x1) -- (x2);
\draw [blue] (x1p) -- (x2p);
\end{tikzpicture}}}}}
= \f{( \delta_{\alpha\alpha'}-\frac1N)( \delta_{\beta\beta'}-\frac1N)}{k^2 + m^2},
\ee
then the triangle diagram (for example) generalizes from Potts to
\be
{\parbox{1.1cm}{{\begin{tikzpicture}[scale=1]
\coordinate (x1) at (0,0) ;
\coordinate (x2) at  (0.7,0) ;
\coordinate (x3) at  (0.35,0.5) ;
\coordinate (x1p) at (-.2,-0.2) ;
\coordinate (x2p) at  (0.9,-.2) ;
\coordinate (x3p) at  (0.35,0.8) ;
\fill (x1) circle (1.5pt);
\fill (x2) circle (1.5pt);
\fill (x3) circle (1.5pt);
\draw [black] (x1) -- (x2) -- (x3) -- (x1);
\draw [black] (x1) -- (x1p) ;
\draw [black] (x2) -- (x2p) ;
\draw [black] (x3) -- (x3p) ;
\end{tikzpicture}}}}
\longrightarrow
{\parbox{1.3cm}{{\begin{tikzpicture}[scale=1]
\coordinate (x1) at (0,0) ;
\coordinate (x2) at  (0.7,0) ;
\coordinate (x3) at  (0.35,0.5) ;
\coordinate (x1p) at (-.2,-0.2) ;
\coordinate (x2p) at  (0.9,-.2) ;
\coordinate (x3p) at  (0.35,0.8) ;
\coordinate (x1R) at (0.2,0.1) ;
\coordinate (x2R) at  (0.9,0.1) ;
\coordinate (x3R) at  (0.55,0.6) ;
\coordinate (x1pR) at (0,-.1) ;
\coordinate (x2pR) at  (1.1,-.1) ;
\coordinate (x3pR) at  (0.55,0.9) ;
\fill (x1) circle (1.5pt);
\fill (x2) circle (1.5pt);
\fill (x3) circle (1.5pt);
\fill (x1R) circle (1.5pt);
\fill (x2R) circle (1.5pt);
\fill (x3R) circle (1.5pt);
\draw [blue] (x1) -- (x2) -- (x3) -- (x1);
\draw [blue] (x1R) -- (x2R) -- (x3R) -- (x1R);
\draw [blue] (x1) -- (x1p) ;
\draw [blue] (x2) -- (x2p) ;
\draw [blue] (x3) -- (x3p) ;
\draw [blue] (x1R) -- (x1pR) ;
\draw [blue] (x2R) -- (x2pR) ;
\draw [blue] (x3R) -- (x3pR) ;
\end{tikzpicture}}}}~,
\ee
from which one can see that the index contractions happen separately in the two ``layers''.
As a result, the group theoretic factor for any given diagram is simply the square of that for the Potts model. In particular\footnote{If we took $X$ to be an $N_1\times N_2$ matrix, the combinatorial factor would be the product of that for the $N_1$-state Potts model and that for the $N_2$-state Potts model.}
\ba\label{eq:constantsgetsquared}
C_\smallpropdiag^\text{MPT} & = \lf C_\smallpropdiag^\text{Potts}  \ri^2, 
&
C_\smalltriagdiag^\text{MPT} & = \lf C_\smalltriagdiag^\text{Potts}  \ri^2.
\end{align}
Given the similarity of (\ref{eq:defineMPTLagrangian}) to the Potts Lagrangian, we might at first sight have expected the two theories to show a similar phenomenology in the ${N\rightarrow 1}$ limit. 
However this is not the case, because the squaring operation in \Eq{eq:constantsgetsquared}
changes the sign of the one-loop term in the beta function:
\ba\label{eq:MPTbetafunction}
 \f{\dd g }{\dd \ell } &  = 
 \f{\epsilon}{2} g + \f{15}{4} g^3
 & 
 & \text{for the MPT Lagrangian}.
\end{align}
The theory of interest to us has real $g$. 
Therefore the beta function above shows that there is no perturbatively accessible fixed point below six dimensions.

However, we \textit{do} have a perturbatively accessible fixed point \textit{above} six dimensions.
The schematic RG  flow for \Eq{eq:MPTbetafunction} is shown in Fig.~\ref{fig:MPTflows}.
Note that the nontrivial fixed point is unstable with respect to variations in $g$.
Since the mass is also a relevant perturbation, this is   a tricritical point, reached by tuning two parameters. 

These flows suggest an interesting possibility for the phenomenology of the measurement phase transition.
The simplest interpretation is as follows. 

To start with, both above and below six dimensions 
we have a stable disentangled phase (positive $m^2$) 
and a stable entangled phase (negative $m^2$):
here we are discussing the transition between these phases, obtained by tuning the renormalized mass to zero.

Above six dimensions,
the Gaussian fixed point ($g=0$) is stable, 
indicating that, 
without fine-tuning, we can have a phase transition with mean-field exponents 
(see below).
The transition will be in this universality class if the bare value of the coupling $g$ is small enough, lying in the basin of attraction of~${g=0}$.

However, the nontrivial fixed point ${g_*^2 = \frac2{15} (d-6)}$
in ${d>6}$ is a \textit{tricritical} point
which separates the mean-field universality class from 
an alternative  ``strong-coupling'' universality class for the transition.
Here we are hypothesising that the flow to large positive $g$ in Fig.~\ref{fig:MPTflows} ends at a distinct strong-coupling fixed point, which is  not accessible in our perturbative calculation.
(Calculations on trees may give hints as to the nature of this phase in high dimensions, but this requires investigation.)
In principle, in an appropriate model with two tuning parameters 
it should be possible to access both the mean-field and the ``strong-coupling'' universality classes for the phase transition, and the tricritical point separating them.
By contrast, for ${d\leq 6}$ the Gaussian fixed point is unstable, so that only the putative strong-coupling fixed point survives.

Interestingly, these RG flows are reminiscent of the flows for a directed polymer in a random medium in $d+1$ dimensions, 
or equivalently for Kardar-Parisi-Zhang (KPZ) surface growth in $d$ spatial dimensions.
The  nonlinearity in the KPZ equation has a flow diagram similar to Fig.~\ref{fig:MPTflows}, but with a critical dimensionality $d$ equal to two \cite{kardar2007statistical}. A difference is that here the mass-squared $m^2$ must be tuned to zero in order to be in the critical plane shown in  Fig.~\ref{fig:MPTflows}, whereas    KPZ growth is scale invariant (corresponding formally to  a massless field theory) without the need to tune a parameter. However  a nonzero $m^2$ can   be given an interpretation in the KPZ context,~see \cite{MukerjeeBonachelaMunozWiese2022,MukerjeeWiese2022} for   quenched~KPZ.

Finally, we briefly discuss exponents. 
The basic  exponents of the 
Gaussian universality class 
(which is stable for $d>6$)
are given by mean-field theory.
For example, the order parameter in the ordered phase scales as 
${\<X\> \sim (r_c-r)^\beta}$, 
where $r$ is the microscopic parameter that ultimately controls the mass in the effective theory, and ${\beta=1}$.
Scaling forms for entanglement entropies just inside the ordered phase  may be obtained by considering systems with domain-wall boundary conditions and are given in \cite{nahum2021measurement} --- for example the entropy density coefficient associated with the entanglement volume law  (or with 
the amount of quantum information preserved at late times \cite{choi2020quantum,gullans2020dynamical})
scales as 
\be
s\sim (r_c-r)^{5/2}.
\ee
Slightly inside the disordered phase,
there is a characteristic length (or time) scale 
${\xi\sim |r_c-r|^{-\nu}}$ 
with ${\nu=1/2}$,
and two-point functions obey Wick's theorem.
These exponents happen to be the same as those in a percolation toy model for the measurement transition in high dimensions \cite{gullans2020dynamical,nahum2021measurement}, since the latter is also governed by a cubic theory that flows to its Gaussian fixed point (albeit a different cubic theory).

For the nontrivial (unstable) fixed point above six dimensions,
the exponents are corrected from their mean-field values at order ${d-6}$.
Exponents at the putative strong coupling fixed point are not known, and would have to be studied by another method.

\section{Random tensor networks/forced measurements: $N=0$ theory}
\label{s:FTforNt00}

The action proposed for the random tensor network contains an $N\times N$ dimensional matrix, and we are interested in the limit $N\rightarrow 0$: 
\bea\label{eq:S-N=0}
\mathcal{L} [Y] &=&  \sum_{a,b=1}^{N}   \half \left[ \nabla Y_{ab}(x) \right]^2 
+ \frac{m^2}2 Y_{ab}(x)^2 + \frac{g}{3!} Y_{ab}^3 \nn\\ 
&&  + \frac{\sigma_{\rm R}}2
\sum_{a,b,c=1}^{N} Y_{ab}(x) Y_{ac}(x) \nn \\
&& +\frac{\sigma_{\rm L}}2 \sum_{a,b,c=1}^{N} Y_{ab}(x) Y_{cb}(x).
\eea
The theory of interest has ${\sigma_{\rm L} =  \sigma_{\rm R} = \sigma}$, but below it is convenient to distinguish the two couplings.
A linear term in $Y$, of the form ${r \sum_{a,b}Y_{ab}}$ is now also allowed by symmetry
(in contrast to the previous theory  where such a term  vanishes). However, we have the freedom to shift $Y$ by a constant, which reduces the number of independent couplings by one.   In writing \Eq{eq:S-N=0} we have assumed that we are either in the disordered phase (${m^2>0}$) or at the critical point (${m^2=0}$), and have shifted the field so that the linear term vanishes.
We   focus on the RG flow for the critical theory: the mean-field behavior will be described elsewhere \cite{NahumWieseToBePublished}.
The microscopic derivation of \Eq{eq:S-N=0} for a particular family of tensor networks in Sec.~\ref{sec:derivefieldtheory} shows that $\sigma$ is positive and $g$ is negative; 
however, the formal RG flows are independent of these signs. 

The fact that we need to retain the quadratic $\sigma$ terms in the critical theory may seem counter-intuitive,
but is closely analogous to what happens in the  replica field theory for the random-field Ising model, which is discussed in  Refs.~\cite{Cardy2003,CardyBook,KavirajRychkovTrevisani2019,KavirajRychkovTrevisani2021,Wiese2021,Rychkov2022}.
In the massless free theory, and in the limit ${N\to 0}$, the effect of  $\sigma$ is to 
split $Y$ into modes with distinct scaling dimensions
(see for example \Eq{eq:Ytwopointfunction} below), 
and this modifies the usual classification of couplings as relevant or irrelevant. 
In Sec.~\ref{s:Supersymmetry and dimensional reduction} we will discuss an explicit change of basis for the fields \cite{Cardy1985} that makes the assignment of engineering dimensions straightforward.
Here we instead perform a direct diagrammatic analysis, but before getting to this we  anticipate two results that are most easily understood in the language of  Sec.~\ref{s:Supersymmetry and dimensional reduction} (though  also obtainable diagrammatically):

First, the critical dimension of the above theory is 10, 
rather than the 6 of more conventional cubic theories such as that in the previous section. 
We will see shortly that this is consistent with the beta function we derive.
Second,  other couplings that could have been written down in Eq.~\eq{eq:S-N=0}, for example the alternate cubic interactions
\ba
& \sum_{a,b}\sum_{a',b'} Y_{ab} Y_{a'b'}^2, 
&
& \sum_{a,b}\sum_{a',b'} \sum_{a'', b''} Y_{ab}Y_{a'b'} Y_{a''b''}, 
\end{align}
are less relevant   near the Gaussian fixed point. 
(See \cite{nahum2021measurement} or  Sec.~\ref{s:Cardy fields and supersymmetry} for the justification.)

The Lagrangian (\ref{eq:S-N=0}) has a remarkable dimensional reduction property: 
after discarding subleading corrections, 
various perturbative calculations in this theory in $d$ dimensions,
including the calculation of the beta function,
 reduce to the analogous calculations in the {\it scalar} Yang-Lee theory (\ref{eq:yangleeLagrangian}) in ${d-4}$ dimensions   with the (real) coupling 
$ g_\mathrm{YL}= g \sigma$.
This dimensional reduction by 4 generalizes the dimensional reduction by 2 that occurs, at least perturbatively,  in various replica-like theories with a single replica index \cite{ParisiSourlas1979,ParisiSourlas1981,Cardy1982,Cardy1983}.
Our aim in this section is to  motivate the result by showing schematically how dimensional reduction works
for the one-loop diagrams that renormalize the coupling  in Eq.~(\ref{eq:S-N=0}), using dimensional regularization. 
In the next section we give a non-diagrammatic proof of dimensional reduction using supersymmetry.

As will be clear below, the effective coupling is $g\sigma$, rather than $g$, because $g$ always appears together with $\sigma$ in the leading diagrams. This is consistent with dimensional analysis: 
in a cubic theory the canonical momentum dimension of $g$ is  ${[g]_{m}= 3-d/2}$, so it is not $g$  but $g\sigma$ which has canonical dimension ${\frac\epsilon2 = \frac{10-d}{2}}$ and is marginal in ${d=10}$,  as required by dimensional reduction to a scalar theory which is marginal in six dimensions.

For later reference, let us first consider the renormalization of the cubic vertex in a simple scalar theory, arising from the diagram
{\parbox{3.5mm}{$\begin{tikzpicture}[scale=0.3]
\coordinate (x1) at (0,0) ;
\coordinate (x2) at  (0.7,0) ;
\coordinate (x3) at  (0.35,0.5) ;
\coordinate (x1p) at (-.2,-0.2) ;
\coordinate (x2p) at  (0.9,-.2) ;
\coordinate (x3p) at  (0.35,0.8) ;
\fill (x1) circle (1.5pt);
\fill (x2) circle (1.5pt);
\fill (x3) circle (1.5pt);
\draw [black] (x1) -- (x2) -- (x3) -- (x1);
\draw [black] (x1) -- (x1p) ;
\draw [black] (x2) -- (x2p) ;
\draw [black] (x3) -- (x3p) ;
\end{tikzpicture}$}} at zero external momentum.
This involves the momentum integral
\be\label{eq:scalartrianglediagram}
\triagdiag  := \int_{k}\frac{1}{(k^2+m^2)^{3}}
=\frac{m^{d-6}}2 \,\Gamma\!\left(3-\frac d2 \right).
\ee
We will review the computation of the momentum integral below. Our convention is to normalize such integrals so that
\be
\int_{k}\rme^{{-a k^{2}}} := \int\frac{\rmd ^d k}{\pi^{d/2}} \rme^{{-a k^{2}}} = a^{{-d/2}}\ .
\ee
Before discussing the analogous diagrams in the matrix theory  (\ref{eq:S-N=0}), let us fix  the diagrammatic conventions.
The propagator is defined using only 
the diagonal part of the quadratic Lagrangian in \Eq{eq:S-N=0}. To avoid clutter we represent it by a single line, which carries both row and  column indices, reflecting the matrix structure of $Y_{ab}$:
\be
\< Y_{ab} Y_{a'b'}\>=
{\fboxsep0mm{\parbox{1.75cm}{{\begin{tikzpicture}[scale=1]
\coordinate (x1) at (0,0) ;
\coordinate (x2) at  (1,0) ;
\coordinate (x1LU) at (-0.2,0.07) ;
\coordinate (x1LD) at (-0.2,-0.1) ;
\coordinate (x2RU) at  (1.2,0.1) ;
\coordinate (x2RD) at  (1.2,-0.1) ;
\node [above] at (x1LD)  {$\scriptstyle a$} ;
\node [below] at (x1LU)  {$\scriptstyle b$} ;
\node [above] at (x2RD)  {$\scriptstyle a'$} ;
\node [below] at (x2RU)  {$\scriptstyle b'$} ;
\draw [black] (x1) -- (x2);
\end{tikzpicture}}}}}
= \f{\delta_{a,a'}\delta_{b,b'}}{k^2 + m^2}.
\ee
The couplings $\sigma_{L}$ and $\sigma_{\rm R}$ are taken into account as Feynman vertices that are off-diagonal in either the row or column index. We represent these with either a red (lighter) or blue (darker) dot,
\begin{align}
&{\fboxsep0mm{\parbox{1.2cm}{{
\begin{tikzpicture}[scale=1]
\coordinate (x1) at (0,0) ;
\coordinate (x2) at  (0.4,0) ;
\coordinate (x1LU) at (-0.2,0.07) ;
\coordinate (x1LD) at (-0.2,-0.1) ;
\coordinate (x2RU) at  (0.6,0.1) ;
\coordinate (x2RD) at  (0.6,-0.1) ;
\coordinate (xdot) at  (0.2,0) ;
\node [above] at (x1LD)  {$\scriptstyle a$} ;
\node [below] at (x1LU)  {$\scriptstyle b$} ;
\node [above] at (x2RD)  {$\scriptstyle a'$} ;
\node [below] at (x2RU)  {$\scriptstyle b'$} ;
\draw [black] (x1) -- (x2);
\fill [red] (xdot) circle (1.5pt);
\end{tikzpicture}}}}}
:= \delta_{a,a'},
&
&{\fboxsep0mm{\parbox{1.2cm}{{
\begin{tikzpicture}[scale=1]
\coordinate (x1) at (0,0) ;
\coordinate (x2) at  (0.4,0) ;
\coordinate (x1LU) at (-0.2,0.07) ;
\coordinate (x1LD) at (-0.2,-0.1) ;
\coordinate (x2RU) at  (0.6,0.1) ;
\coordinate (x2RD) at  (0.6,-0.1) ;
\coordinate (xdot) at  (0.2,0) ;
\node [above] at (x1LD)  {$\scriptstyle a$} ;
\node [below] at (x1LU)  {$\scriptstyle b$} ;
\node [above] at (x2RD)  {$\scriptstyle a'$} ;
\node [below] at (x2RU)  {$\scriptstyle b'$} ;
\draw [black] (x1) -- (x2);
\fill [blue] (xdot) circle (1.5pt);
\end{tikzpicture}}}}}
:= \delta_{b,b'}.
\end{align}
For later convenience we do not include the factors of the couplings in the definitions of the diagrammatic vertices.

The interaction vertex is diagonal in all indices, 
\be
\vertex= 
\delta_{a,a',a''}\,
\delta_{b,b',b''}\, ,
\ee
where $\delta_{a,a',a''}$ is one if all indices agree and zero otherwise.

Now consider the diagram 
\be\label{eq:scalarliketriangle}
{\fboxsep0mm{\parbox{2.1cm}{{
\begin{tikzpicture}[scale=1]
\coordinate (x1) at (0,0) ;
\coordinate (x2) at  (0.7,0) ;
\coordinate (x3) at  (0.35,0.5) ;
\coordinate (x1p) at (-.2,-0.2) ;
\coordinate (x2p) at  (0.9,-.2) ;
\coordinate (x3p) at  (0.35,0.8) ;
\fill (x1) circle (1.5pt);
\fill (x2) circle (1.5pt);
\fill (x3) circle (1.5pt);
\draw [black] (x1) -- (x2) -- (x3) -- (x1);
\draw [black] (x1) -- (x1p) ;
\draw [black] (x2) -- (x2p) ;
\draw [black] (x3) -- (x3p) ;
\coordinate (x1L) at (0.33,0.8) ;
\coordinate (x2L) at  (-0.26,-0.4) ;
\coordinate (x3L) at  (1.21,-0.4) ;
\node  at (x1L)  {$\scriptstyle a\;b$} ;
\node  at (x2L)  {$\scriptstyle a'b'$} ;
\node  at (x3L)  {$\scriptstyle a''b''$} ;
\end{tikzpicture}}}}}
=  \vertex ~~\times~~ \triagdiag
\ee
which does not include any $\sigma$  vertices (red or blue dots).
This diagram is present in the theory, but it is subdominant  in the IR limit, i.e.\ for small momentum, or equivalently at small mass which we consider here.
The dominant diagrams (for a given topology) are those with the  
\textit{maximal} number of red and blue dots.  
This can be seen by  dimensional analysis: $\sigma_{\rm L,R}$ has the same dimension as $m^2$ or $k^2$, so each additional factor of $\sigma_{\rm L,R}$ brings with it an additional dimensional factor of $1/m^2$.

However, 
the maximal number of dots is limited by the fact that 
${\sum_i 1=N}$ vanishes in the replica limit.
For example, inserting two red dots on the same propagator line gives 
a free index sum, which vanishes.
If we are interested in the renormalization of the coupling $g$, then we must also ensure that the index structure of the resulting diagram agrees with the RHS of \Eq{eq:scalarliketriangle}
(i.e.\ is proportional to ${\delta_{a,a',a''}
\delta_{b,b',b''}}$).
Then, we can add at most one red and one blue dot {\em per loop}.

The dominant 1-loop diagrams therefore give the following momentum  integral:
\ba
\label{triangle-with-dots-1}
{6 {{\parbox{0.8cm}{{\begin{tikzpicture}[scale=1]
\coordinate (x1) at (0,0) ;
\coordinate (x12) at (0.35,0) ;
\coordinate (x2) at  (0.7,0) ;
\coordinate (x3) at  (0.35,0.5) ;
\coordinate (x13) at (0.175,0.25) ;
\fill (x1) circle (1.5pt);
\fill (x2) circle (1.5pt);
\fill (x3) circle (1.5pt);
\fill [blue] (x12) circle (1.5pt);
\fill [red] (x13) circle (1.5pt);
\draw [black] (x1) -- (x2) -- (x3) -- (x1);
\end{tikzpicture}}}}}
+6{{\parbox{0.8cm}{{\begin{tikzpicture}[scale=1]
\coordinate (x1) at (0,0) ;
\coordinate (x12a) at (0.25,0) ;
\coordinate (x12b) at (0.45,0) ;
\coordinate (x2) at  (0.7,0) ;
\coordinate (x3) at  (0.35,0.5) ;
\fill (x1) circle (1.5pt);
\fill (x2) circle (1.5pt);
\fill (x3) circle (1.5pt);
\fill [blue] (x12a) circle (1.5pt);
\fill [red] (x12b) circle (1.5pt);
\draw [black] (x1) -- (x2) -- (x3) -- (x1);
\end{tikzpicture}}}}}
 = \int_{k}\frac{12}{(k^2+m^2)^{5}}}
 = \frac{m^{d-10}}2 \,\Gamma\!\left(5-\frac d2 \right).
\end{align}
In the first of the two diagrams in 
(\ref{triangle-with-dots-1}) the 
combinatorial factors are $3 \times 2$ to place the red and blue dot, 
and in the second they are
$3$ to choose the line where the two dots are placed, times 2 for their order.

\Eqs{eq:scalartrianglediagram} and 
\eqref{triangle-with-dots-1} are identical up to a shift in dimension by 4,
\be\label{eq:triangleequivalence}
\bigg[6 {{\parbox{0.8cm}{{\begin{tikzpicture}[scale=1]
\coordinate (x1) at (0,0) ;
\coordinate (x12) at (0.35,0) ;
\coordinate (x2) at  (0.7,0) ;
\coordinate (x3) at  (0.35,0.5) ;
\coordinate (x13) at (0.175,0.25) ;
\fill (x1) circle (1.5pt);
\fill (x2) circle (1.5pt);
\fill (x3) circle (1.5pt);
\fill [blue] (x12) circle (1.5pt);
\fill [red] (x13) circle (1.5pt);
\draw [black] (x1) -- (x2) -- (x3) -- (x1);
\end{tikzpicture}}}}}
+6{{\parbox{0.8cm}{{\begin{tikzpicture}[scale=1]
\coordinate (x1) at (0,0) ;
\coordinate (x12a) at (0.25,0) ;
\coordinate (x12b) at (0.45,0) ;
\coordinate (x2) at  (0.7,0) ;
\coordinate (x3) at  (0.35,0.5) ;
\fill (x1) circle (1.5pt);
\fill (x2) circle (1.5pt);
\fill (x3) circle (1.5pt);
\fill [blue] (x12a) circle (1.5pt);
\fill [red] (x12b) circle (1.5pt);
\draw [black] (x1) -- (x2) -- (x3) -- (x1);
\end{tikzpicture}}}}}\bigg] _{d-4}
= \bigg[ \triagdiag \bigg]_d .
\ee
To see why this has happened, consider the representation of the momentum integrals in terms of  Schwinger parameters, using
\be
\f{1}{(k^2+m^2)^q} =  \int_{s>0}    \f{s^{q-1}}{(q-1)!}   e^{-s(k^2+m^2)}
\ee
with  ${\int_{s>0} = \int_0^\infty \dd s}$.  For the scalar diagram, 
\ba \nn
\triagdiag  & 
= \int_{k}\frac{1}{(k^2+m^2)^{3}}
\\
\nn
& =  \int_{s>0} \frac{s^{2}}{2!}\int_{k}\rme^{{-s (k^ 2+m^2)}}
\\
& = \frac{1}{2}\int_{s>0} s^{2-d/2}\rme^{{-s m^2}} ,
\end{align}
which gives the result in \Eq{eq:scalartrianglediagram}.
For the matrix diagram,
\bea
\label{triangle-with-dots}
{6 {{\parbox{0.8cm}{{\begin{tikzpicture}[scale=1]
\coordinate (x1) at (0,0) ;
\coordinate (x12) at (0.35,0) ;
\coordinate (x2) at  (0.7,0) ;
\coordinate (x3) at  (0.35,0.5) ;
\coordinate (x13) at (0.175,0.25) ;
\fill (x1) circle (1.5pt);
\fill (x2) circle (1.5pt);
\fill (x3) circle (1.5pt);
\fill [blue] (x12) circle (1.5pt);
\fill [red] (x13) circle (1.5pt);
\draw [black] (x1) -- (x2) -- (x3) -- (x1);
\end{tikzpicture}}}}}
+6{{\parbox{0.8cm}{{\begin{tikzpicture}[scale=1]
\coordinate (x1) at (0,0) ;
\coordinate (x12a) at (0.25,0) ;
\coordinate (x12b) at (0.45,0) ;
\coordinate (x2) at  (0.7,0) ;
\coordinate (x3) at  (0.35,0.5) ;
\fill (x1) circle (1.5pt);
\fill (x2) circle (1.5pt);
\fill (x3) circle (1.5pt);
\fill [blue] (x12a) circle (1.5pt);
\fill [red] (x12b) circle (1.5pt);
\draw [black] (x1) -- (x2) -- (x3) -- (x1);
\end{tikzpicture}}}}}}
 &=& 12 \int_{k}\frac{1}{(k^2+m^2)^{5}}\nn\\
&=& 12 \int_{s>0} \frac{s^{4}}{4!}\int_{k}\rme^{{-s  (k^ 2+m^2)}}\nn\\
&=&\frac{1}{2}\int_{s>0} s^{4-d/2}\rme^{{-s m^2}},
\eea
which gives the result of \Eq{triangle-with-dots-1}.
Interestingly, the combinatorial factors due to the different ways to insert the two dots into the triangle are compensated by the factorial factors in the representation of the propagator (the factors of $s^4/4!$ and $s^2/2!$ above).

Let us also consider the factors of the couplings. 
In the scalar theory with coupling $g_\mathrm{YL}$, the triangle diagram has a factor $g_\mathrm{YL}^2$ relative to the bare coupling.
In the matrix theory the diagram  again has a factor of $g^2$ relative to the bare vertex, but it also comes  with a factor $\sigma_{\rm L}\sigma_{\rm R}$ from the red/blue dots.
 Recalling that ${\sigma_{\rm L}=\sigma_{\rm R}=\sigma}$, we see that $g\sigma$ in the matrix theory maps to  $g_\mathrm{YL}$ in the scalar theory.

The other diagrams required for the 1-loop $\beta$-function 
are given by the leading renormalization of the propagator, namely   the LHS of
\be\label{eq:circequiv}
\Bigg[2\;{\parbox{1cm}{{\begin{tikzpicture}[scale=1]
\coordinate (x1) at (0,0) ;
\coordinate (x2) at  (0.6,0) ;
\coordinate (x1p) at  (-.2,0) ;
\coordinate (x2p) at  (0.8,0) ;
\coordinate (x3) at  (0.3,0.3) ;
\coordinate (x4) at  (0.3,-0.3) ;
\fill (x1) circle (1.5pt);
\fill (x2) circle (1.5pt);
\fill [blue] (x3) circle (1.5pt);
\fill [red] (x4) circle (1.5pt);
\draw (.3,0) circle (3mm);
\draw [black] (x1) -- (x1p);
\draw [black] (x2) -- (x2p);
\end{tikzpicture}}}
+4\;
\parbox{1cm}{{\begin{tikzpicture}[scale=1]
\coordinate (x1) at (0,0) ;
\coordinate (x2) at  (0.6,0) ;
\coordinate (x1p) at  (-.2,0) ;
\coordinate (x2p) at  (0.8,0) ;
\coordinate (x3) at  (0.25,-0.3) ;
\coordinate (x4) at  (0.35,-0.3) ;
\fill (x1) circle (1.5pt);
\fill (x2) circle (1.5pt);
\fill [blue] (x3) circle (1.5pt);
\fill [red] (x4) circle (1.5pt);
\draw (.3,0) circle (3mm);
\draw [black] (x1) -- (x1p);
\draw [black] (x2) -- (x2p);
\end{tikzpicture}}}}\Bigg]_{d-4} =\Bigg[\parbox{1cm}{{\begin{tikzpicture}[scale=1]
\coordinate (x1) at (0,0) ;
\coordinate (x2) at  (0.6,0) ;
\coordinate (x1p) at  (-.2,0) ;
\coordinate (x2p) at  (0.8,0) ;
\fill (x1) circle (1.5pt);
\fill (x2) circle (1.5pt);
\draw (.3,0) circle (3mm);
\draw [black] (x1) -- (x1p);
\draw [black] (x2) -- (x2p);
\end{tikzpicture}}}\Bigg]_d .
\ee
This identity holds for any value of the external momentum.
In more detail: 
in the Schwinger representation,
the scalar diagram gives
\bea
\lefteqn{\parbox{1cm}{{\begin{tikzpicture}[scale=1]
\coordinate (x1) at (0,0) ;
\coordinate (x2) at  (0.6,0) ;
\coordinate (x1p) at  (-.2,0) ;
\coordinate (x2p) at  (0.8,0) ;
\fill (x1) circle (1.5pt);
\fill (x2) circle (1.5pt);
\draw (.3,0) circle (3mm);
\draw [black] (x1) -- (x1p);
\draw [black] (x2) -- (x2p);
\end{tikzpicture}}}
=  \int_{{s_1,s_{2}>0}} \int_k   
\rme^{-s_1[(k+p)^2+m^2] -s_2 (k^2+m^2) }}
\nn\\
&=&  \int_{{s_1,s_{2}}}(s_{1}+s_{2})^{{-d/2}} \rme^{-\frac{s_{1}s_{2} }{s_{1}+s_{2}} p^{2}-(s_{1}+s_{2})m^2}.
\label{86}
\eea
The diagrams of the matrix theory are
\bea
\lefteqn{2\;\parbox{1cm}{{\begin{tikzpicture}[scale=1]
\coordinate (x1) at (0,0) ;
\coordinate (x2) at  (0.6,0) ;
\coordinate (x1p) at  (-.2,0) ;
\coordinate (x2p) at  (0.8,0) ;
\coordinate (x3) at  (0.3,0.3) ;
\coordinate (x4) at  (0.3,-0.3) ;
\fill (x1) circle (1.5pt);
\fill (x2) circle (1.5pt);
\fill [blue] (x3) circle (1.5pt);
\fill [red] (x4) circle (1.5pt);
\draw (.3,0) circle (3mm);
\draw [black] (x1) -- (x1p);
\draw [black] (x2) -- (x2p);
\end{tikzpicture}}}
+4\;
\parbox{1cm}{{\begin{tikzpicture}[scale=1]
\coordinate (x1) at (0,0) ;
\coordinate (x2) at  (0.6,0) ;
\coordinate (x1p) at  (-.2,0) ;
\coordinate (x2p) at  (0.8,0) ;
\coordinate (x3) at  (0.25,-0.3) ;
\coordinate (x4) at  (0.35,-0.3) ;
\fill (x1) circle (1.5pt);
\fill (x2) circle (1.5pt);
\fill [blue] (x3) circle (1.5pt);
\fill [red] (x4) circle (1.5pt);
\draw (.3,0) circle (3mm);
\draw [black] (x1) -- (x1p);
\draw [black] (x2) -- (x2p);
\end{tikzpicture}}}} \nn\\
&=&
\int_k \int_{s_1,s_2}  \left[ 2\frac{s_1 s_2}{1!1!}{+}\frac{2 s_1^2{+}2 s_2^2}{1! 2!}\right]\! \rme^{-s_1[(k+   p)^2+m^2]-s_2( k^2+m^2)}\nn\\
&=&
\int_k \int_{s_1,s_2} (s_1+s_2)^2 \rme^{-s_1[(k+   p)^2+m^2]-s_2( k^2+m^2)}\nn\\
&=&\int_{{s_1,s_{2}}}(s_{1}+s_{2})^{{2-d/2}} \rme^{-\frac{s_{1}s_{2} }{s_{1}+s_{2}} p^{2}-(s_{1}+s_{2})m^2}.
\label{85}
\eea
We see that the same nontrivial function of $p^2/m^2$ arises in each case, up to the dimensional shift. 
(The supersymmetric approach in the next section shows that this extends to higher correlations, so long as all points live in the same $d-4$-dimensional hyperplane.)

As a consequence, at 1-loop order the theory \eqref{eq:S-N=0} has the same diagrams and combinatorics as the Yang-Lee theory, 
with $g\sigma$ playing the role of $g_\mathrm{YL}$.
Let us write the one-loop Yang-Lee RG equation \eq{beta-YL} as
\ba
\f{\dd  }{\dd \ell } \, g_{\rm YL} 
 & = 
 \f{{6-d }}{2} \,  g_{\rm YL} + \f{3}{4} g_{\rm YL}^3,
\end{align}
Then the diagrammatic argument above shows that in the matrix theory, 
\ba\label{eq:gflowpreliminary}
\f{\dd  }{\dd \ell } \, g
 & = 
 \f{{6-d }}{2} \,  g + \f{3}{4} (g \sigma)^2 g,
\end{align}
where the first term is given by the usual engineering dimension of $g$.
However, this is not a closed equation for $g$.
A closed equation is instead obtained if we consider   $g\sigma$, which as discussed is the natural definition of the coupling constant in this theory.
By a one-loop calculation that we report in App.~\ref{sec:sigmaoneloop}, one finds that $\sigma$ renormalizes trivially:
\ba\label{eq:sigmaflow}
\f{\dd  }{\dd \ell } \, \sigma 
& = 
2 \sigma.
\end{align}
Combining this with \Eq{eq:gflowpreliminary}
yields the RG equation for the coupling $g\sigma$,
\ba\label{eq:gflowDR}
\f{\dd  }{\dd \ell } \, (g \sigma)
 & = 
 \f{{10-d }}{2} \,  (g \sigma) + \f{3}{4} (g \sigma)^3.
\end{align}
This shows the dimensional reduction property at one-loop order,\footnote{As a further check, in App.~\ref{s:dim-red-2-loop-example} 
we confirm the dimensional reduction property (analogous to 
\Eq{eq:triangleequivalence}) for one of the two-loop diagrams renormalizing the coupling.} with the flow of $(g\sigma)$ mapping to the flow of $g_{\rm YL}$ in four fewer dimensions.

According to \Eq{eq:sigmaflow},  $\sigma$ flows to infinity.
This closely parallels the random field Ising model,
where the analogous quantity is the variance $\Delta$ of the random field.
There, the flow of $\Delta$
to infinity is equivalent to the flow of the temperature $T$ to zero,
reflecting the fact that the universality class of the phase transition is the same irrespective of whether the physical (bare) temperature is zero or nonzero.

In known examples of dimensional reduction,
the  underlying structure is supersymmetry. 
This motivates looking for a superfield  formulation of the action in \Eq{eq:S-N=0}. In the next section we derive a supersymmetric action following Cardy's approach \cite{Cardy1983}.
As an aside, we note that one can   see a hint of SUSY by rewriting the low-order diagrams in terms  of Grassman integrals. 
One may check that the 
 combinatorial factors in the diagrams above are properly accounted for by replacing the propagator  by 
\be\label{Susy-prop}
\frac1{k^2+m^2} \longrightarrow \frac1{k^2 + m^2 + {\sigma_{\rm L} }  \theta_k^{\rm L}\bar \theta_k^{\rm L} +  {\sigma_{\rm R}}  \theta_k^{\rm R}   \bar \theta_k^{\rm R}  },
\ee
where the $\theta$ are Grassmanian variables, 
and performing Grassman integrals over ${(\bar\theta_k^{\rm L},\theta_k^{\rm L},\bar\theta_k^{\rm R},\theta_k^{\rm R})}$ alongside the integral over  $k$. The Grassman integrals then effectively reduce the dimensionality of the $k$ integral.\footnote{In this formalism, 
${\int_k \rme^{-s (k^2+m^2)} = s^{-d/2} \rme^{-s m^2}}$
is replaced by
\bea\label{super-integral}
\int_\theta
\int_k \rme^{-s ({k^2 + m^2 + {\sigma_{\rm L}}    \theta_k^{\rm L} \bar\theta_k^{\rm L} + {\sigma_{\rm R}}    \theta_k^{\rm R} \bar\theta_k^{\rm R} })}
= {\sigma_{\rm L}\sigma_{\rm R}}\,  s^{2-d/2} \rme^{-s m^2},
\eea
with ${\int_\theta = \int \rmd \bar \theta^{\rm L}_k \rmd \theta^{\rm L}_k
\rmd \bar \theta^{\rm R}_k \rmd \bar \theta^{\rm R}_k}$.}
In contrast to the Parisi-Sourlas theory \cite{ParisiSourlas1979}, we have ``left'' and ``right'' Grassman variables, giving dimensional reduction by 4 rather than 2.

The flow diagram for $g\sigma$ is qualitatively similar to that shown in Fig.~\ref{fig:MPTflows}, with $d_c=10$ rather than 6.
The most basic  consequence is similar to the $N\to 1$ case discussed at the end of Sec.~\ref{sec:Nequals1RG}:
below $d_c$ the flow is to strong coupling, while above $d_c$ 
there are (at least) two distinct stable universality classes for the random tensor network phase transition.
One is described by the Gaussian fixed point at $g\sigma=0$, while the other is described by a putative strong coupling fixed point (which we would need some other method to analyze).
The ``tricritical''  point separating the two kinds of transition 
could be accessed in a model with two tuning parameters, and has nontrivial exponents that can be computed in perturbation theory.

While the topology of the flows is similar, the Gaussian universality class for ${N\to 0}$ differs qualitatively  from that in Sec.~\ref{sec:Nequals1RG}:
even in the Gaussian theory the two-point function 
of the field $Y$ contains pieces with distinct index structures that decay with different powers of distance. (We will discuss the physical interpretation of these pieces separately \cite{NahumWieseToBePublished}.)
Again this is analogous to the random field Ising model, but it is in contrast to   the $N\to 1$ theory where there is a single power law for $\<XX\>$.

\section{Supersymmetry \& dimensional reduction}
\label{s:Supersymmetry and dimensional reduction}
\subsection{Summary}

In this Section we argue that the action (\ref{eq:S-N=0}) 
may be related to an action for a ``superfield'' ${\Upsilon (x,\bar \theta_{\rm L}, \theta_{\rm L}, \bar \theta_{\rm R}, \theta_{\rm R})}$
(i.e.\ a field with Grassman coordinates as well as physical spatial coordinates) 
that is of the form
\bea\label{L-Susy-top}
\ca L[\Upsilon] 
&=&
\frac 12 \Upsilon \lf - \nabla^2_{\rm S} \ri   \Upsilon + \frac{m^2}2  \Upsilon^2 + \frac{g\sigma}{6} \Upsilon^3.
\eea
The superspace Laplacian appearing here is 
\be
\nabla^2_{\rm S} = \nabla^2 +
\frac{\partial}{\partial \bar \theta_{\rm R} } \frac{\partial}{\partial  \theta_{\rm R} }
+ \frac{\partial}{\partial \bar \theta_{\rm L} } \frac{\partial}{\partial  \theta_{\rm L} },
\ee
and the action is obtained by integrating $\ca L[\Upsilon]$   over both $x$ and the Grassman variables, as discussed below.
The superspace formulation 
may be used to argue for dimensional reduction  \cite{ParisiSourlas1979,Cardy1985,BrydgesImbrie2003}.
The relation between (\ref{L-Susy-top}) and the original theory (\ref{eq:S-N=0})  is that a subset of correlation functions are equal between the two.

The following derivation has three steps, which  parallel Cardy's approach to the random field Ising model and branched polymers \cite{Cardy1983,Cardy1985,Cardy2001} (recently developed further in Refs.~\cite{KavirajRychkovTrevisani2019,KavirajRychkovTrevisani2021,KavirajTrevisani2022}). 
First, we make a change of basis for the fields using the ``Cardy transform'', and eliminate some less-relevant terms. 
Second, we show that in the resulting action, 
 some of the fields --- 
whose multiplicity is negative in the replica limit --- can be exchanged for fermionic fields,
so that in a sense the replica limit can be taken explicitly at the level of the action.
The final step is a simple check that the resulting action for bosons and fermions matches the superfield action (\ref{L-Susy-top}), if the superfield is expressed as a sum of conventional fields by Taylor expanding in its Grassman arguments.
The intermediate Lagrangians that we encounter below look complicated, but the proliferation of fields eventually reduces to the Susy Lagrangian (\ref{L-Susy-top}).

In the following, we aim to demonstrate the mapping only at the level of perturbation theory, so we will not worry when we encounter divergent path integrals. We leave open the question of whether these path integrals can be defined beyond perturbation theory.

\subsection{Cardy transformation}
\label{sec:cardytransform}

As mentioned in Sec.~\ref{s:FTforNt00}, the field $Y_{ab}$ in \Eq{eq:S-N=0} does not have a definite scaling dimension in the Gaussian theory with nonzero $\sigma$ \cite{CardyBook}. 
To see this, we may compute the propagator in the theory with ${m^2=g=0}$ but with nonzero $\sigma$. Taking the limit ${N\to 0}$ gives
\be\label{eq:Ytwopointfunction}
\< Y_{ab}(k) Y_{a'b'}(-k)\> 
=
\f{2\sigma^2}{k^6}
-
\f{\sigma\, (\delta_{aa'}+ \delta_{bb'})}{k^4} + \f{\delta_{aa'}\delta_{bb'}}{k^2}.
\ee
It is convenient to swith to a basis of fields with definite scaling dimensions in the limit ${N\to 0}$.
In this basis, terms such as $\sum_{ab} Y_{ab}^3$ 
 decompose into terms with different engineering dimensions, allowing us to simplify the action by discarding less relevant terms \cite{nahum2021measurement}.

The Cardy transformation is a change of basis for a field that carries an $S_N$ index ${a=1,\ldots,N}$:
it amounts to splitting fields not into representations of $S_N$, but into representations of the $S_{N-1}$ subgroup that acts nontrivially on index values ${a=2,\ldots,N}$.
(The full symmetry of the original Lagrangian is   hidden in the new basis, but it nevertheless constrains the RG flows
\cite{KavirajRychkovTrevisani2019,KavirajRychkovTrevisani2021,KavirajTrevisani2022}.)
Here our field has two $S_N$ indices, corresponding to the left and right $S_N$ factors in \Eq{eq:symmgp},
so we make the transformation for each index separately. 
We introduce a field $y_{IJ}$ whose indices take values
\be\label{eq:newindexset}
I,J\in \{ + , - , 1, \ldots, N-2 \},
\ee
and which is related to $Y$ by
\ba \label{Yfromy}
Y_{ab} & = \vec{v}_a \cdot y \cdot  \vec{v}_b
 = {v}_a^I  y_{IJ}  {v}_b^J.
\end{align}
The vectors specifying the change of basis are
\ba\label{v1}
\vec{v}_1 &=  \left(\sqrt{\sigma},\frac1{2\sqrt \sigma},\vec 0  \right),
&
\vec{v}_{a>1} &= \left(\sqrt{\sigma},\frac{-1}{2\sqrt{\sigma}},  \vec e_a \right),
\end{align}
where $\vec 0$ is the zero vector of length ${N-2}$, 
and ${\vec e_2, \ldots \vec e_{N}}$ are a collection of ${N-1}$ vectors, of length ${N-2}$, satisfying
\be\label{base-change}
\sum_{a=2}^N \vec e_a  = 0, \qquad 
\sum_{a=2}^N e_a^i e_a^j = \delta^{ij}.
\ee
We   use $i, j$ for indices that run over the numerical values in the set (\ref{eq:newindexset}), i.e.\ ${i,j = 1,\ldots, N-2}$. 
(Note that in the replica limit there are $-2$ possible values for these indices.)
We have included factors of $\sqrt{\sigma}$ in the transformation to simplify the Lagrangian.

There are two ways to  determine the engineering dimension of the new fields: either one recalls that  $\sigma$ introduced in the change of variables \eq{Yfromy}-\eq{v1} 
has the same engineering dimension as $\nabla^2$, or one reads  off the dimensions from the free theory given below in \Eq{eq:quadraticaftercardy}.
Either method yields
\be\label{y-dim}
\mbox{dim} (y_{IJ}) = \frac{d-2}2 - \#\{\rm plus\} + \#\{\rm minus\}.
\ee
Therefore the field with the lowest dimension is $y_{++}$, and that with the highest is $y_{--}$.

We   now substitute the transformation into the Lagrangian, and simplify the coefficients by taking  the limit ${N\to 0}$:
The  quadratic terms, at ${m^2=0}$, become
\ba\label{eq:quadraticaftercardy}
\ca L_2 & = 
\nabla y_{-+}
  \nabla y_{+-}+\nabla y_{--} \nabla y_{++}
  \nn \\
 & +
  \nabla y_{i-}
   \nabla y_{i+}+  \nabla y_{-i}
   \nabla y_{+i}  +\frac{1}{2}
 (\nabla  y_{ij})(\nabla  y_{ij}) \nn\\
&+   y_{--}\left( y_{-+}+y_{+-}\right) 
+  \half \left( y_{-i}y_{-i} +y_{i-} y_{i-}  \right) ,
\end{align}
where free $i$ and $j$ indices are summed.
The final line comes from the $\sigma$-term, namely ${\f{\sigma}{2}\sum (Y_{ab}Y_{ac}+Y_{ab}Y_{cb})}$, written out in the new basis (the factor of $\sigma$ has been absorbed into the field redefinitions).
We have not written the mass term (with coefficient $m^2$) since it has exactly the same form as the first two lines but without the derivatives.
If present, a linear term, ${r \sum_{ab}Y_{ab}}$, becomes ${(r/\sigma) (y_{--})}$.

Assigning engineering dimensions to fields in the usual way, on the basis of this quadratic term, gives \Eq{y-dim}.
We   now use these dimensions to organize the interactions.
When we write a given term such as ${\mathcal{L}_3=\f{g}{3!} \sum_{ab} Y_{ab}^3}$   in the new basis,
we will obtain terms with varying values for 
${\#\{\rm plus\} - \#\{\rm minus\}}$
and therefore with different scaling dimensions. 
In view of the field transformation (\ref{v1}),
the value of ${\#\{\rm plus\} - \#\{\rm minus\}}$ is also the power of $\sigma^{1/2}$ that accompanies the term. 
The most relevant terms are those with the largest value of ${\#\{\rm plus\} - \#\{\rm minus\}}$, or equivalently the highest power of $\sigma$.
The interaction term ${\mathcal{L}_3}$ becomes
\begin{widetext}
\be\label{eq:cubictransformed}
\ca L_3 = {g\sigma}  \bigg[\frac{1}{2}y_{++}^2 y_{--} +  y_{++} \Big(y_{-+}
   y_{+-}+  
   y_{i -} y_{i +}+  y_{-i } y_{+i }  +\frac{1}{2} 
   y_{i j }y_{i j } \Big)
   +\frac{1}{2}   \big(y_{+-} y_{i +}y_{i +}
   +  y_{-+} y_{+i}y_{+i } \big)
   + 
   y_{i +} y_{i j } y_{+j }  \bigg] ,
\ee
\end{widetext}
plus terms of order $g \sqrt \sigma$
 (and lower orders in $\sigma$)
which are less relevant. 
By standard dimension counting, the leading term shown above is marginal in 10 dimensions, consistent with the beta function for ${g\sigma}$ in Sec.~\ref{s:FTforNt00}.

Note that the replica-like indices are now the $i,j$ indices on $y_{ij}$. Each of these runs over ${1,\ldots, N-2}$.
We   write this as 
${i=1,\ldots, m_{\rm L}}$
for the row index and 
${j=1,\ldots, m_{\rm R}}$
for the column index, with the replica limit being ${m_{\rm L},m_{\rm R}\rightarrow -2}$.
Once the subleading terms are dropped, the
remaining terms in ${\mathcal{L}_2+\mathcal{L}_3}$
have an ${\rm O}(m_{\rm L})\times {\rm O}(m_{\rm R})$ symmetry under rotations on the $i,j$ indices associated with rows and columns respectively.

\subsection{Eliminating   replicas via fermions}
\label{sec:introducefermions}

If we retain only the quadratic terms and the most relevant cubic terms, it is possible (at least formally) to exchange the replica Lagrangian above for a Lagrangian without   replicas   but with both commuting and anticommuting fields. 
The needed replacements are straightforward at the level of the quadratic theory, so we describe this case first as motivation before generalizing the mapping to the interacting theory.

There are 3 types of fields: 
fields (e.g.\ $y_{+-}$) without a replica index;
fields with either  a left (e.g.\ $y_{i+}$) or a right (e.g.\ $y_{+i}$) replica index; and the matrix $y_{ij}$ with both.
Nothing needs to be done with the fields of the first type.
The other types of field must be replaced with fields without a replica index in such a way that the path integral over these fields is unchanged.\footnote{The complete path integral over all the fields is trivial in the limit ${N\to 0}$. However this is not true if we integrate over only a subset of the fields, or if we include appropriate sources in order to compute correlators.}

At the level of the quadratic theory, the matrix $y_{ij}$ simply gives  ${m_{\rm L} \times m_{\rm R}\rightarrow (-2)^2=4}$ free scalars,
so formally may be replaced with 4 real fields, 
or equivalently two complex fields which we denote $B$ and $M$. We normalize these so that
\be
\f{1}{2} (\nabla y_{ij}) (\nabla y_{ij}) 
\rightarrow \nabla \bar B \nabla B + 
\nabla \bar M \nabla M. 
\ee
Next consider the fields ${(y_{i+},y_{i-})}$ with a left replica index. These appear with multiplicity ${m_{\rm L}\rightarrow - 2}$.
The Gaussian path integral over these fields gives a determinant raised to the power ${-m_{\rm L}/2\rightarrow +1}$, and is therefore equivalent to a path integral over Grassman fields 
$(\lambda_+,\lambda_-)$ and their ``conjugates'' (in fact  independent Grassman variables) $(\bar\lambda_+,\bar \lambda_-)$.\footnote{That is, using $\int_\lambda$ to denote the integrals over the Grassman fields,
\be\notag
\prod_{i=1}^{-2} \int \mathcal{D}  y_{i\pm} \rme^{- \frac 12 (y_{i+} , y_{i-}) \ca A (y_{i+} , y_{i-} ) } 
= \int_\lambda 
\rme^{- ( \bar \lambda_{-}, \bar \lambda_{+}) \ca A  (  \lambda_{-}, \lambda_{+})  }. 
\ee}
We normalize the Grassman fields so that for example
\be
(\nabla y_{i-})(\nabla y_{i+}) \rightarrow  
\nabla \bar \lambda_- \nabla\lambda_+ + \nabla\bar \lambda_+ \nabla\lambda_-.
\ee
The fields $y_{\pm i}$ with a right replica index are replaced similarly with Grassman variables denoted $\rho$.

This replacement of $-2$ bosons with a pair of fermions is a well-known trick  in the context not only of the random field Ising model but also many other  replica field theories \cite{Wegner2016}: for example it may be used to argue that in the ${n\rightarrow -2}$ limit the partition function of the $O(n)$ Landau-Ginsburg theory is equivalent to that of a free Grassman (fermionic) theory   ---  this mapping only gives access to a subsector of the correlators of the $O(-2)$ model, but may be generalized \cite{WieseFedorenko2018,WieseFedorenko2019,HelmuthShapira2020,ShapiraWiese2020}.\footnote{The ${O(n)}$  model has more fields and is much richer than the free fermionic theory. To capture more of this structure, the ${O(-2)}$ model may be mapped onto a system of two complex fermions (Grassmann fields) and one complex boson \cite{WieseFedorenko2018,WieseFedorenko2019,HelmuthShapira2020,ShapiraWiese2020}. It then describes the fractal dimension of loop-erased random walks and charge-density waves at depinning.}

Altogether the quadratic terms in $\mathcal{L}_2$ (\Eq{eq:quadraticaftercardy}) give rise to 
\begin{align}\label{S-bos-fer-quadratic}
\widetilde {\ca L}_2 &=  
\nabla y_{-+} \nabla y_{+-}+\nabla y_{--}\nabla y_{++}
\nn\\
 & + \nabla \bar{\lambda }_- \nabla \lambda _+ +\nabla\bar{\lambda}_+ \nabla\lambda _- + \nabla\bar{\rho }_-\nabla\rho _+  +\nabla\bar{\rho}_+\nabla\rho _-   
 \nn\\ 
& +\nabla\bar{B} \nabla B +  \nabla\bar{M}\nabla M 
   \nn\\
& + \bar{\lambda }_-  \lambda _-  + \bar\rho_-\rho _-  + y_{--}( y_{+-}+y_{-+} ) .
\end{align}
So far we have considered only the quadratic theory.
Let us now turn to the interacting theory which includes the 
 leading interaction terms given by  (\ref{eq:cubictransformed}).
We can no longer use   the same simple logic
(based on counting powers of determinants)
to eliminate the replica fields in favor of bosons and fermions without a replica index:
this approach would work for almost all the terms, 
but we meet a problem in the term
${y_{i+}y_{ij}y_{+j}}$,
which has a new index structure compared to the terms already discussed.

As an aside, we note that a formal way to see that it is still possible to eliminate replicas is first to write a generalized action, 
in which a ``left'' index $i$
is replaced by a graded index taking an arbitrary number of $m_{\rm L}^{\rm B}$   ``bosonic'' values and an arbitrary even number of $m_{\rm L}^{\rm F}$   ``fermionic'' values, 
and similarly for every ``right'' index, with boson and fermion numbers $m_{\rm R}^{\rm B}, \, m_{\rm R}^{\rm F}$.
In this process fields such as $y_{-i}$ become supervectors and $y_{ij}$ becomes a supermatrix.
Index contractions such as $y_{-i}y_{-i}$ are replaced by  inner products that are invariant under the natural
global supersymmetry,
${{\rm Osp}(m_{\rm L}^{\rm B}|m_{\rm L}^{\rm F})\times {\rm Osp}(m_{\rm R}^{\rm B}|m_{\rm R}^{\rm F})}$, that generalizes ${{\rm O}(m_{\rm L})\times {\rm O}(m_{\rm R})}$.
By examining the integral 
over only the fields with a left index, which is Gaussian, we may   argue that the integral over these fields is invariant under the shift
\ba
m_{\rm L}^{\rm B} & \rightarrow m_{\rm L}^{\rm B}  + 2,
&
m_{\rm L}^{\rm F} & \rightarrow m_{\rm L}^{\rm F}  + 2,
\end{align}
for any fixed values of ${(m_{\rm R}^{\rm B},m_{\rm R}^{\rm F})}$.
The same property holds with left and right indices exchanged.
As a result, we can go from
${(m_{\rm L}^{\rm B}=m_{\rm R}^B=-2,m_{\rm L}^{\rm F}=m_{\rm R}^{\rm F}=0)}$
to 
${(m_{\rm L}^{\rm B}=m_{\rm R}^{\rm B}=0,m_{\rm L}^{\rm F}=m_{\rm R}^{\rm F}=2)}$ without changing the path integal over these fields.
Ultimately, this leaves an action where the fields that carried a single replica index   become fermions, and $y_{ij}$ is again bosonic.

Here  we   follow a more direct approach. 
In a first step, we consider the integral over the fields carrying a left index. 
The integral \textit{only} over these fields is Gaussian, so we can use the above trick to
exchange them for fermions in the limit ${m_L \rightarrow -2}$.
The remaining replica indices are right indices.

The fields carrying a right index now include both fermions and bosons
(because of the fermionization in the first step).
However these fields again appear only quadratically in the action, 
so, for the second step,
we can use a generalization of the above trick for Gaussian superintegrals:
taking the limit of a negative number of values for the right index is equivalent to 
replacing all bosons with fermions and all fermions with bosons.
Having done this, there are no more replica indices.
During this process, is is convenient to group pairs of real fields into complex fields 
--- the details are in App.~\ref{s:Cardy fields and supersymmetry}.
The final result is \smallskip
\begin{widetext}
\begin{eqnarray}\label{S-bos-fer}
\phantom{\Big|} \ca L &=&  
\lf \nabla y_{-+} \nabla y_{+-}+\nabla y_{--}
   \nabla y_{++} \ri
+
\lf 
\nabla \bar{\lambda }_- \nabla \lambda _+ +\nabla\bar{\lambda}_+ \nabla\lambda _- + \nabla\bar{\rho }_-\nabla\rho _+  +\nabla\bar{\rho}_+\nabla\rho _-  \ri
+
\lf 
\nabla\bar{B} \nabla B +  \nabla\bar{M}\nabla M 
\ri
   \nn\\
&+& m^2 \big( \bar{\lambda }_- \lambda _+ +\bar{\lambda }_+\lambda_- +\bar\rho _+ {\rho }_-+\bar\rho_-  {\rho }_+   +\bar B {B}+\bar M  {M}+y_{-+}
   y_{+-}+y_{--}
   y_{++}\big) \nn\\
 &+&   
 \Big[
 \bar{\lambda }_-  \lambda _-  + \bar\rho_-\rho _-  + y_{--}( y_{+-}+y_{-+} ) 
\Big]
\nn \\
&+& {g\sigma} \Big[
\,
\frac{1}{2} 
\, 
y_{--} y_{++}^2
+ y_{++} \, \big( \bar{\lambda }_- \lambda _+ +\bar{\lambda }_+\lambda_- +\bar\rho _+ {\rho }_-+\bar\rho_-  {\rho }_+   +\bar B {B}+\bar M  {M}+y_{-+}
   y_{+-}\big)   \nn\\
   &&  \qquad \qquad \qquad \,\,\,\hspace{0.2mm}
    + \lf 
   y_{-+} \, \bar\lambda _+ {\lambda }_+
   +
   y_{+-} \,
   \bar\rho _+ {\rho }_+  \ri 
   + 
   \bar{B}
   \lambda _+ \rho _+  +B \bar{\rho}_+ \bar{\lambda }_+ +\bar M  \bar{\lambda }_+\rho _+
   + {M} \lambda_+ \bar{\rho }_+
   \Big].
\end{eqnarray}
\end{widetext}
The quadratic terms agree with the previous discussion of the quadratic theory. We note that if less-relevant interactions are included this rewriting in terms of bosons and fermions is not expected to be possible   \cite{KavirajRychkovTrevisani2019}.

\subsection{Superfield representation}

The Lagrangian \eq{S-bos-fer} admits a compact superspace representation. To this aim, we  combine the above fields 
$y_{++},\, \bar \lambda_+, \cdots,\, y_{--}$
into a superfield by defining
\bea\label{eq:definesuperfield}
&& \!\!\! \Upsilon (x,\bar \theta_{\rm L}, \theta_{\rm L}, \bar \theta_{\rm R}, \theta_{\rm R}) =  
y_{++} \nn\\ 
&& +\theta _{\rm R} \bar{\lambda}_+
+\bar{\theta }_{\rm R} \lambda_+
+\theta _{\rm L} \bar{\rho}_+
+\bar{\theta }_{\rm L} \rho _+
\nn\\
&&
+\bar{\theta }_{\rm R} \theta_{\rm R} y_{+-}
+\bar{\theta }_{\rm L} \theta_{\rm L} y_{-+} \nn\\
&&
+\theta_{\rm R} \theta _{\rm L} \bar{B}
+\bar{\theta}_{\rm L} \bar{\theta }_{\rm R} B +\theta_{\rm L} \bar{\theta}_{\rm R} \bar{M}
+\bar{\theta }_{\rm L} \theta_{\rm R} M \nn\\
&&
+\theta _{\rm R} \bar{\theta}_{\rm L} \theta _{\rm L} \bar{\lambda}_-
+\bar{\theta }_{\rm L} \theta_{\rm L} \bar{\theta }_{\rm R} \lambda_-
 +\bar{\theta }_{\rm R} \theta _{\rm R} \theta_{\rm L} \bar{\rho }_-
+\bar{\theta}_{\rm R} \theta _{\rm R} \bar{\theta }_{\rm L} \rho_-\nn\\
&&+\bar{\theta }_{\rm L} \theta_{\rm L} \bar{\theta }_{\rm R} \theta_{\rm R} y_{--}\;.
\eea
On the right-hand side $\Upsilon$ has been Taylor expanded in its Grassman ``coordinates''. 
This expansion truncates since each Grassmann variable squares to 0.

The scaling dimensions of the fields $y_{++},\, \bar \lambda_+, \cdots,\, y_{--}$
were given in \Eq{y-dim}:
these assignments are consistent with \Eq{eq:definesuperfield}, if we give a momentum dimension $(d-6)/2$ to $\Upsilon$ and a dimension $-1$ to the Grassman coordinates (so that  $\theta$ and $x$ have the same dimension).

Next, defining
\be
\int_{\theta}:= \int\rmd \bar \theta_{\rm L} \rmd  \theta_{\rm L} \rmd \bar \theta_{\rm R}    \theta_{\rm R},
\ee
the Lagrangian \eqref{S-bos-fer} can be written as
\bea\label{L-Susy}
\ca L[\Upsilon] 
&=&
\int _\theta \bigg[  \frac 12 (\nabla \Upsilon)^2 + \frac{m^2}2  \Upsilon^2 + \frac{g\sigma}{3!} \Upsilon^3 \nn\\
&& \qquad ~~~+  \half    \Upsilon\Big(\frac{\partial}{\partial \theta_{\rm R} } \frac{\partial}{\partial \bar \theta_{\rm R} }
+\frac{\partial}{\partial \theta_{\rm L} } \frac{\partial}{\partial \bar \theta_{\rm L} }
 \Big) \Upsilon\bigg].\qquad 
\eea
In this form, dimensional reduction to the Yang-Lee theory in \Eq{eq:yangleeLagrangian} may be proven (at the level of perturbation theory)
 following the treatment for random-field systems \cite{ParisiSourlas1979,Wiese2021}, or branched polymers \cite{ParisiSourlas1981,Cardy2003}.
The general idea 
was given above in \Eqs{Susy-prop} to \eq{super-integral}. 

The upshot is that, as long as we are   interested in perturbative results involving observables expressible in the dimensionally reduced theory, i.e.\ Yang-Lee, we can use the latter for our computations. This is advantageous, since 5-loop results are available there
\cite{BorinskyGraceyKompanietsSchnetz2021,KompanietsPikelner2021}.
It remains to be seen whether the superfield formulation can be used for nonperturbative results \cite{CardyMcKane1985,Cardy1985}.

\subsection{Aside:  signs}

Let us briefly comment on the signs of couplings in relation to other examples of dimensional reduction.

The random-field Ising model
may be formulated in terms of an  Ising spin $\phi_a$, with replica index ${a=1,\ldots,N}$. 
The analog of our $\sigma$ term
arises from averaging over the random magnetic field $h$: schematically, 
$\overline {\rme^{\sum_a h \phi_a} }= \rme^{\frac\sigma2 \sum_{ab}\phi_a\phi_b}$, with ${\sigma>0}$. 
The action also includes a quartic coupling ${g>0}$. 
The quartic coupling in the \textit{dimensionally-reduced} theory
(a standard scalar $\phi^4$ theory)
is given by $g{\sigma}$, and is real and positive.

The branched polymer theory 
again involves a field $\phi_a$ with ${a=1,\ldots,N}$:
the branched polymers correspond to the real-space Feynman diagrams for this field.
The action has a 
cubic coupling $g$, 
together with the above-mentioned term 
$\rme^{\sigma\sum_{ab} \phi_a \phi_b}$,
but with ${\sigma<0}$.
(This term arises, after absorbing a constant into the field, from a quartic term   $e^{-c\sum_{ab} \phi_a^2 \phi_b^2}$ encoding self-repulsion.)
The cubic coupling in the dimensionally reduced theory is  given by $g\sqrt{\sigma}$, and since $\sigma$ is here negative this coupling is imaginary. 
As a result, the branched polymer theory maps onto the standard Yang-Lee problem with an imaginary cubic coupling.

For the random tensor network, dimensional reduction again gives a scalar $\phi^3$ theory, but now with a real coupling $g\sigma$ in contrast to the Yang-Lee problem.

\begin{figure}
\begin{tikzpicture}
\foreach \i in {1,...,12}
{
	\newcommand{\boxsize}{1.5}
    \pgfmathtruncatemacro{\yy}{(\i - 1) / 4};
    \pgfmathtruncatemacro{\x}{\i - 4 * \yy};
    \pgfmathtruncatemacro{\label}{\x + 4 * (2 - \yy)};
    \node    (\label) at (\boxsize*\x,\boxsize*\yy) {$T$};
    \draw [blue] (\boxsize*\x-.5,\boxsize*\yy-.5)--(\boxsize*\x+.5,\boxsize*\yy-.5)
    	--(\boxsize*\x+0.5,\boxsize*\yy+.5)--(\boxsize*\x-.5,\boxsize*\yy+.5)--(\boxsize*\x-.5,\boxsize*\yy+-.5);
	\draw [thick,black] (\boxsize*\x,\boxsize*\yy-.5)--(\boxsize*\x,\boxsize*\yy-0.5*\boxsize);
	\draw [thick,black] (\boxsize*\x,\boxsize*\yy+.5)--(\boxsize*\x,\boxsize*\yy+\boxsize*.5);
	\draw [thick,black] (\boxsize*\x+0.5,\boxsize*\yy)--(\boxsize*\x+\boxsize*.5,\boxsize*\yy);
	\draw [thick,black] (\boxsize*\x-0.5,\boxsize*\yy)--(\boxsize*\x-\boxsize*.5,\boxsize*\yy);
}
\end{tikzpicture}
\caption{Part of the square-lattice tensor network $\mathcal{T}$.}
\label{f:TN}
\end{figure}
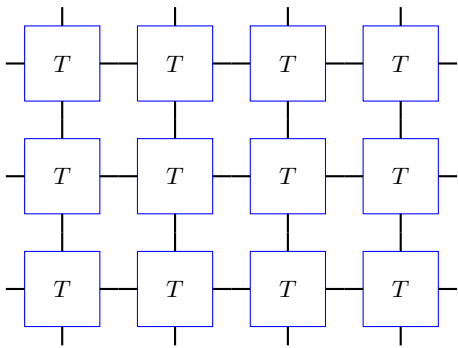

\section{From microscopic models to field theory}
\label{sec:derivefieldtheory}

The field theories discussed above have the  $G_N$ replica symmetry of the quantum problems of interest.
But symmetry is not enough on its own to guarantee that these field theories describe any physical measurement problem or random tensor network. 
Therefore, here we demonstrate that  the  above field theories do describe --- at least some --- microscopic models.
We show that both theories can be derived, in a controlled way, for a  family of simple random tensor networks, by considering separately the $N\to 0$ limit and the $N\to 1$ limit.

These networks generalize the model of Ref.~\cite{nahum2021measurement} by adding an arbitrary number $ N_{\rm f}$ of   ``flavors''. 
When $ N_{\rm f}$ is large, we obtain the continuum Lagrangians in the weak-coupling regime, ensuring that the perturbative RG treatment is   appropriate.

For the tensor network problem, 
this shows
(modulo the lack of rigor in the replica approach) 
that the flows which we found in the previous sections on the $N\to 0$
theory are relevant to microscopic random tensor networks.
For example, 
the Gaussian fixed point is accessible in high dimensions. 
We also expect that the corresponding mean-field exponents are accessible in appropriate tensor networks with an ``all-to-all' connectivity (we will discuss this case in  \cite{NahumWieseToBePublished}).

The tensor networks that we study may   be interpreted as nonunitary quantum circuits of a certain type.
If we consider the ${N\to 1}$ limit (instead of the ${N\to 0}$ limit), we obtain a replica structure similar to that required to handle quantum circuits with measurement. 
We show that in this limit we obtain the field theory in Sec.~\ref{sec:Nequals1RG}.

However, we caution the reader that the ${N\to 1}$ limit in the model below does not correspond to a simple monitored dynamics with random unitaries and Born-rule measurements.
Formally, this is because the ensemble of tensor networks does not form a Kraus decomposition of a quantum channel (App.~\ref{app:replicarecap}).\footnote{We expect that the $N\to 1$ limit can be interpreted in terms of monitored dynamics that is subject to partial postselection, reweighting the Born probabilities.}
We study the $N\to 1$ limit of the tensor network below as a toy  model
that has the same replica symmetry as   the more natural   monitored dynamics, 
and therefore plausibly shares 
the same critical theory.
(We will discuss microscopic derivations for models with genuine Born rule measurements in separate work.)

\begin{figure}
\begin{tikzpicture}
{\newcommand{\boxsize}{1.7}
\newcommand{\hgatesize}{0.35}
\newcommand{\yy}{0}
\foreach \i in {1,...,3}
{ \pgfmathtruncatemacro{\x}{\i};
   \draw [black] (\boxsize*\x-\hgatesize,\boxsize*\yy-\hgatesize)--(\boxsize*\x+\hgatesize,\boxsize*\yy-\hgatesize)
    	--(\boxsize*\x+\hgatesize,\boxsize*\yy+\hgatesize)--(\boxsize*\x-\hgatesize,\boxsize*\yy+\hgatesize)--(\boxsize*\x-\hgatesize,\boxsize*\yy+-\hgatesize);
		\draw [thick,black] (\boxsize*\x,\yy-\hgatesize) -- (\boxsize*\x,\yy-\boxsize/2-\hgatesize);
		\draw [thick,black] (\boxsize*\x,\yy+\hgatesize) -- (\boxsize*\x,\yy+\boxsize/2);
}
\renewcommand{\yy}{\boxsize}
\foreach \i in {1,...,3}
{ \pgfmathtruncatemacro{\x}{\i};
   \draw [black] (\boxsize*\x-\hgatesize,\yy-\hgatesize)--(\boxsize*\x+\hgatesize,\yy-\hgatesize)
    	--(\boxsize*\x+\hgatesize,\yy+\hgatesize)--(\boxsize*\x-\hgatesize,\yy+\hgatesize)--(\boxsize*\x-\hgatesize,\yy+-\hgatesize);
	\draw [thick,black] (\boxsize*\x,\yy-\hgatesize) -- (\boxsize*\x,\yy-\boxsize/2);
		\draw [thick,black] (\boxsize*\x,\yy+\hgatesize) -- (\boxsize*\x,\yy+\boxsize/2);
}
\draw [line width=3pt, black] (\boxsize/2,-\boxsize/2) -- (3.5*\boxsize,-\boxsize/2);
	\foreach \i in {1,...,3}{
	\node at (\boxsize*\i,-\boxsize/2) [circle,fill,inner sep=3pt]{};
	}
\draw [line width=3pt, black] (\boxsize/2,\boxsize/2) -- (3.5*\boxsize,\boxsize/2);
\foreach \i in {1,...,3}{
	\node at (\boxsize*\i,\boxsize/2) [circle,fill,inner sep=3pt]{};
	}
	    \node at (2*\boxsize,\boxsize) {${\vec g.\vec \sigma}$};
	        \node at (2.5*\boxsize,0.67*\boxsize) {${J\sigma^z \sigma^z}$};
	        \node at (0.1*\boxsize,1.1*\boxsize) {$t$};
	  \draw [-stealth, thick, black] (\boxsize/4 ,-0.65* \boxsize) -- (\boxsize/4,1.29*\boxsize);
}
\end{tikzpicture}
\caption{If we interpret the diagonal direction in Fig.~\ref{f:TN} as time, 
the tensor network is a non-unitary quantum circuit.
 In the case $N_{\rm f}=1$, the circuit acts on a chain of qubits.
After regrouping gates (App.~\ref{app:fieldtheoryderivations}), it   can be drawn as  above (two time-steps are shown).
  The single-site gates are of the form $\exp ( \vec g . \vec \sigma)$, where ${\vec g\in \mathbb{C}}$ is a random ``magnetic field'', and the two-site gates (indicated by horizontal bars) are of the form ${\exp( J \sigma_z\otimes \sigma_z )}$ with random~${J\in \mathbb{C}}$. This is a nonunitary version of the kicked Ising model. For $ N_{\rm f}>1$ it is a generalization acting on a chain with $ N_{\rm f}$ qubits at each site.}
\label{f:TNtoQC}
\end{figure}
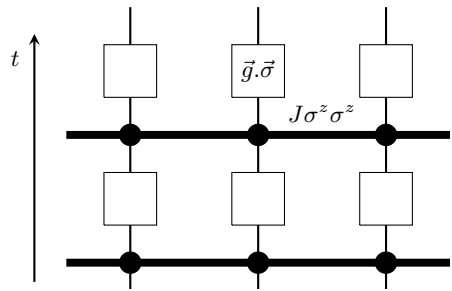

It is striking that the same lattice model can give rise to the two quite different field theories in Secs.~\ref{sec:Nequals1RG} and \ref{s:FTforNt00}, simply by changing the replica limit under consideration.

For concreteness, we describe the following derivations for   a 2D tensor network. 
However, these derivations extend  immediately to higher dimensions, with the only changes being in some finite geometrical factors. 
The resulting field theories are expected to have a broader applicability than the specific models discussed (because perturbing these models does not lead to new RG relevant couplings).
In the main text we summarize the key points, with  details  given in  App.~\ref{app:DerivationfieldtheoryforNto0} and App.~\ref{app:detailsTNtoQC}.

\subsection{Tensor network with $N_{\rm f}$ flavors}
\label{sec:defineTN}

We start with a square-lattice random tensor network, see Fig.~\ref{f:TN}.
We label the bonds of the tensor network by $j,k,\ldots$.
Each bond of the tensor network has bond dimension $2^{N_{\rm f}}$, and the state of the $j$-th bond is represented as the state of a collection of $N_{\rm f}$ Ising spins,
\ba
&S_{j\mu}=\pm 1, 
&
\mu & = 1, \ldots, N_{\rm f}.
\end{align}
We refer to $\mu$ as the flavor. 
Note that the ``bond index'' of a given bond $j$ is the entire set ${\{S_{j\mu}\}_{\mu=1}^{N_{\rm f}}= \mathbf{S}_j}$.

Each node tensor $T$ is taken to be independently random. For a moment, let us denote the bonds involved in a given node by $i=1,2,3,4$, labelled cyclically around the node, and $i=5$ identified with $i=1$. The node tensors are taken to be of the form
\begin{eqnarray}
\nonumber
\lefteqn{T_{\mathbf{S}_1,\mathbf{S}_2,\mathbf{S}_3,\mathbf{S}_4} }  \\
&=&  \exp \!
\Big(  
\sum_{i=1}^4 \sum_{\mu=1}^{N_{\rm f}} \tilde h_{i }^\mu S_{i\mu} 
{+} 
\sum_{i=1}^4 \sum_{\mu,\nu=1}^{N_{\rm f}} J_{i,i+1}^{\mu\nu}  S_{i\mu} S_{i+1,\nu} \Big),\nn
\label{eq:Ttensor}
\end{eqnarray}%
\begin{figure}[t]
\begin{tikzpicture}
\foreach \i in {1,...,12}
{
	\newcommand{\boxsize}{1.5}
    \pgfmathtruncatemacro{\yy}{(\i - 1) / 4};
    \pgfmathtruncatemacro{\x}{\i - 4 * \yy};
    \pgfmathtruncatemacro{\label}{\x + 4 * (2 - \yy)};
    \node    (\label) at (\boxsize*\x,\boxsize*\yy) {};
    \draw [blue] (\boxsize*\x-.5,\boxsize*\yy-.5)--(\boxsize*\x+.5,\boxsize*\yy-.5)
    	--(\boxsize*\x+0.5,\boxsize*\yy+.5)--(\boxsize*\x-.5,\boxsize*\yy+.5)--(\boxsize*\x-.5,\boxsize*\yy+-.5);
	\draw [thick,black] (\boxsize*\x,\boxsize*\yy-.5)--(\boxsize*\x,\boxsize*\yy-0.5*\boxsize);
	\draw [thick,black] (\boxsize*\x,\boxsize*\yy+.5)--(\boxsize*\x,\boxsize*\yy+\boxsize*.5);
	\draw [thick,black] (\boxsize*\x+0.5,\boxsize*\yy)--(\boxsize*\x+\boxsize*.5,\boxsize*\yy);
	\draw [thick,black] (\boxsize*\x-0.5,\boxsize*\yy)--(\boxsize*\x-\boxsize*.5,\boxsize*\yy);
		\draw [very thick,red,dashed] (\boxsize*\x-0.5*\boxsize,\boxsize*\yy)--(\boxsize*\x,\boxsize*\yy+0.5*\boxsize);
	\draw [very thick,red,dashed] (\boxsize*\x-0.5*\boxsize,\boxsize*\yy)--(\boxsize*\x,\boxsize*\yy-\boxsize*.5);
	\draw [very thick,red,dashed] (\boxsize*\x+0.5*\boxsize,\boxsize*\yy)--(\boxsize*\x,\boxsize*\yy+0.5*\boxsize);
	\draw [very thick,red,dashed] (\boxsize*\x+0.5*\boxsize,\boxsize*\yy)--(\boxsize*\x,\boxsize*\yy-\boxsize*.5);
	\fill [red] (\boxsize*\x-0.5*\boxsize,\boxsize*\yy) circle (2pt);
	\fill [red] (\boxsize*\x+0.5*\boxsize,\boxsize*\yy) circle (2pt);
	\fill [red] (\boxsize*\x,\boxsize*\yy+0.5*\boxsize) circle (2pt);
	\fill [red] (\boxsize*\x,\boxsize*\yy-0.5*\boxsize) circle (2pt);
}
\end{tikzpicture}
\caption{The Ising spins ${\bf S}_j$ (red dots) associated to the tensor network on Fig.~\ref{f:TN}. They live on the red dashed lattice.}
\label{f:dual-Ising-for-TN}
\end{figure}
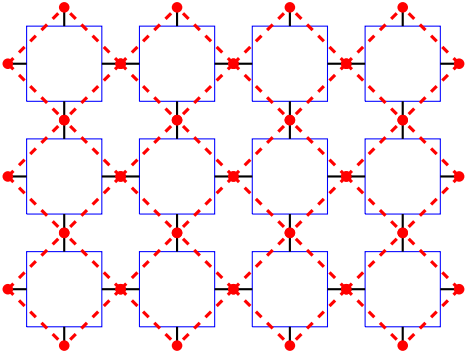%
with Gaussian random \textit{complex} couplings $\tilde h$ and $J$, whose variances are specified below.
Schematically, the terms in the exponential above are
\begin{equation}
 {\newcommand{\boxsize}{1.5}\parbox{1.1cm}{\begin{tikzpicture}
    \pgfmathtruncatemacro{\yy}{0};
    \pgfmathtruncatemacro{\x}{0};
    \pgfmathtruncatemacro{\label}{0};
    \node    (\label) at (0,0) {$T$};
    \draw [blue] (-.35,-.35)--(.35,-.35)--(0.35,+.35)--(-.35,+.35)--(-.35,+-.35);
	\draw [thick,black] (0,-.35)--(0,-0.35*\boxsize);
	\draw [thick,black] (0,+.35)--(0,+\boxsize*.35);
	\draw [thick,black] (0.35,0)--(0+\boxsize*.35,0);
	\draw [thick,black] (-0.35,0)--(0-\boxsize*.35,0);
\end{tikzpicture}}}
=
\exp\left(\newcommand{\boxsize}{1.5}
{\parbox{1.1cm}{\begin{tikzpicture}
    \pgfmathtruncatemacro{\yy}{0};
    \pgfmathtruncatemacro{\x}{0};
    \pgfmathtruncatemacro{\label}{0};
    \node    (\label) at (0,0) {};
    \draw [blue] (-.35,-.35)--(.35,-.35)--(0.35,+.35)--(-.35,+.35)--(-.35,+-.35);
\draw [thick,black] (0,-.35)--(0,-0.35*\boxsize);
	\draw [thick,black] (0,+.35)--(0,+\boxsize*.35);
	\draw [thick,black] (0.35,0)--(0+\boxsize*.35,0);
	\draw [thick,black] (-0.35,0)--(0-\boxsize*.35,0);
	\fill [red] (-0.5,0) circle (3pt);
\end{tikzpicture}}}\black~
+ \mbox{3 more} + 
{\parbox{1.1cm}{\begin{tikzpicture}
    \pgfmathtruncatemacro{\yy}{0};
    \pgfmathtruncatemacro{\x}{0};
    \pgfmathtruncatemacro{\label}{0};
    \node    (\label) at (0,0) {};
    \draw [blue] (-.35,-.35)--(.35,-.35)--(0.35,+.35)--(-.35,+.35)--(-.35,+-.35);
\draw [thick,black] (0,-.35)--(0,-0.35*\boxsize);
	\draw [thick,black] (0,+.35)--(0,+\boxsize*.35);
	\draw [thick,black] (0.35,0)--(0+\boxsize*.35,0);
	\draw [thick,black] (-0.35,0)--(0-\boxsize*.35,0);
		\draw [ultra thick,red,dashed] (0,-.5)-- (-.5,0) ;
	\fill [red] (-0.5,0) circle (3pt);
	\fill [red] (0,-0.5) circle (3pt);
	\end{tikzpicture}}}\black~
+ \mbox{3 more} 
\right).
\end{equation}
The contraction of such random tensors gives a tensor network 
 ${\mathcal{T}}$ (see Fig.~\ref{f:TN})
which has bond indices associated with any uncontracted bonds on the boundary of the system.
The physical interpretation of $\mathcal{T}$ will depend on the context.
For example, we can think of one of the directions as a  ``time'' coordinate, 
and consider  a network in the form of an $L\times T$ cylinder that is periodic in the spatial coordinate.
This cylinder has two boundaries, one at the final and one at the initial time, each with  $L$ dangling bonds.
In this setting we can think of $\mathcal{T}$
 as  a nonunitary  ``evolution operator'' that   acts on a spin configuration for $L$ sites.

A natural way to do this is to take the north-east direction in Fig.~\ref{f:TN} to be the time coordinate. 
Then the tensor network can be reinterpreted as a  nonunitary quantum circuit acting on a one-dimensional chain. 
There is some freedom about how this circuit is drawn, because we can use the standard trick of regrouping  ``gates''. Fig.~\ref{f:TNtoQC} shows one way to draw it. 
The horizontal bonds are commuting gates, acting on two sites. The boxes are gates acting on single sites. 
In the case ${N_\mathrm{f}=1}$, this can be seen to be a version of the ``kicked Ising model'' \cite{akila2016particle,GopalakrishnanLamacraft2019,BertiniKosProsen2018} in which the gates are nonunitary (App.~\ref{app:detailsTNtoQC}).

From now on we   stick with the tensor network language. 
For the most part the boundary conditions will not concern us, since we are    interested in the bulk Lagrangian. 
However, for completeness, we give a concrete example of how to construct physical quantities, using the replica trick, in App.~\ref{app:replicarecap}.
The key object in the replica approach is the product of $N$ copies of the tensor network $\mathcal{T}$ and $N$ copies of its complex conjugate, $\mathcal{T}^*$ 
(with appropriate boundary conditions).

\subsection{Replicated action on the lattice} 
\label{sec:latticeS}

When the tensors $T$ defined above are joined together and the contracted bond indices are summed, we obtain a partition function for an Ising model with complex couplings.
Geometrically, this is a standard 2D square lattice Ising model:
the  Ising lattice sites are located at the middles of the tensor network bonds, see Fig.~\ref{f:dual-Ising-for-TN}. 
That is, we may   think of the bond label $j$ as  labelling a \textit{site} of this square-lattice Ising model.

The partition function is
\begin{equation}\label{eq:ZIsing}
Z[h,J] 
= \sum_{\{S\}}  \exp \! \Big( 
\sum_j\sum_\mu h_i^\mu S_{j\mu} {+}
\sum_{\langle i,j\rangle} \sum_{\mu,\nu} J_{ij}^{\mu\nu} S_{i\mu}S_{j\nu}
\Big). 
\end{equation}
On each Ising site $j$, we have grouped together two ``magnetic fields'' $\tilde h$, one coming from each of  the adjacent tensors, to give the field $h$.\footnote{Since we take the fields to be Gaussian, this grouping   doubles the variance of $h$ as compared with $\tilde h$.}
Above, we have suppressed the dependence of $Z$ on possible  boundary bond indices  that are not summed over. Boundary conditions will be neglected below, but see  App.~\ref{app:replicarecap}. 

We take $h$ and $J$ to be random Gaussian variables with mean zero. If the real and complex parts of $h$ are $h_{\Re}$ and $h_{\Im}$, and similarly for $J$, we choose the variances to be
\begin{eqnarray}
\overline{ h_{i\Re }^\mu h_{j\Re }^{\mu'}} &=& \overline{ h_{i\Im }^\mu h_{j\Im }^{\mu'} } 
 = \f{h^2}{2N_{\rm f}^{1/2}} \delta_{ij}  \delta^{\mu\mu' } , 
\\
\overline{ J_{ij\Re }^{\mu\nu} J_{ij\Re }^{\mu'\nu'} } &=& 
\overline {J_{ij\Im }^{\mu\nu} J_{i'j'\Im }^{\mu'\nu'} }
 = \f{J^2}{2 N_{\rm f}}\  \delta_{ii'}\delta_{jj'}     \delta^{\mu\mu' } \delta^{\nu\nu' } .~~~ 
\end{eqnarray}
The quantities $h^2$ and $J^2$ on the RHS are nonrandom and should not be confounded with the random variables on the LHS.
The choice of equal variances for the real and imaginary parts simplifies the effective Lagrangian obtained below. However, moving away from this 
point does not change the universal behavior, at least if the perturbation is weak.

Next we introduce replicas. We take $N$ powers of $Z$, and $N$ powers of its complex conjugate  $Z^*$, and average this product over disorder. 
Schematically,
\begin{equation}\label{eq:replicasofpartitionfn}
\overline{\underbrace{Z\dots Z}_N \underbrace{Z^*\cdots Z^*}_N} 
= 
\sum_{\{S\} }
\exp \lf 
- \mathcal{S}_0[{\mathbf S}]
\ri,
\end{equation}
where $\mathcal{S}_0[{\mathbf S}]$ is   a lattice action for spins ${\mathbf S}$ that carry replica indices as well as site and flavor indices.
We denote the spin variables that come from forward replicas $Z$ by $S$, and those that come from backward replicas $Z^*$ by $\bar S$ (note that this bar should not be confused with complex conjugation; $S$ is a binary real variable).
 Defining the overlap field
 \be\label{Y-def}
Y_j^{aA} := \f{1}{\sqrt{N_{\rm f}}} \sum_{\mu=1}^{N_\mathrm{f}} S_{j\mu}^a \bar{S}_{j\mu}^A,
\ee
this action is 
\be
\mathcal{S}_0 [Y]= 
- h^2 \sum_j \sum_{aA} Y_j^{aA} 
- J^2 \sum_{\langle jk\rangle} \sum_{aA} Y_j^{aA} Y_k^{aA}.
\ee
However, the 
partition function is still written as a sum over the spins $S$ rather than over the field $Y$. 
In the limit of large $N_\mathrm{f}$ we can exchange these discrete spins sums for an integral over $Y$ with an appropriate weight:
\be\label{eq:spinsum}
\sum_{ \{S^a_\mu, \bar{S}^A_\mu \}} 
\longrightarrow 
\int \dd Y  {\rm e}^{ - \mathcal{W}(Y) }.
\ee
The weight $\mathcal{W}$ must   be included in the action for $Y$.
At leading order in $1/N_\mathrm{f}$ this weight is Gaussian, simply because $Y_{aA}$ is the  sum over a large number of terms in \Eq{Y-def}, but there are corrections at order $1/ N_{\rm f}$: 
\be
\mathcal{W}(Y) = \f{1}{2} \sum_{aA} Y_{aA}^2 + \f{1}{N_\mathrm{f}} \Delta \mathcal{W}(Y) + \ldots .
\ee
$\Delta\mathcal{W}$ has a term of order $Y^2$ and a term of order $Y^4$  with a certain index structure.\footnote{It is even in $Y$,  because the measure defined by  the left-hand side of \Eq{eq:spinsum} is invariant under the transformation ${(S,\bar S)\to (S, -\bar S)}$, which changes the sign of $Y$.} 
We compute it in App.~\ref{app:fieldtheoryderivations}, but for brevity we do not write it out here.

Including $\mathcal{W}$, 
the full lattice action has the form
\ba\notag
\mathcal{S}[Y] &= 
\f{J^2}{2} \sum_{\<jk\>}\sum_{aA} (Y^{aA}_j{-} Y^{aA}_k)^2  
+ \f{1{-}z J^2}{2}\sum_{j} \sum_{aA} (Y_j^{aA})^2  \\
& - h^2 \sum_{j}\sum_{aA} Y_j^{aA} 
 + \f{1}{N_\mathrm{f}} \sum_j \Delta \mathcal{W}(Y_j),
\label{eq:actionbeforeshiftmaintext}
\end{align}
where $z=4$ arose from the coordination number of the square lattice.
The term 
$\Delta \mathcal{W}(Y_j)$, although it has a small prefactor, is important in generating the nontrivial interaction terms.

The next step depends crucially on the replica limit that we wish to consider. 
First we consider the $N\to 0$ limit.
In this limit we work with the unconstrained matrix $Y$. 
Subsequently we   consider the $N\to 1$ limit, 
where we must proceed differently, splitting $Y$ into massless and massive modes and eliminating the massive modes. 
(That is not possible at $N\to 0$, because in that limit all the modes are massless at the critical point \cite{nahum2021measurement}.)

\subsection{Obtaining the continuum action for ${N\to 0}$ theory}

As it stands, this action is not written in a very convenient form, because of the linear term. 
The next step is to   shift the field by a constant, ${Y^{aA}\rightarrow Y^{aA}+u}$,
 in order to eliminate this term.  In general, after making such a shift, we are still left with a nonzero mass term $\sum_{a,A} (Y^{aA})^2$.
However, there is a massless line in the $(h,J)$ phase diagram where this mass vanishes. 
For simplicity, let us focus on this massless line,  
which at large $N_\mathrm{f}$ is given by  (App.~\ref{app:DerivationfieldtheoryforNto0})
\be
J_c(h)^2  =\f{1}{z} \lf   
1 - \lf \f{9h^4}{2N_{\rm f}} \ri^{1/3}
+ \ldots  \ri.
\ee
The entangled phase is at ${J^2>J_c(h)^2}$.

When the shift is made, the nontrivial $\Delta \mathcal{W}$ term in Eq.~\eqref{eq:actionbeforeshiftmaintext}  generates linear, quadratic and cubic terms.
In App.~\ref{app:DerivationfieldtheoryforNto0} we find that after making the shift, 
and rescaling the field by a factor of $J$, the action takes the form
\begin{align}
\label{eq:actiondesiredform-crit-linemaintext}
\mathcal{S}[Y] = & 
\f{1}{2} \sum_{\<jk\>}\sum_{aA} (Y^{aA}_j- Y^{aA}_k)^2  
\nonumber\\
&+ \frac{\sigma}2 \sum_j \lf \sum_{aAB} Y_j^{aA} Y_j^{aB} + \sum_{aBA} Y_j^{aA} Y_j^{bA} \ri \nn\\
&+ g \sum_j \sum_{aA} (Y_j^{aA})^3 + \ldots,
\end{align}
with constants given by 
\ba
\sigma & = \frac1{J^2}\lf \f{9 h^4}{2 N_{\rm f}} \ri^{\!1/3},
& 
g & = - \frac1{J^3}\lf \f{2 h^2}{9 N_{\rm f}^2} \ri^{\!1/3},
\end{align}
where we took the limit ${N\to 0}$ to simplify the expressions.
As discussed in Sec.~\ref{s:FTforNt00}, the coupling relevant for perturbation theory is not $g$  but the combination
\be
g \sigma =  -\f{h^2}{ N_{\rm f} J^5}.
\ee
In \Eq{eq:actiondesiredform-crit-linemaintext} we have omitted terms that either  have more index sums, 
or higher powers of the field. 
This is because we now shift our focus from the case $d=2$ to higher dimensions:
at least for $d>6$, these omitted terms are less relevant according to the discussion of scaling dimensions in Sec.~\ref{s:Supersymmetry and dimensional reduction}.
While we have described the derivation for a  square lattice in $d=2$, it may be repeated for any lattice and in any dimension $d$. 
There will be modified geometrical constants in converting sums to integrals and lattice differences to derivatives, but otherwise the structure will be the same.
The key point is that when $N_\mathrm{f}$ is large, the expected continuum action is obtained in the regime of small $g\sigma$ where perturbative RG  is justified.

In Eq.~\eqref{eq:actiondesiredform-crit-linemaintext} the equation is still written on the lattice. The continuum Lagrangian follows immediately from a derivative expansion: 
\begin{align}\label{eq:continuumLNzeromaintext}
\mathcal{L}[Y] = & 
\f{1}{2}  \sum_{aA} (\nabla Y^{aA})^2 
\nonumber\\
&+ \frac{\sigma}2  \lf \sum_{aAB} Y^{aA} Y^{aB} + \sum_{abA} Y^{aA} Y^{bA} \ri \nn\\
&+ g  \sum_{aA} (Y^{aA})^3.
\end{align}
Keeping the lowest order in the derivative expansion is  justified here, as discussed in App.~\ref{app:DerivationfieldtheoryforNto0}.

Finally let us comment on boundary conditions.
Recall that using the replica trick to compute entropies requires pairwise index contractions between layers at the boundaries of the circuit (see App.~\ref{app:replicarecap} for a summary).
For the $N$-layer circuit, there are $N!$ ways to pair the $\mathcal{T}$ and $\mathcal{T}^*$ layers, 
and a given choice may be represented by a permutation ${\sigma\in S_N}$.
Such a choice imposes a symmetry-breaking boundary condition for $Y$.
Indeed the   microscopic boundary condition for $Y$ in \Eq{eq:continuumLNzeromaintext} is given in App.~\ref{app:replicarecap} as
\be\label{eq:Ybcmaintext}
Y^{aA} = J \sqrt{N_{\rm f}} \lf R^{(\sigma)}_{aA}  
+ O(N_{\rm f}^{-1/6})
\ri,
\ee
where $R^{(\sigma)}$ is the permutation $\sigma$ represented as a matrix of ones and zeros,
and where the dominant subleading term inside the brackets is due to the shift of the field ${Y\mapsto Y+u}$ that we performed.

\subsection{Continuum action for the $N\to 1$ theory}
\label{sec:deriveNto1action}

Finally we return to \Eq{eq:actionbeforeshiftmaintext} in order to consider the ${N\to 1}$ limit.
We separate the field $Y$ into pieces that transform in distinct representations of the global $G_N$ symmetry group:
\be\label{eq:fielddecomposition}
Y^{aA} :=  X^{aA} + L^a + R^A + \phi.
\ee
On the right hand side, any field with an index vanishes if that index is summed. 
This is a decomposition into four irreducible representations of the ${S_N\times S_N}$ symmetry, 
but  $L$ and $R$ combine into a single irrep of the full $G_N$ symmetry.\footnote{The global symmetry group is ${G_N=(S_N\times S_N)\rtimes \mathbb{Z}_2}$ as discussed in Sec.~\ref{sec:introduction}. The $\mathbb{Z}_2$ generator can be taken to be transposition of $X$. Since this operation exchanges $L$ and $R$,
these two fields combine into a single irrep of $G_N$. The fact that $L$ and $R$ are in the same irrep ensures that they have the same mass.}

We substitute this decomposition into the lattice action  in \Eq{eq:actionbeforeshiftmaintext}
 using the form for $\Delta \mathcal{W}$ given in App.~\ref{app:fieldtheoryderivations}, 
and shift $\phi$ to the minimum of the potential.
The key observation is  that,
when ${N>0}$,
the fields $L$, $R$ and $\phi$ remain massive at the point where $X$ becomes massless. 
As a result, they can simply be integrated out of the critical theory.

(We must be careful because the difference in masses is small in absolute terms when ${N_{\rm f}\gg 1}$, namely of order $N_{\rm f}^{-1/3}$.
Despite this,  the interactions are sufficiently weak at large $N_{\rm f}$ 
to ensure that integrating out the massive fields only gives subleading renormalizations of the couplings in the resulting theory for $X$.)

On the critical line, integrating out the massive fields and dropping terms of order higher than $O(X^4)$ leads to 
(App.~\ref{app:deriveNto1theory})
\ba \label{eq:SwidetildeXctm}
\mathcal{L}[  X] &=\f{1}{2} \sum_{aA} (  \nabla X^{aA})^2  
+ \f{g}{3!}\sum_{aA}   (X^{aA})^3,
\end{align}
with
\bea
g &=& - \frac1{J^3}\lf \f{192 h^2}{N _{\rm f}^2} \ri^{1/3}.
\eea
Here we have taken the ${N\to 1}$ limit, and we have rescaled $X$ 
so that the derivative term is normalized conventionally.
At leading order in $N_{\rm f}$, the critical line in the $(h,J)$ plane is again given by 
${J^2_c(h)= 1/z + \cdots}$.
Varying $J^2$ will induce a squared mass $m^2$ in the above Lagrangian which at leading order is given by ${z(J_c^2-J^2)}$, where again $z=4$ is the coordination number of the square lattice.

The boundary condition on $X$ corresponding to a permutation $\sigma$ is as in \Eq{eq:Ybcmaintext}, except that only the traceless part of the RHS is taken.

\section{Conclusions}
\label{sec:conclusions}

Our aim has been to develop ``Landau-Ginsburg-Wilson'' theories for entanglement transitions, based on the pairing order parameter $Y_{ab}$. 
We found that the associated Lagrangians 
can be derived from lattice models in a controlled way, 
 establishing the existence of two distinct theories for the ${N\to 0}$ and ${N\to 1}$ limits. 
The RG flows for these Lagrangians indicate the existence of  weak and strong coupling universality classes for the transition in high dimensions.

At the formal (perturbative) level, a  supersymmetric structure appears in the continuum theory of the random tensor network, leading to dimensional reduction by four, rather than by two as in famous examples of dimensional reduction in statistical mechanics.  

High-dimensional systems are not likely to be   accessible, but the existence of simple mean-field theories for the two kinds of problem is   relevant to more realistic models which have all-to-all connectivity rather than being spatially local (we will describe this in more detail separately).  

It will be interesting to further explore mean field scaling on trees \cite{lopez2020mean,deterministictreeinprep,moretrees} and  the phase transition to the   universality classes that govern trees made from more strongly random tensor components.\footnote{For example, trees built from $k$-legged tensors $T_{a_1, \ldots, a_k}$ whose probability distribution is invariant under unitary transformations on each  leg (i.e. $T_{a_1,a_2, \ldots, a_k}\rightarrow U_{a_l,a_l'} T_{a_1, \ldots, a_l', \ldots a_k}$)  show a distinct universal behavior.} (These tree universality classes  may be analogs of the finite-dimensional strong coupling universality classes.)

The present methods generalize to cases where the tensor networks or the quantum dynamics have a global symmetry. 
(The simplest case is a $\mathbb{Z}_2$ symmetry, which leads to quartic rather than cubic interactions: preliminary investigations of the ${N=1}$ theory again show a flow to strong coupling in low dimensions.)
In settings with global symmetries the field theories may be a useful tool for analysing more complex phase diagrams even when the critical behavior is not solvable. 

Finally, a question for the future is whether the field theories allow any approximate treatment
of the strong coupling regimes that are relevant for the phase transitions in low dimensions, for example using functional renormalization group methods.

\acknowledgements

We thank M.~Fava for discussions. AN and KJW were supported by the CNRS and the ENS.

\section*{Appendices}
\appendix
\section{Combinatorial factors in the Potts model}
\label{s:Potts-comb}

The standard treatment of the Potts model
\cite{ZiaWallace1975,Amit1976} 
involves rewriting the field $\Phi_\alpha$,
satisfying the constraint ${\sum_{\alpha=1}^N \Phi_\alpha = 0}$,
in terms of $N-1$ unconstrained fields $\phi_i$, with ${i=1,\ldots, N-1}$, 
via $\Phi_\alpha = \vec e_\alpha \cdot \vec \phi$.
Here $\{\vec e_\alpha\}$ are $N$ vectors $\vec e_\alpha$, of dimension $N-1$, s.t. 
\bea\label{4} 
   \vec e_\a \cdot \vec e_\beta &:=& 
\sum_{i=1}^{N-1}  e_\alpha^i    e_\beta^i =   \delta_{\alpha\beta} - \frac1N,  \\
e^i \circ e^j &:=& \sum_{\alpha=1}^N e^i_\alpha  e^j_\alpha = \delta ^{ij},\\
&& \sum_{\alpha=1}^N e_\alpha^i  =  0.
\label{0-sum}
\eea
The Potts Lagrangian\footnote{Our considerations remain valid if  the cubic interaction  is  replaced by $\sum_{\alpha} \Phi_\alpha^4$. This is the case studied by Amit \cite{Amit1976}.} is   given by \Eq{eq:definePottsLagrangian}. Writing the kinetic term in terms of $\vec\phi$ shows that the propagator for the original $\Phi_\alpha$ field is 
 $( \delta_{\alpha\alpha'}-\frac1N)/(k^2+m^2)$  (in analogy to \Eq{eq:matrixtheorypropagator} for the matrix theory).

As an explicit example  of a perturbative correction in the Potts model, we  
recall how the renormalization of the cubic vertex works to leading order. 
We split the propagator into the difference of two terms:  $ \delta_{\alpha\alpha'}/(k^2+m^2)$, which we denote with a thick solid line, and $1/[N(k^2+m^2)]$, which we denote with  a  dashed line (times $1/N$).

Graphically, the 1-loop diagram can be written as 
\bea
\fboxsep0mm
\lefteqn{\parbox{1.3cm}{{\begin{tikzpicture}[scale=1]
\coordinate (x1) at (0,0) ;
\coordinate (x2) at  (0.7,0) ;
\coordinate (x3) at  (0.35,0.5) ;
\coordinate (x1p) at (-.3,-0.1) ;
\coordinate (x2p) at  (1,-.1) ;
\coordinate (x3p) at  (0.35,0.8) ;
\node [below] at (x1)  {$x$} ;
\node [below] at (x2)  {$y$} ;
\node [above,left] at (x3)  {$z$} ;
\fill (x1) circle (1.5pt);
\fill (x2) circle (1.5pt);
\fill (x3) circle (1.5pt);
\draw [blue] (x1) -- (x2) -- (x3) -- (x1);
\draw [blue] (x1) -- (x1p) ;
\draw [blue] (x2) -- (x2p) ;
\draw [blue] (x3) -- (x3p) ;
\end{tikzpicture}}}
= \parbox{1.3cm}{{\begin{tikzpicture}[scale=1]
\coordinate (x1) at (0,0) ;
\coordinate (x2) at  (0.7,0) ;
\coordinate (x3) at  (0.35,0.5) ;
\coordinate (x1p) at (-.2,-0.1) ;
\coordinate (x2p) at  (0.9,-.1) ;
\coordinate (x3p) at  (0.35,0.75) ;
\fill (x1) circle (1.5pt);
\fill (x2) circle (1.5pt);
\fill (x3) circle (1.5pt);
\draw [blue,thick] (x1) -- (x2) -- (x3) -- (x1);
\draw [blue] (x1) -- (x1p) ;
\draw [blue] (x2) -- (x2p) ;
\draw [blue] (x3) -- (x3p) ;
\end{tikzpicture}}}}\nn\\
&& -\frac 1 N \Bigg[ ~\parbox{1.3cm}{{\begin{tikzpicture}[scale=1]
\coordinate (x1) at (0,0) ;
\coordinate (x2) at  (0.7,0) ;
\coordinate (x3) at  (0.35,0.5) ;
\coordinate (x1p) at (-.2,-0.1) ;
\coordinate (x2p) at  (0.9,-.1) ;
\coordinate (x3p) at  (0.35,0.75) ;
\fill (x1) circle (1.5pt);
\fill (x2) circle (1.5pt);
\fill (x3) circle (1.5pt);
\draw [blue,thick] (x1) -- (x2) -- (x3) ;
\draw [blue,thick,dashed] (x1) -- (x3) ;
\draw [blue] (x1) -- (x1p) ;
\draw [blue] (x2) -- (x2p) ;
\draw [blue] (x3) -- (x3p) ;
\end{tikzpicture}}}+\parbox{1.3cm}{{\begin{tikzpicture}[scale=1]
\coordinate (x1) at (0,0) ;
\coordinate (x2) at  (0.7,0) ;
\coordinate (x3) at  (0.35,0.5) ;
\coordinate (x1p) at (-.2,-0.1) ;
\coordinate (x2p) at  (0.9,-.1) ;
\coordinate (x3p) at  (0.35,0.75) ;
\fill (x1) circle (1.5pt);
\fill (x2) circle (1.5pt);
\fill (x3) circle (1.5pt);
\draw [blue,thick]  (x2) -- (x3) -- (x1);
\draw [blue,thick,dashed] (x1) -- (x2) ;
\draw [blue] (x1) -- (x1p) ;
\draw [blue] (x2) -- (x2p) ;
\draw [blue] (x3) -- (x3p) ;
\end{tikzpicture}}} 
+\parbox{1.3cm}{{{\begin{tikzpicture}[scale=1]
\coordinate (x1) at (0,0) ;
\coordinate (x2) at  (0.7,0) ;
\coordinate (x3) at  (0.35,0.5) ;
\coordinate (x1p) at (-.2,-0.1) ;
\coordinate (x2p) at  (0.9,-.1) ;
\coordinate (x3p) at  (0.35,0.75) ;
\fill (x1) circle (1.5pt);
\fill (x2) circle (1.5pt);
\fill (x3) circle (1.5pt);
\draw [blue,thick] (x3) -- (x1) -- (x2) ;
\draw [blue,thick,dashed] (x3) -- (x2) ;
\draw [blue] (x1) -- (x1p) ;
\draw [blue] (x2) -- (x2p) ;
\draw [blue] (x3) -- (x3p) ;
\end{tikzpicture}}}}
\Bigg] \nn\\
&& + \frac1{N^2}\Bigg[ ~\parbox{1.3cm}{{{\begin{tikzpicture}[scale=1]
\coordinate (x1) at (0,0) ;
\coordinate (x2) at  (0.7,0) ;
\coordinate (x3) at  (0.35,0.5) ;
\coordinate (x1p) at (-.2,-0.1) ;
\coordinate (x2p) at  (0.9,-.1) ;
\coordinate (x3p) at  (0.35,0.75) ;
\fill (x1) circle (1.5pt);
\fill (x2) circle (1.5pt);
\fill (x3) circle (1.5pt);
\draw [blue,thick] (x1) -- (x2);
\draw [blue,thick,dashed] (x1)--(x3) -- (x2);
\draw [blue] (x1) -- (x1p) ;
\draw [blue] (x2) -- (x2p) ;
\draw [blue] (x3) -- (x3p) ;
\end{tikzpicture}}}}+\parbox{1.3cm}{{{\begin{tikzpicture}[scale=1]
\coordinate (x1) at (0,0) ;
\coordinate (x2) at  (0.7,0) ;
\coordinate (x3) at  (0.35,0.5) ;
\coordinate (x1p) at (-.2,-0.1) ;
\coordinate (x2p) at  (0.9,-.1) ;
\coordinate (x3p) at  (0.35,0.75) ;
\fill (x1) circle (1.5pt);
\fill (x2) circle (1.5pt);
\fill (x3) circle (1.5pt);
\draw [blue,thick] (x3) -- (x2);
\draw [blue,thick,dashed] (x2)--(x1)--(x3);
\draw [blue] (x1) -- (x1p) ;
\draw [blue] (x2) -- (x2p) ;
\draw [blue] (x3) -- (x3p) ;
\end{tikzpicture}}}} +\parbox{1.3cm}{{{\begin{tikzpicture}[scale=1]
\coordinate (x1) at (0,0) ;
\coordinate (x2) at  (0.7,0) ;
\coordinate (x3) at  (0.35,0.5) ;
\coordinate (x1p) at (-.2,-0.1) ;
\coordinate (x2p) at  (0.9,-.1) ;
\coordinate (x3p) at  (0.35,0.75) ;
\fill (x1) circle (1.5pt);
\fill (x2) circle (1.5pt);
\fill (x3) circle (1.5pt);
\draw [blue,thick] (x3) -- (x1);
\draw [blue,thick,dashed] (x1)--(x2) -- (x3);
\draw [blue] (x1) -- (x1p) ;
\draw [blue] (x2) -- (x2p) ;
\draw [blue] (x3) -- (x3p) ;
\end{tikzpicture}}}}
\Bigg] \nn\\
 \nn\\
&& -\frac1{N^3} ~\parbox{1.3cm}{{{\begin{tikzpicture}[scale=1]
\coordinate (x1) at (0,0) ;
\coordinate (x2) at  (0.7,0) ;
\coordinate (x3) at  (0.35,0.5) ;
\coordinate (x1p) at (-.2,-0.1) ;
\coordinate (x2p) at  (0.9,-.1) ;
\coordinate (x3p) at  (0.35,0.75) ;
\fill (x1) circle (1.5pt);
\fill (x2) circle (1.5pt);
\fill (x3) circle (1.5pt);
\draw [blue,thick,dashed] (x1) -- (x2) -- (x3) -- (x1);
\draw [blue] (x1) -- (x1p) ;
\draw [blue] (x2) -- (x2p) ;
\draw [blue] (x3) -- (x3p) ;
\end{tikzpicture}}}}  \nn\\
&=& \left(1-\frac3{N}\right) \times{\parbox{.8cm}{{\begin{tikzpicture}[scale=1]
\coordinate (x1) at (0,0) ;
\coordinate (x2) at  (0.7,0) ;
\coordinate (x3) at  (0.35,0.5) ;
\fill (x1) circle (1.5pt);
\fill (x2) circle (1.5pt);
\fill (x3) circle (1.5pt);
\draw [thick] (x1) -- (x2) -- (x3) -- (x1);
\end{tikzpicture}}}}~\times ~
{\parbox{0.4cm}{{\begin{tikzpicture}[scale=1]
\coordinate (x1) at (0,0) ;
\coordinate (x1p) at (-.2,-0.1) ;
\coordinate (x1pp) at (.2,-0.1) ;
\coordinate (x1ppp) at (0,0.25) ;
\fill (x1) circle (1.5pt);
\draw [blue] (x1) -- (x1p) ;
\draw [blue] (x1pp) -- (x1)--(x1ppp) ;
\end{tikzpicture}}}}~~.
\label{33}
\eea
One sees that a solid line, which forces the adjacent indices to be equal, comes with a factor of $1$, whereas a broken line, which does not force the adjacent indices to be equal, comes with a factor of $-1/N$.
The   diagrams in the third and fourth line vanish due to \Eq{0-sum}, thus are not   reported in the final result. The remaining terms force all field indices to be equal, thus renormalize the interaction ${\parbox{0.4cm}{{\begin{tikzpicture}[scale=1]
\coordinate (x1) at (0,0) ;
\coordinate (x1p) at (-.2,-0.1) ;
\coordinate (x1pp) at (.2,-0.1) ;
\coordinate (x1ppp) at (0,0.25) ;
\fill (x1) circle (1.5pt);
\draw [blue] (x1) -- (x1p) ;
\draw [blue] (x1pp) -- (x1)--(x1ppp) ;
\end{tikzpicture}}}}=\sum_\alpha \Phi_\alpha^3$. This leads to the   combinatorial factor of
$C_\smalltriagdiag^{\text{Potts}}   = 1-3/N $ reported in \Eq{eq:CvalsforPotts} for the $N$-state Potts model as compared to YL. 

For the correction to $\Gamma^{(2)}$, the leading contribution is 
\be
\propdiagBlue
=
{\parbox{1.cm}{{\begin{tikzpicture}[scale=1]
\coordinate (x1) at (0,0) ;
\coordinate (x2) at  (0.6,0) ;
\coordinate (x1p) at  (-.2,0) ;
\coordinate (x2p) at  (0.8,0) ;
\fill (x1) circle (1.5pt);
\fill (x2) circle (1.5pt);
\draw [blue,thick] (.3,0) circle (3mm);
\draw [blue] (x1) -- (x1p);
\draw [blue] (x2) -- (x2p);
\end{tikzpicture}}}}
-\frac{2}{N} \,
{\parbox{1.cm}{{\begin{tikzpicture}[scale=1]
\coordinate (x1) at (0,0) ;
\coordinate (x2) at  (0.6,0) ;
\coordinate (x1p) at  (-.2,0) ;
\coordinate (x2p) at  (0.8,0) ;
\fill (x1) circle (1.5pt);
\fill (x2) circle (1.5pt);
\draw [thick,blue,dashed] (.6,0) arc (0:180:3mm);
\draw [thick,blue] (0,0) arc (-180:0:3mm);
\draw [blue] (x1) -- (x1p);
\draw [blue] (x2) -- (x2p);
\end{tikzpicture}}}}
+\frac1{N^2}\,
{\parbox{1.cm}{{\begin{tikzpicture}[scale=1]
\coordinate (x1) at (0,0) ;
\coordinate (x2) at  (0.6,0) ;
\coordinate (x1p) at  (-.2,0) ;
\coordinate (x2p) at  (0.8,0) ;
\fill (x1) circle (1.5pt);
\fill (x2) circle (1.5pt);
\draw [blue,thick,dashed] (.3,0) circle (3mm);
\draw [blue] (x1) -- (x1p);
\draw [blue] (x2) -- (x2p);
\end{tikzpicture}}}}~~.
\ee
The last term does not contribute due to rule \eq{0-sum}, resulting in  
\be
\propdiagBlue
=  \left(1-\frac2N\right)  \times   \propdiagAmp ~
\times  {\parbox{0.4cm}{{\begin{tikzpicture}[scale=1]
\coordinate (x1) at (0,0) ;
\coordinate (x1p) at  (-.2,0) ;
\coordinate (x1pp) at  (.2,0) ;
\fill (x1) circle (1.5pt);
\draw [blue] (x1) -- (x1p)--(x1pp);
\end{tikzpicture}}}}~~.
\ee
This leads to the combinatorial factor of 
$C_\smallpropdiag^{\text{Potts}}  = 1-2/N$ given in \Eq{eq:CvalsforPotts}.

For the MPT, we have to repeat our analysis for a matrix field $ X_{\alpha\beta}$. The key observation is that each of the index contractions  (for row/column indices) works independently, so that the only thing we have to do is to square the combinatorial factor for each diagram, leading to the rules of \Eq{eq:constantsgetsquared}.

Finally we observe that results for Yang-Lee are recovered for $N\to \infty$.

\section{The dimensionally reduced theory}

\subsection{Correction to $\sigma_{\rm L}$ and $\sigma_{\rm R}$}
\label{sec:sigmaoneloop}

In \Eq{eq:S-N=0} there is a term proportional to  $\sigma_{\rm L}$,  and another one proportional to $\sigma_{\rm R}$. Let us discuss the perturbative corrections to $\sigma_{\rm R}$. Write $
\Gamma_{\sigma_{\rm R}}[Y]:=\int_x \half {(1+\delta Z_{\sigma}) \sigma_{\rm R}} \sum_{a,A,B} Y_{aA}(x) Y_{aB}(x) $, where $\delta Z_{\sigma}$ contains its perturbative corrections we wish to calculate. 
Analogously,  we define $\Gamma_{\rm diag}[Y]:=\int_x \sum_{a,A}\frac12  (1+\delta Z) (\nabla Y_{aA})^2$. 

Taking into account the field renormalization, the correction to $\sigma_{\rm L}$ is 
\be
\frac{\delta \sigma_{\rm L}}{\sigma_{\rm L}} = \frac{1+\delta Z_\sigma}{1+ \delta Z} -1 = \delta Z_\sigma - \delta Z + ...,
\ee
Another way to interprete this ratio is to observe that each insertion of a red or blue dot comes with an additional propagator, of which the perturbative correction is given in the denominator. 

 The {\em field renormalization} $\delta Z$ is contained in   diagram  \eq{85}, 
\bea
&& \left[ 2\;\parbox{1cm}{{\begin{tikzpicture}[scale=1]
\coordinate (x1) at (0,0) ;
\coordinate (x2) at  (0.6,0) ;
\coordinate (x1p) at  (-.2,0) ;
\coordinate (x2p) at  (0.8,0) ;
\coordinate (x3) at  (0.3,0.3) ;
\coordinate (x4) at  (0.3,-0.3) ;
\fill (x1) circle (1.5pt);
\fill (x2) circle (1.5pt);
\fill [blue] (x3) circle (1.5pt);
\fill [red] (x4) circle (1.5pt);
\draw (.3,0) circle (3mm);
\draw [black] (x1) -- (x1p);
\draw [black] (x2) -- (x2p);
\end{tikzpicture}}}
+4\;
\parbox{1cm}{{\begin{tikzpicture}[scale=1]
\coordinate (x1) at (0,0) ;
\coordinate (x2) at  (0.6,0) ;
\coordinate (x1p) at  (-.2,0) ;
\coordinate (x2p) at  (0.8,0) ;
\coordinate (x3) at  (0.25,-0.3) ;
\coordinate (x4) at  (0.35,-0.3) ;
\fill (x1) circle (1.5pt);
\fill (x2) circle (1.5pt);
\fill [blue] (x3) circle (1.5pt);
\fill [red] (x4) circle (1.5pt);
\draw (.3,0) circle (3mm);
\draw [black] (x1) -- (x1p);
\draw [black] (x2) -- (x2p);
\end{tikzpicture}}} \right] \nn\\
&&=\int_{{s_1,s_{2}}}(s_{1}+s_{2})^{2-\frac{d}{2}} \rme^{-\frac{s_{1}s_{2} }{s_{1}+s_{2}} p^{2}-s_{1}-s_{2}}.
\eea
To extract  $\delta Z$, we take a $p^2$-derivative, 
\begin{align}
&  \delta Z = -\frac{g^2 \sigma_{\rm L} \sigma_{\rm R}}{2}  \frac{\rmd}{\rmd p^2}
\left[ 2\;\parbox{1cm}{{\begin{tikzpicture}[scale=1]
\coordinate (x1) at (0,0) ;
\coordinate (x2) at  (0.6,0) ;
\coordinate (x1p) at  (-.2,0) ;
\coordinate (x2p) at  (0.8,0) ;
\coordinate (x3) at  (0.3,0.3) ;
\coordinate (x4) at  (0.3,-0.3) ;
\fill (x1) circle (1.5pt);
\fill (x2) circle (1.5pt);
\fill [blue] (x3) circle (1.5pt);
\fill [red] (x4) circle (1.5pt);
\draw (.3,0) circle (3mm);
\draw [black] (x1) -- (x1p);
\draw [black] (x2) -- (x2p);
\end{tikzpicture}}}
+4\;
\parbox{1cm}{{\begin{tikzpicture}[scale=1]
\coordinate (x1) at (0,0) ;
\coordinate (x2) at  (0.6,0) ;
\coordinate (x1p) at  (-.2,0) ;
\coordinate (x2p) at  (0.8,0) ;
\coordinate (x3) at  (0.25,-0.3) ;
\coordinate (x4) at  (0.35,-0.3) ;
\fill (x1) circle (1.5pt);
\fill (x2) circle (1.5pt);
\fill [blue] (x3) circle (1.5pt);
\fill [red] (x4) circle (1.5pt);
\draw (.3,0) circle (3mm);
\draw [black] (x1) -- (x1p);
\draw [black] (x2) -- (x2p);
\end{tikzpicture}}} \right] \nn\\
&{=\,} \frac{g^2 \sigma_{\rm L} \sigma_{\rm R}}{2} \left( - \frac{\rmd}{\rmd p^2}\right) \int_{{s_1,s_{2}}}(s_{1}+s_{2})^{{2-\frac{d}2}} \rme^{-\frac{s_{1}s_{2} }{s_{1}+s_{2}} p^{2}-s_{1}-s_{2}} \nn\\
& {=\,} \frac{g^2 \sigma_{\rm L} \sigma_{\rm R}}{2} \int_{{s_1,s_{2}}}(s_{1}+s_{2})^{{1-\frac{d}2}} s_1s_2\,\rme^{-\frac{s_{1}s_{2} }{s_{1}+s_{2}} p^{2}-s_{1}-s_{2}} . 
\end{align}
The correction $\delta Z_\sigma$ is
\bea
&&\!\!\! \delta Z_\sigma= \frac{g^2 \sigma_{\rm L} \sigma_{\rm R} }2 \left[ 4\;\parbox{1cm}{{\begin{tikzpicture}[scale=1]
\coordinate (x1) at (0,0) ;
\coordinate (x2) at  (0.6,0) ;
\coordinate (x1p) at  (-.2,0) ;
\coordinate (x2p) at  (0.8,0) ;
\coordinate (x3) at  (0.25,-0.3) ;
\coordinate (x4) at  (0.35,-0.3) ;
\coordinate (x6) at  (0.3,0.3) ;
\fill (x1) circle (1.5pt);
\fill (x2) circle (1.5pt);
\fill [blue] (x3) circle (1.5pt);
\fill [red] (x4) circle (1.5pt);
\fill [red] (x6) circle (1.5pt);
\draw (.3,0) circle (3mm);
\draw [black] (x1) -- (x1p);
\draw [black] (x2) -- (x2p);
\end{tikzpicture}}}\right] \nn\\
  &&\!\!\!{=\,}\frac{g^2 \sigma_{\rm L} \sigma_{\rm R} }2\!\!\int\limits_k \!\!\! \int\limits_{s_1}\!\!\!\int\limits_{s_2}    \frac{2(s_1{+}s_2) s_1 s_2 }{1! 2!}  \rme^{{-}s_1(k{+}\frac p2)^2-s_2(k{-}\frac p2)^2{-}s_1{-}s_2}\nn\\
&&\!\!\!{=\,}  \frac{g^2 \sigma_{\rm L} \sigma_{\rm R} }2 \int_{{s_1,s_{2}}}(s_{1}+s_{2})^{{1-\frac{d}2}} s_1s_2\,\rme^{-\frac{s_{1}s_{2} }{s_{1}+s_{2}} p^{2}-s_{1}-s_{2}} .
\eea
There is an equivalent correction to $\sigma_{\rm L}$ (the blue dot). 
In summary
\bea
&&\!\!\! \frac{\delta \sigma_{\rm L}}{\sigma_{\rm L}} = 
\frac{\delta \sigma_{\rm R}}{\sigma_{\rm R}}  \nn\\
&&= 
\frac{g^2 \sigma_{\rm L} \sigma_{\rm R}}{2} \left\{4\;\parbox{1cm}{{\begin{tikzpicture}[scale=1]
\coordinate (x1) at (0,0) ;
\coordinate (x2) at  (0.6,0) ;
\coordinate (x1p) at  (-.2,0) ;
\coordinate (x2p) at  (0.8,0) ;
\coordinate (x3) at  (0.25,-0.3) ;
\coordinate (x4) at  (0.35,-0.3) ;
\coordinate (x6) at  (0.3,0.3) ;
\fill (x1) circle (1.5pt);
\fill (x2) circle (1.5pt);
\fill [blue] (x3) circle (1.5pt);
\fill [red] (x4) circle (1.5pt);
\fill [red] (x6) circle (1.5pt);
\draw (.3,0) circle (3mm);
\draw [black] (x1) -- (x1p);
\draw [black] (x2) -- (x2p);
\end{tikzpicture}}}  + \frac{\rmd}{\rmd p^2}
\left[ 2\;\parbox{1cm}{{\begin{tikzpicture}[scale=1]
\coordinate (x1) at (0,0) ;
\coordinate (x2) at  (0.6,0) ;
\coordinate (x1p) at  (-.2,0) ;
\coordinate (x2p) at  (0.8,0) ;
\coordinate (x3) at  (0.3,0.3) ;
\coordinate (x4) at  (0.3,-0.3) ;
\fill (x1) circle (1.5pt);
\fill (x2) circle (1.5pt);
\fill [blue] (x3) circle (1.5pt);
\fill [red] (x4) circle (1.5pt);
\draw (.3,0) circle (3mm);
\draw [black] (x1) -- (x1p);
\draw [black] (x2) -- (x2p);
\end{tikzpicture}}}
+4\;
\parbox{1cm}{{\begin{tikzpicture}[scale=1]
\coordinate (x1) at (0,0) ;
\coordinate (x2) at  (0.6,0) ;
\coordinate (x1p) at  (-.2,0) ;
\coordinate (x2p) at  (0.8,0) ;
\coordinate (x3) at  (0.25,-0.3) ;
\coordinate (x4) at  (0.35,-0.3) ;
\fill (x1) circle (1.5pt);
\fill (x2) circle (1.5pt);
\fill [blue] (x3) circle (1.5pt);
\fill [red] (x4) circle (1.5pt);
\draw (.3,0) circle (3mm);
\draw [black] (x1) -- (x1p);
\draw [black] (x2) -- (x2p);
\end{tikzpicture}}} \right] \right\} \nn\\
&&=0.
\eea
We note a similar relation for the   subdominant diagram correcting 
$\sum_{abAB}Y_{aA} Y_{bB}$ (without prefactor), 
\bea
&&4 \parbox{1cm}{{\begin{tikzpicture}[scale=1]
\coordinate (x1) at (0,0) ;
\coordinate (x2) at  (0.6,0) ;
\coordinate (x1p) at  (-.2,0) ;
\coordinate (x2p) at  (0.8,0) ;
\coordinate (x3) at  (0.25,-0.3) ;
\coordinate (x4) at  (0.35,-0.3) ;
\coordinate (x5) at  (0.25,0.3) ;
\coordinate (x6) at  (0.35,0.3) ;
\fill (x1) circle (1.5pt);
\fill (x2) circle (1.5pt);
\fill [blue] (x3) circle (1.5pt);
\fill [red] (x4) circle (1.5pt);
\fill [blue] (x5) circle (1.5pt);
\fill [red] (x6) circle (1.5pt);
\draw (.3,0) circle (3mm);
\draw [black] (x1) -- (x1p);
\draw [black] (x2) -- (x2p);
\end{tikzpicture}}} = 4 \int_{{s_1,s_{2}}}(s_{1}+s_{2})^{{-\frac{d}2}} \frac{s_1^2 s_2^2}{2!2!}\,\rme^{-\frac{s_{1}s_{2} }{s_{1}+s_{2}} p^{2}-s_{1}-s_{2}}  \nn\\
&& =\left(- \frac{\rmd }{\rmd p^2}\right)^{\!2}  \left[ 2\;\parbox{1cm}{{\begin{tikzpicture}[scale=1]
\coordinate (x1) at (0,0) ;
\coordinate (x2) at  (0.6,0) ;
\coordinate (x1p) at  (-.2,0) ;
\coordinate (x2p) at  (0.8,0) ;
\coordinate (x3) at  (0.3,0.3) ;
\coordinate (x4) at  (0.3,-0.3) ;
\fill (x1) circle (1.5pt);
\fill (x2) circle (1.5pt);
\fill [blue] (x3) circle (1.5pt);
\fill [red] (x4) circle (1.5pt);
\draw (.3,0) circle (3mm);
\draw [black] (x1) -- (x1p);
\draw [black] (x2) -- (x2p);
\end{tikzpicture}}}
+4\;
\parbox{1cm}{{\begin{tikzpicture}[scale=1]
\coordinate (x1) at (0,0) ;
\coordinate (x2) at  (0.6,0) ;
\coordinate (x1p) at  (-.2,0) ;
\coordinate (x2p) at  (0.8,0) ;
\coordinate (x3) at  (0.25,-0.3) ;
\coordinate (x4) at  (0.35,-0.3) ;
\fill (x1) circle (1.5pt);
\fill (x2) circle (1.5pt);
\fill [blue] (x3) circle (1.5pt);
\fill [red] (x4) circle (1.5pt);
\draw (.3,0) circle (3mm);
\draw [black] (x1) -- (x1p);
\draw [black] (x2) -- (x2p);
\end{tikzpicture}}} \right]
\eea
It looks like there is an exact compensation for the subdominant term in the 2-point function as well.

\subsection{Another 1-loop example}
In a cubic theory, the quartic vertex is generated at fourth order in the cubic coupling, and is proportional to 
\bea
\lefteqn{\parbox{8mm}{{\begin{tikzpicture}[scale=1]
\coordinate (x1) at (0,0) ;
\coordinate (x2) at  (0.7,0) ;
\coordinate (x3) at  (0.7,0.7) ;
\coordinate (x4) at (0,0.7) ;
\fill (x1) circle (1.5pt);
\fill (x2) circle (1.5pt);
\fill (x3) circle (1.5pt);
\fill (x4) circle (1.5pt);
\draw [black] (x1) -- (x2) -- (x3) -- (x4) -- (x1);
\end{tikzpicture}}}= \int_k \frac 1{(k^2+1)^4} 
=\int_{s>0} \int_k\frac{s^3}{3!} \rme^{-s (k^2+1)}
}\nn\\
&=& \frac1{3!}\int_{s>0} s^{3-d/2}\rme^{-s}
=
\frac1{3!} \Gamma\Big(4-\frac{d}2\Big)
\label{4-bis}
\eea
In the disordered theory, this diagram becomes
\bea
12\;
\parbox{8mm}{{\begin{tikzpicture}[scale=1]
\coordinate (x1) at (0,0) ;
\coordinate (x2) at  (0.7,0) ;
\coordinate (x3) at  (0.7,0.7) ;
\coordinate (x4) at (0,0.7) ;
\coordinate (x12a) at (0.35,0) ;
\coordinate (x12b) at (0,0.35) ;
\fill (x1) circle (1.5pt);
\fill (x2) circle (1.5pt);
\fill (x3) circle (1.5pt);
\fill (x4) circle (1.5pt);
\fill [blue] (x12a) circle (1.5pt);
\fill [red] (x12b) circle (1.5pt);
\draw [black] (x1) -- (x2) -- (x3) -- (x4) -- (x1);
\end{tikzpicture}}}
+8\;
\parbox{8mm}{{\begin{tikzpicture}[scale=1]
\coordinate (x1) at (0,0) ;
\coordinate (x2) at  (0.7,0) ;
\coordinate (x3) at  (0.7,0.7) ;
\coordinate (x4) at (0,0.7) ;
\coordinate (x12a) at (0.25,0) ;
\coordinate (x12b) at (0.45,0) ;
\fill (x1) circle (1.5pt);
\fill (x2) circle (1.5pt);
\fill (x3) circle (1.5pt);
\fill (x4) circle (1.5pt);
\fill [blue] (x12a) circle (1.5pt);
\fill [red] (x12b) circle (1.5pt);
\draw [black] (x1) -- (x2) -- (x3) -- (x4) -- (x1);
\end{tikzpicture}}} &=& 20 \int_k \frac1{(k^2+1)^6} \nn\\
&=&\frac{20}{5!}  \int_s s^{5-d/2} \rme^{-s} \nn\\
&=& \frac1{3!} \Gamma\Big(6- \frac d2 \Big) 
\eea
This is the dimensionally reduced version of \Eq{4-bis}.

\subsection{An example for dimensional reduction  at 2-loop order}
\label{s:dim-red-2-loop-example}

The first 2-loop diagram of the pure theory reads (with parameters $s_1$, $s_2$ and $s_3$ from top to bottom)
\newcommand{\hut}{\coordinate (x1) at (0,0) ;
\coordinate (x2) at  (0.7,0) ;
\coordinate (x3) at  (0,0.5) ;
\coordinate (x4) at (0.7,0.5) ;
\coordinate (x5) at (0.35,1) ;
\fill (x1) circle (1.5pt);
\fill (x2) circle (1.5pt);
\fill (x3) circle (1.5pt);
\fill (x4) circle (1.5pt);
\fill (x5) circle (1.5pt);
\draw [black] (x1) -- (x2) -- (x4) -- (x3) -- (x1);
\draw [black] (x4)--(x5)--(x3);}
\newcommand{\rco}{\fill [red] (r1) circle (1.5pt);}
\newcommand{\rct}{\fill [red] (r2) circle (1.5pt);}
\newcommand{\bco}{\fill [blue] (b1) circle (1.5pt);}
\newcommand{\bct}{\fill [blue] (b2) circle (1.5pt);}
\newcommand{\cccc}{\rco \rct \bco \bct}
\bea\label{Hut1}
\parbox{8mm}{{\begin{tikzpicture}[scale=1]
\hut
\end{tikzpicture}}} &=& \int_{k,p} \int_{s_1,s_2,s_3} 
\frac{s_1s_3^2}{2}{\rme^{-s_1 k^2-s_2(k+p)^2-s_3 p^2 -s_1-s_2-s_3}} \nn\\
&=& \int_{s_1,s_2,s_3} 
\frac{s_1s_3^2}{2} \frac{\rme^{-s_1-s_2-s_3}} {(s_1 s_2 + s_1 s_3 + s_2 s_3)^{d/2}}.
\eea
For illustration, we first evaluate the corresponding diagram 
for a disordered vector model ($\sigma_{\rm L}=0$, and $N_{\rm R}=1$)
\bea\label{eq:S-vector-N=0}
\mathcal{L} [\Phi] &=&  \sum_{i=1}^{N}   \half \left[ \nabla \Phi_{i}(x) \right]^2 + \frac{m^2}2 \Phi_{i}(x)^2 + \frac{g}{3!} \Phi_{i}^3 \nn\\ 
&&   - \frac{\sigma}2
\sum_{i,j=1}^{N} \Phi_{i}(x) \Phi_{j}(x)
\eea
Combinatorics for the vector theory: lines are upper ``u", middle ``m", and lower ``l", allowing for $(1)=\rm (ul)$, $(2)=(\rm ml)$, $(3)=(\rm um)$, 
\bea
&&\!\!\!6 \;\parbox{8mm}{{\begin{tikzpicture}[scale=1]
\hut
\coordinate (r1) at (0.35,0) ;
\coordinate (r2) at (0.18,0.75) ;
\rco\rct
\end{tikzpicture}}}
+
3\;\parbox{8mm}{{\begin{tikzpicture}[scale=1]
\hut
\coordinate (r1) at (0.35,0) ;
\coordinate (r2) at (0.35,0.5) ;
\rco\rct
\end{tikzpicture}}}
+
2\;\parbox{8mm}{{\begin{tikzpicture}[scale=1]
\hut
\coordinate (r1) at (0.35,0.5) ;
\coordinate (r2) at (0.18,0.75) ;
\rco\rct
\end{tikzpicture}}} 
\nn\\
&&=\int_{k,p} \int_{s_1,s_2,s_3} \frac{s_1s_3^2}{2}
 {\rme^{-s_1 k^2-s_2(k+p)^2-s_3 p^2 -s_1-s_2-s_3}} \times
\nn\\
&& \qquad \qquad \times\left[ 6 \times \frac{s_1 s_3}{2\times 3} + 3\times{\frac{s_2 s_3}{3}} + 2 \frac{s_1 s_2}{2} \right] \nn\\
 &&= \int_{s_1,s_2,s_3} 
\frac{s_1s_3^2}{2} \frac{\rme^{-s_1-s_2-s_3}} {(s_1 s_2 + s_1 s_3 + s_2 s_3)^{d/2-1}}\nn\\
&&= \parbox{8mm}{{\begin{tikzpicture}[scale=1]
\hut
\end{tikzpicture}}}\,\Bigg|_{d\to d-2}
\eea
For the tensor theory we mark again with  blue and red dots; we do not draw twice if obtained as all red and blue exchanged, that are accounted for by an explicit factor of 2. 
We give diagrams $[(1)\times (1)]$, twice $[(1)\times (2)]$, twice $[(1)\times(3)]$, $[(2)\times (2)]$, twice $[(2)\times(3)]$ and $[(3)\times(3)]$:
\bea
&&\left[4\times 6\;\parbox{8mm}{{\begin{tikzpicture}[scale=1]
\hut
\coordinate (b1) at (0.45,0) ;
\coordinate (r1) at (0.25,0) ;
\coordinate (b2) at (0.15,0.7) ;
\coordinate (r2) at (0.22,0.8) ;
\cccc
\end{tikzpicture}}}
+
2\times 6\;\parbox{8mm}{{\begin{tikzpicture}[scale=1]
\hut
\coordinate (b1) at (0.45,0) ;
\coordinate (r1) at (0.25,0) ;
\coordinate (b2) at (0.52,0.75) ;
\coordinate (r2) at (0.18,0.75) ;
\cccc
\end{tikzpicture}}}
+4\times 6\;\parbox{8mm}{{\begin{tikzpicture}[scale=1]
\hut
\coordinate (b1) at (0,0.25) ;
\coordinate (r1) at (0.35,0) ;
\coordinate (b2) at (0.15,0.7) ;
\coordinate (r2) at (0.22,0.8) ;
\cccc
\end{tikzpicture}}}
+
2\times 6\;\parbox{8mm}{{\begin{tikzpicture}[scale=1]
\hut
\coordinate (b1) at (0,0.25) ;
\coordinate (r1) at (0.35,0) ;
\coordinate (b2) at (0.52,0.75) ;
\coordinate (r2) at (0.18,0.75) ;
\cccc
\end{tikzpicture}}}\,\right]
\nn\\
&&+2   \left[2\times 6 \;\parbox{8mm}{{\begin{tikzpicture}[scale=1]
 \hut
\coordinate (b1) at (0.45,0) ;
\coordinate (r1) at (0.25,0) ;
\coordinate (b2) at (0.35,0.5) ;
\coordinate (r2) at (0.18,0.75) ;
\cccc
\end{tikzpicture}}}+2\times 6\;\parbox{8mm}{{\begin{tikzpicture}[scale=1]
\hut
\coordinate (b1) at (0.35,0) ;
\coordinate (b2) at (0.35,0.5) ;
\coordinate (r1) at (0,0.25) ;
\coordinate (r2) at (0.18,0.75) ;
\cccc
\end{tikzpicture}}}\,\right]\nn\\
&&+2\left[4\times 3\;\parbox{8mm}{{\begin{tikzpicture}[scale=1]
\hut
\coordinate (b1) at (0.35,0.5) ;
\coordinate (r1) at (0.35,0) ;
\coordinate (b2) at (0.15,0.7) ;
\coordinate (r2) at (0.22,0.8) ;
\cccc
\end{tikzpicture}}}
+
2\times 3\;\parbox{8mm}{{\begin{tikzpicture}[scale=1]
\hut
\coordinate (b1) at (0.35,0.5) ;
\coordinate (r1) at (0.35,0) ;
\coordinate (b2) at (0.52,0.75) ;
\coordinate (r2) at (0.18,0.75) ;
\cccc
\end{tikzpicture}}}\,\right]\nn\\
&&+\left[2\times 6\;\parbox{8mm}{{\begin{tikzpicture}[scale=1]
\hut
\coordinate (b1) at (0.45,0) ;
\coordinate (r1) at (0.25,0) ;
\coordinate (b2) at (0.45,0.5) ;
\coordinate (r2) at (0.25,0.5) ;
\cccc
\end{tikzpicture}}}
+
2\times 6\;\parbox{8mm}{{\begin{tikzpicture}[scale=1]
\hut
\coordinate (b1) at (0,0.25) ;
\coordinate (r1) at (0.35,0) ;
\coordinate (b2) at (0.45,0.5) ;
\coordinate (r2) at (0.25,0.5) ;
\cccc
\end{tikzpicture}}}\,\right] + 2\left[2\times 2\times 3\;\parbox{8mm}{{\begin{tikzpicture}[scale=1]
\hut
\coordinate (b1) at (0.45,0.5) ;
\coordinate (r1) at (0.35,0) ;
\coordinate (b2) at (0.18,0.75) ;
\coordinate (r2) at (0.25,0.5) ;
\cccc
\end{tikzpicture}}}\,\right]\nn\\
&&+\left[4\times 2\;\parbox{8mm}{{\begin{tikzpicture}[scale=1]
\hut
\coordinate (b1) at (0.45,0.5) ;
\coordinate (r1) at (0.25,0.5) ;
\coordinate (b2) at (0.15,0.7) ;
\coordinate (r2) at (0.22,0.8) ;
\cccc
\end{tikzpicture}}}
+
2\times 2\;\parbox{8mm}{{\begin{tikzpicture}[scale=1]
\hut
\coordinate (b1) at (0.45,0.5) ;
\coordinate (r1) at (0.25,0.5) ;
\coordinate (b2) at (0.52,0.75) ;
\coordinate (r2) at (0.18,0.75) ;
\cccc
\end{tikzpicture}}}\,\right]\nn\\
&&=  \int_{k,p} \int_{s_1,s_2,s_3} 
\frac{s_1s_3^2}{2}{\rme^{-s_1 k^2-s_2(k+p)^2-s_3 p^2 -s_1-s_2-s_3}}\times \nn \\
&& \qquad \times \bigg[ 72  \frac{s_1^2 s_3^2}{6 \cd   12 } + 48 \frac{s_1 s_2 s_3^2}{2\cd 1\cd 12 } +36 \frac{s_1^2 s_2 s_3}{6\cd 1\cd 3} + 24 \frac{s_2^2 s_3^2}{2 \cd 12} \nn\\
&& \qquad \qquad + 24 \frac{s_1 s_2^2 s_3}{2\cd 2 \cd 3}+ 12 \frac{s_1^2 s_2^2}{6\cd 2}\bigg] \nn\\
&&=  \int_{k,p} \int_{s_1,s_2,s_3} 
\frac{s_1s_3^2}{2}{\rme^{-s_1 k^2-s_2(k+p)^2-s_3 p^2 -s_1-s_2-s_3}}\times \nn \\
&& \qquad \times \Big[{s_1^2 s_3^2}  {+} 2  {s_1 s_2 s_3^2}  {+}2  {s_1^2 s_2 s_3}  {+}{s_2^2 s_3^2}  {+} 2  {s_1 s_2^2 s_3} {+}{s_1^2 s_2^2} \Big] \nn\\
&&=  \int_{k,p} \int_{s_1,s_2,s_3} 
\frac{s_1s_3^2}{2}{\rme^{-s_1 k^2-s_2(k+p)^2-s_3 p^2 -s_1-s_2-s_3}}\times \nn \\
&& \qquad \times [  s_1 s_2 + s_1 s_3 + s_2 s_3 ]^2 
\nn
\eea
\bea
&&=\int_{s_1,s_2,s_3} 
\frac{s_1s_3^2}{2} \frac{ \rme^{ -s_1-s_2-s_3}}{(s_1 s_2 + s_1 s_3 + s_2 s_3)^{d/2-2}}
\nn\\
&&= \,\parbox{8mm}{{\begin{tikzpicture}[scale=1]
\hut
\end{tikzpicture}}}\,\Bigg|_{d\to d-4}
\label{89}.
\eea

\section{Replacing Cardy fields with fermions}
\label{s:Cardy fields and supersymmetry}

Our starting point is the replica Lagrangian ${\mathcal{L}=\mathcal{L}_2+ \mathcal{L}_3}$
(Eqs.~\ref{eq:quadraticaftercardy},~\ref{eq:cubictransformed})
written in the Cardy field basis
\ba
& y_{IJ}, 
& 
I,J & \in \{ +, - , 1,\ldots, m \},
\end{align}
with ${m\equiv N-2}$, and ${m\to -2}$ in the limit $N\to 0$.
Lowercase indices $i,j$  run over the numerical values ${1,\ldots, m}$: 
in this Appendix we refer to these as the replica indices.
Our aim here is to map the Lagrangian to one for fields that carry no such indices.

It is sufficient to consider   the terms in $\mathcal{L}$ that involve $i,j$ indices.
As many of the terms repeat the same index structures, it is enough to discuss a subset of them: the quadratic terms in
\ba\label{eq:quadraticpart}
\ca L_2' & = 
  \nabla y_{i-}
   \nabla y_{i+}+  \nabla y_{-j}
   \nabla y_{+j}  +\frac{1}{2}
 (\nabla  y_{ij})(\nabla  y_{ij}) \nn\\
& +  \half \left( y_{-j}y_{-j} +y_{i-} y_{i-}  \right) ,
\end{align}
and the cubic term with the distinct index structure
\be\label{eq:cubictransformedpart}
\ca L_3' =  {g\sigma}  \bigg[   y_{i +} y_{i j } y_{+j }  \bigg] .
\ee
For clarity we  use $i$ for left indices and $j$ for right indices, though they run over the same set of values.

First, we exchange the (real) fields that carry replica indices for half the number of complex fields. 
Any field $A_i$ with a single left replica index ${i=1,\ldots,m}$ 
(for example $y_{i+}$)
is replaced by  a pair of complex conjugate fields, denoted $A'_\alpha$, $A'_{\bar \alpha}$, which carry an index  ${\alpha=1,\ldots, m/2}$:
\ba
A_{\alpha}' & = \f{1}{\sqrt 2} (A_{2\alpha-1} + i A_{2\alpha} ), 
&
A_{\bar \alpha}' & = \f{1}{\sqrt 2} (A_{2\alpha-1} - i A_{2\alpha} ).
\end{align}
The equivalent replacement is made for fields with right indices (which after the transformation we label by $\beta$).
For the field $y_{ij}$ we make the transformations for both indices, so that (for example)
\ba
y_{\alpha\beta}' = \f{1}{2} ( y_{2\alpha-1,2\beta-1} + i y_{2\alpha-1,2\beta} + i y_{2\alpha, 2\beta -1} - y_{2\alpha, 2\beta}). 
\end{align}
From now on we drop the primes on the new fields.
Note that  complex conjugating a field replaces unbarred  indices with barred ones and vice versa.

With the above convention (repeated indices summed)
\ba
  A_i B_i & = A_{\bar \alpha} B_\alpha + B_{\bar \alpha} A_{\alpha},  
\end{align}
if $A_i$, $B_i$ are fields carrying a single replica index, and 
\ba
\f{1}{2} y_{ij} y_{ij} & = y_{\bar \alpha \bar \beta} y_{\alpha \beta} + y_{\alpha \bar \beta} y_{\bar \alpha \beta}.
\end{align}
The terms under examination  become
\ba\label{eq:lagrangiancomplex}
\ca L_2' & = 
(  \nabla y_{\bar \alpha -}
   \nabla y_{\alpha +}  + \nabla y_{\bar \alpha +}
   \nabla y_{\alpha -}  ) \nn \\
    &  +  (  \nabla y_{- \bar \beta }
   \nabla y_{+ \beta }  + \nabla y_{+ \bar \beta }
   \nabla y_{- \beta }  ) \nn \\
  &   +
( \nabla y_{\bar \alpha \bar \beta} \nabla y_{\alpha \beta} + \nabla y_{\bar \alpha \beta}
\nabla y_{\alpha \bar \beta} 
) 
   \nn\\
&
+  \left( y_{-\bar \beta }y_{- \beta} +y_{\bar \alpha -} y_{\alpha -}  \right) ,
\\
\ca L_3'
   & = 
 {g\sigma}  \bigg[   
y_{\bar \alpha +}  y_{+ \beta} y_{\alpha \bar \beta }
+
y_{\bar \alpha +}  y_{+  \bar \beta} y_{\alpha  \beta }
\nn
\\
& \quad\quad\quad +
 y_{\bar \alpha \bar \beta } y_{+ \beta}  y_{ \alpha +} 
+
y_{\bar \alpha  \beta }  y_{+  \bar \beta}  y_{ \alpha +}
  \bigg]  .
\end{align}
We have placed the terms with an  $\bar \alpha$ index to the left and those with an $\alpha$ index to the right, to match the rewriting of the action in \Eq{eq:Sschematic1} below.
The ordering does not matter at this point (since the fields are commuting) but becomes important when we introduce fermions.

The second step is the ``fermionization'' of the fields that carry a left replica index (denoted by $\alpha$ in the above formulas). 
These fields make up a collection of ${m/2 \to -1}$ complex vectors, indexed by $\alpha$,
of the form
\be
w^\alpha = (y_{\alpha+}, y_{\alpha-},y_{\alpha,1},\ldots, y_{\alpha,m/2},y_{\alpha,\bar 1},\ldots, y_{\alpha,\overline{m/2}})^T,
\ee
together with their complex conjugates
\be
(w^\alpha)^\dag = (y_{\bar \alpha+}, y_{\bar \alpha-},y_{\bar \alpha,\bar 1},\ldots, y_{\bar \alpha,\overline{m/2}},y_{\bar \alpha, 1},\ldots, y_{\bar \alpha, m/2}).
\ee
The action may be written in the schematic form (spatial dependence is suppressed) 
\be\label{eq:Sschematic1}
\mathcal{S}  = \sum_{\alpha=1}^{m/2} (w^\alpha)^\dag \mathcal{A} w^\alpha + \mathcal{S}_\text{rest},
\ee
where $\mathcal{S}_\text{rest}$ and the kernel $\mathcal{A}$ depend only on the fields \textit{without} a left index.
The integral over the fields $w^\alpha$, $(w^\alpha)^\dag$
 is a Gaussian integral over complex bosons, giving ${(\det \mathcal{A})^{-m/2}\to(\det \mathcal{A})}$ in the replica limit.
 Therefore it is equivalent to the corresponding integral in which we replace the multiplet $\{w^\alpha\}_{\alpha=1}^{m/2}$ of vectors with a single fermionic vector, and similarly for the conjugates.
 When we make this replacement we use the notation:
\bea
y_{\alpha,\pm} &\to& \lambda_\pm, \nn \\
y_{\bar \alpha,\pm} &\to& \bar \lambda_\pm, \nn\\
y_{\alpha,\beta} &\to& \eta^{B}_\beta, \nn \\ 
y_{\bar \alpha,\bar \beta} &\to& \bar\eta^{B}_{ \beta}, \nn\\
y_{\alpha,\bar \beta} &\to& { \bar \eta^{M}_{ \beta}}, \nn\\ 
y_{\bar \alpha, \beta} &\to&{ -  \eta^{M}_{ \beta} }.
\eea
Note that $\eta^B$, $\bar \eta^B$,  $\eta^M$ and $\bar \eta^M$ are independent Grassman variables: 
the above naming convention with barred and unbarred variables is   for notational convenience in the next step. The additional minus sign in the last line ensures that $  \nabla     \bar \eta_\beta^M  \nabla \eta^M_\beta  $ in the   equation for $\ca L_2'  $ has the same sign as the preceding terms. 

The replacement gives the Lagrangian terms
\ba
\ca L_2' 
& \to 
 (  \nabla  \bar \lambda_-
   \nabla  \lambda_+
   + \nabla \bar \lambda_+
   \nabla  \lambda_-
    ) 
   \nn \\
    &  +  (  \nabla y_{- \bar \beta }
   \nabla y_{+ \beta }  + \nabla y_{+ \bar \beta }
   \nabla y_{- \beta }  ) \nn \\
  &   +
( \nabla    \bar \eta^B_\beta
 \nabla  \eta^B_\beta
 +   \nabla     \bar \eta_\beta^M  \nabla \eta^M_\beta  
 ) 
   \nn\\
&
+  \left( y_{-\bar \beta }y_{- \beta} +\bar\lambda_- \lambda_-  \right) ,
\\
\ca L_3' & \to 
     {g\sigma}  \bigg[   
  - \bar \eta^M_\beta
 \bar\lambda_+
 y_{+ \beta} 
+
y_{+  \bar \beta}
 \bar \lambda_+
\eta^B_\beta \nn \\
& \quad \quad \quad + \bar \eta^B_\beta
 \lambda_+
  y_{+ \beta} 
+
 y_{+  \bar \beta}  
 \lambda_+
 \eta^M_\beta
  \bigg]  .
\end{align}
We have ordered the fields with a $\bar \beta$ index on the left and those with a $\beta$ index on the right (taking account of minus signs from exchanging Grassman fields).
As a result, the action is   in an analogous form to \Eq{eq:Sschematic1}.
Defining, for each value of $\beta$,
 a supervector and its conjugate
\ba\label{eq:supervectorpsi}
\psi^\beta &  = ( y_{+\beta}, y_{-\beta}, \eta^B_\beta, \eta^M_\beta)^T,
\\
( \psi^\beta)^\dag &  = ( y_{+\bar \beta}, y_{-\bar \beta}, \bar\eta^B_\beta, \bar \eta^M_\beta),
\end{align}
the action is   in the form
\be
\mathcal{S}'=
\sum_{\beta=1}^{m/2} ( \psi^\beta)^\dag \mathcal{B}\, \psi^\beta + \mathcal{S}_\text{rest}',
\ee
where $\mathcal{S}_\text{rest}'$ and  $\mathcal{B}$ are independent of the fields with a right ($\beta$) index.
The kernel $\mathcal{B}$ is   a supermatrix. 
Standard formulas for integration over supervectors 
\cite{HaakeBook,Wegner2016}
show (in analogy to the previous step) that taking the limit ${m/2 \to -1}$ for the multiplicity is equivalent to replacing bosons with fermions and fermions with bosons.
The names of the new fields after the replacement are chosen to be
\bea
\bar \eta_\beta^B &\to & \bar B, \nn \\
 \eta_\beta^B &\to &   B,\nn \\
  y_{+,\beta} &\to& \rho_+,\nn \\
  y_{+,\bar\beta} &\to&\bar \rho_+,\nn \\
  \eta_{\beta}^M  &\to&  -  M, \nn \\
\bar \eta_{\beta}^M  &\to& - \bar M.  
\eea
(and similarly for $y_{-,\beta}$ etc.)
Note that $B$, $\bar B$, $M$, $\bar M$ are bosonic, while $\rho_\pm$ and $\bar \rho_\pm$ are Grassmanian.

This finally gives
\ba
\ca L_2 
 & = 
(  \nabla  \bar \lambda_-
   \nabla  \lambda_+
   + \nabla \bar \lambda_+
   \nabla  \lambda_-
    ) 
   \nn \\
    &  +  ( 
  \nabla  \bar\rho_-
   \nabla   \rho_+
   + 
   \nabla  \bar\rho_+
   \nabla  \rho_-
   ) \nn \\
  &   +
( \nabla  \bar B
 \nabla  B
 +  
  \nabla  \bar M
  \nabla M
 ) 
   \nn\\
&
+ 
 \left( 
\bar \rho_-
 \rho_-
 +
 \bar\lambda_- \lambda_-  \right) 
\end{align}
and 
\ba
\ca L_3
 & = 
 {g\sigma}  \bigg[   
 +
 \bar B
 \lambda_+
\rho_+
+
B
\bar \rho_+
 \bar \lambda_+
+
\bar M
 \bar\lambda_+
\rho_+
+
M
 \lambda_+
\bar\rho_+
  \bigg]  ,
\end{align}
matching the corresponding terms in  \Eq{S-bos-fer} of the main text.  
The full \Eq{S-bos-fer} is obtained by restoring the remaining terms in ${\mathcal{L}_2+\mathcal{L}_3}$.
This is straightforward since these terms either do not involve $i$, $j$ indices 
(i.e.\ they depend only on $y_{\pm \pm}$ and are left unchanged by the transformation in this appendix)
or they involve the same index structures that appear in $\mathcal{L}_2'$ and are transformed similarly.

\section{Brief recap of the replica approach}
\label{app:replicarecap}

We start with the random tensor network, made of uncorrelated random tensors.
For a concrete example, we consider the geometry mentioned in the main text, where   the second coordinate is treated as  ``time'' and the tensor network forms an  $L\times T$ cylinder.
Let the bond indices for initial time boundary  be 
${\mathbf{S}_{1}^{\rm i}, \ldots, \mathbf{S}_{L}^{\rm i}}$,
and those for the final time boundary be
${\mathbf{S}_{1}^{\rm f}, \ldots, \mathbf{S}_{L}^{\rm f}}$.
The full network is then a tensor $\mathcal{T}$ 
 with elements 
\be
 \mathcal{T}({\mathbf{S}_{1}^{\rm f}, \ldots, \mathbf{S}_{L}^{\rm f}};
 {\mathbf{S}_{1}^{\rm i}, \ldots, \mathbf{S}_{L}^{\rm i}}).
 \ee
It can be viewed as a transition amplitude for  a nonunitary evolution operator 
that acts on a spin configuration on $L$ sites. 
Let us use a matrix notation: 
 the list ${\mathbf{S}_{1}^{\rm f}, \ldots, \mathbf{S}_{L}^{\rm f}}$ together is the ``row'' index of $\mathcal{T}$, 
 and ${\mathbf{S}_{1}^{\rm i}, \ldots, \mathbf{S}_{L}^{\rm i}}$ its ``column'' index. 

 Next, let us recall that the replica limit may be used to study the entanglement properties of this tensor network \cite{vasseur2019entanglement,nahum2021measurement}. 
The basic object is the product of $N$ copies of $\mathcal{T}$ and $N$ copies of the complex conjugate, $\mathcal{T}^*$. 
 By forming appropriate index contractions of such a product, and then taking the replica limit $N\to 0$, we may express any desired average.
 
 As a concrete example of this, we can quantify correlations between the initial and final time \cite{gullans2020dynamical} using the  R\'enyi entropies $S_n$   defined by 
 \be
 \exp\lf   - (n-1) S_n\ri := \f{\Tr (\mathcal{T} \mathcal{T}^\dag)^n}{(\Tr \mathcal{T} \mathcal{T}^\dag)^n}.
 \ee
The limit ${S_\text{vN} := \lim_{n\to 1} S_n}$ is the von Neumann entropy.
Formally, these entropies characterize the spectrum of singular values  of the matrix 
$\mathcal{T}$. 
If the singular values of $\mathcal{T}$ 
form the set 
${c \{\sqrt{p_1}, \sqrt{p_2}, \ldots \}}$,
where $c$ is an overall constant which normalizes the $p_i$ so that they sum up to 1, then 
\bea
&& \exp\lf   - (n-1) S_n\ri  = \sum_i p_i^n, \\
&& S_\mathrm{vN} = - \sum_i p_i \ln p_i.
 \eea
Physically, if  we think of the action of $\mathcal{T}$ as a nonunitary evolution of an $L$-site system which starts out in a maximum entropy state, then $S_\text{vN}$ quantifies how much entropy the system has left at the final time \cite{gullans2020dynamical}.

In the replica approach, we write the generating function for $S_n$ in the form
 \be\label{eq:genfnreplica}
 \overline{\exp\lf   - k (n-1) S_n\ri}
 = 
 \lim_{N\rightarrow 0}
\overline{
{\lf  \Tr (\mathcal{T} \mathcal{T}^\dag)^{n} \ri^k }{(\Tr \mathcal{T} \mathcal{T}^\dag)^{N-kn}}
}.
 \ee
 Here the overline is an average over the random tensors. 
 In principle, if we can compute the limit ${N\to 0}$ on the right hand side, 
 then the resulting generating function (with argument $k$) contains information on all the cumulants of the random variable $S_n$:
 \be
 \overline{\exp\lf   - k (n-1) S_n\ri}
 =
 \exp \lf
\sum_{r=1}^\infty \f{k^r(n-1)^r}{r!} \overline{S_n^r}^{\rm{c}}
 \ri.
\ee 
Having obtained the cumulants, it is also possible to take the subsequent limit ${n\to 1}$ to obtain the von Neumann entropy.
 See  Ref.~\cite{fava2023nonlinear} for an example where this approach was used to obtain exact results.
 
 By taking $N$ to be an integer, with $N\geq kn$, the object appearing inside the average on the RHS can be written in terms of index contractions of the tensor product
 \be
\overset{N}{\overbrace{\mathcal{T}\otimes \cdots \otimes \mathcal{T}}}
\otimes
\overset{N}{\overbrace{\mathcal{T}^*\otimes \cdots \otimes \mathcal{T}^*}}.
 \ee
 Let us think of the $2N$ terms in this product as $2N$ ``layers'' of a multi-layer tensor network. 
Concretely, these index contractions mean that boundary indices are contracted pairwise between $\mathcal{T}$ and $\mathcal{T}^*$ layers, in a manner prescribed by the matrix products and traces in \Eq{eq:genfnreplica}. To see this, let us simplify notation by writing
${\mathcal{T}_{\mathbf{S}_{\rm f} \mathbf{S}_{\rm i}}}$  in place of  ${
 \mathcal{T}(\mathbf{S}^{\rm f}, \mathbf{S}^{\rm i})}$.
Two simple   cases are 
\ba
\Tr \mathcal{T}\mathcal{T}^\dag & = \sum_{\mathbf{S}_{\rm f}, \mathbf{S}_{\rm i}}
\mathcal{T}_{\mathbf{S}_{\rm f} \mathbf{S}_{\rm i}}\mathcal{T}_{\mathbf{S}_{\rm f} \mathbf{S}_{\rm i}}^*,
\\
 \Tr (\mathcal{T}\mathcal{T}^\dag)^2 & = 
 \sum_{\mathbf{S}_{\rm f} \mathbf{S}_{\rm i},\mathbf{S}_{\rm f}' \mathbf{S}_{\rm i}'}
\mathcal{T}_{\mathbf{S}_{\rm f}' \mathbf{S}_{\rm i}}
\mathcal{T}_{\mathbf{S}_{\rm f} \mathbf{S}_{\rm i}}^*
\mathcal{T}_{\mathbf{S}_{\rm f} \mathbf{S}_{\rm i}'}
\mathcal{T}_{\mathbf{S}_{\rm f}' \mathbf{S}_{\rm i}'}^*. \label{eq:contractionexample}
\end{align}
Once we write each $\mathcal{T}$ in terms of its constituent tensors, we obtain the partition function of an ``Ising model'' with $2N$ layers.
The Ising spins in the bulk are summed over freely, with the weight described in the main text.
The Ising spins at the two temporal boundaries are identified between layers in a pairwise fashion (generalizing \Eq{eq:contractionexample}).

To connect with the standard notation, we recall that
the pattern of pairing at a given boundary can be expressed in terms of a permutation ${\sigma\in S_N}$.
Let the $\mathcal{T}$ layers be labelled ${a=1,\ldots, N}$
 and let the $\mathcal{T}^*$ layers be numbered ${b=1,\ldots, N}$.
 Then the permutation $\sigma$ pairs the $\mathcal{T}$ layer numbered $a$  with the $\mathcal{T}^*$ layer numbered $b=\sigma(a)$.
At the initial-time boundary we can take $\sigma$ to be the identity permutation.
At the final time boundary we can take $\sigma$ to be the product of $k$ disjoint $n$-cycles.

In terms of our overlap order parameter $Y^{ab}$, these boundary conditions impose a large boundary value for some of the elements of $Y$.
On the boundary links of the tensor network, the corresponding Ising spins satisfy
\be
S_\mu^a = \bar S_\mu^{\sigma(a)},
\ee
where $\sigma$ is the appropriate permutation and we have suppressed the position index.
Therefore,
by the definition of $Y$ in \Eq{Y-def}, components of $Y$ of the form $Y^{a\, \sigma(a)}$ are given by sums of terms of the form ${(S^a)^2=1}$,
giving
\be
Y^{a\,\sigma(a)} = \sqrt{N_{\rm f}}.
\ee
By contrast, other components of $Y$ are sums of terms that can be positive or negative and (by standard central-limit theorem reasoning) are naively smaller by order $N_{\rm f}^{-1/2}$. Therefore (as stated in the main text)
\be
Y_{aA} = \sqrt{N_{\rm f}} \lf R^{(\sigma)}_{aA}  
+ O (N_{\rm f}^{-1/2})
\ri,
\ee
where ${R^{(\sigma)}_{ab}=\delta_{b,\sigma(a)}}$ is the  permutation written as a matrix.
So this is essentially a fixed boundary condition for $Y$.
After the shift $Y^{ab}\mapsto Y^{ab}+u$  discussed in App.~\ref{app:DerivationfieldtheoryforNto0}, there is a  subleading term inside the brackets that is independent of the indices and of order $N_{\rm f} ^{-1/6}$ compared to the leading term.
(In the expression \Eq{eq:Ybcmaintext} in the main text we have also taken into account a rescaling by $J$ performed below to put the kinetic term in the standard form.)

Next we briefly consider the effect of taking $N\to 1$ rather than $N\to 0$. 
Above, the physical average was defined   as the Gaussian average over the fields appearing in the random tensors,  
\be
\mathbb{E}_\text{phys} ( \cdots )
:= \overline{(\cdots) }.
\ee
Let us   modify this definition to
\be\label{eq:modifiedav}
\mathbb{E}_\text{mod} ( \cdots ) 
:= \mathcal{N} \times {\overline{( \cdots )\Tr\mathcal{T}\mathcal{T}^\dag}},
\ee
with normalization constant
\be\mathcal{N} = \lf {\overline{\Tr\mathcal{T}\mathcal{T}^\dag}} \ri^{-1}.
\ee
In other words, we are  reweighting the probability measure for the random tensors by ${\Tr\mathcal{T}\mathcal{T}^\dag}$, which itself depends on the tensors.
When we apply the replica trick, the additional factor of ${\Tr \mathcal{T}\mathcal{T}^\dag}$ in \Eq{eq:modifiedav} means that we need to take ${N\to 1}$ in \Eq{eq:genfnreplica} rather than ${N\to 0}$.\footnote{In terms of our lattice field theory, the normalization factor $\mathcal{N}$ in 
\Eq{eq:modifiedav}
is   the partition function at $N=1$. This is taken into account automatically in our derivation of the continuum field theory: after we discard the massive mode $C$, the partition function is trivial when~${N=1}$.}

The structure in \Eq{eq:modifiedav} also appears in monitored dynamics. For comparison, consider  the example of a circuit made up of unitaries and projection operators representing measurements of $\hat \sigma^z$ for qubits.\footnote{In that setting, we could take  $\overline{(\cdots)}$ to represent the average over the random unitaries, together with a ``flat'' sum over measurement outcomes, i.e.~without taking the nontrivial probabilities from Born's rule into account (note that this normalization convention gives ${\overline{1}=2^M}$ where $M$ is the number of measurements). 
 \label{footnote:normalizationofmeasurementaverage}} 
In that context, the nonunitary circuit $\mathcal{T}$ can be viewed as a Kraus operator 
(in the Kraus decomposition of the quantum channel defined by the measurement process),
and the factor of ${\Tr \mathcal{T} \mathcal{T}^\dag}$ in \Eq{eq:modifiedav} is the probability of the measurement outcomes.
The difference is that, here, the operators $\mathcal{T}$ do not make up a Kraus decomposition\footnote{In the  true measurement process, we would have (using the normalization convention in Footnote~\ref{footnote:normalizationofmeasurementaverage})
\be
\overline{\mathcal{T}^\dag \mathcal{T}} =\mathbb{1},
\ee
so that 
\be\label{eq:freeenergyconstraint}
{\ca Z}_1 = \Tr \overline{\mathcal{T}^\dag \mathcal{T}} = 2^L.
\ee 
We have used the symbol $\ca Z_1$ because this is the partition function of the replica theory, with certain boundary conditions, when ${N=1}$.
\Eq{eq:freeenergyconstraint} implies that the free energy of the replicated theory is trivial when $N$ is set equal to 1.
(This can be compared with the lattice action which we obtained for our microscopic model. 
In that lattice action, setting ${N = 1}$ leaves only the massive mode $\phi$ --- see App.~\ref{app:deriveNto1theory} --- which we found to be unimportant for the critical behavior.)}
\be
\overline{\mathcal{T}^\dag \mathcal{T}} \not = \mathbb{1}.
\ee
Therefore, while it is certainly possible to interpret the average defined in  
\Eq{eq:modifiedav}
in terms of a physical process, 
this process is not simply given by applying random unitaries together with weak Born-rule  measurements. 
It can be thought of as a partially postselected dynamics, where the probabilities arising from Born's rule are modified by a postselection process.
We   use it as an example of a problem that shares the  replica symmetry group with more physically natural monitored systems 
(i.e.\ the symmetry group $G_N$ with $N\rightarrow 1$).
This makes it plausible that more realistic microscopic models for monitored dynamics share the same continuum description. Direct derivations of pure Born-rule dynamics will be discussed elsewhere.

\section{Derivation of field theories from lattice}
\label{app:fieldtheoryderivations}

As discussed in Sec.~\ref{sec:derivefieldtheory}, the flavor number $N_{\rm f}$ can be used for a controlled derivation of the continuum field theories 
(in the weak coupling regime where perturbative RG is appropriate).
Here we give a full description of this derivation.
For simplicity and concreteness we work with a 2D tensor network, 
but the derivation generalizes to any   tensor-network geometry, 
using tensors of the same structure as those below.

The first part of the discussion applies to both  limits, $N\to 0$ and $N\to 1$.
In App.~\ref{app:DerivationfieldtheoryforNto0} we derive the field  theory for the limit $N\rightarrow 0$, with an unconstrained matrix field $Y^{aA}$.
Note that in this appendix we put the replica indices as superscripts, 
to allow room for other indices, 
and to distinguish row from column indices we use lowercase for the former and uppercase for the latter.
For the measurement limit $N\rightarrow 1$, there is another step, which is to isolate the massless part of the field that satisfies the constraints ${\sum_\alpha X^{\alpha \beta}=0}$, ${\sum_\beta X^{\alpha \beta}=0}$: we do this subsequently in App.~\ref{app:deriveNto1theory}.

\subsection{Replicated action on the lattice}

The tensor network has been defined in Sec.~\ref{sec:derivefieldtheory}.
We repeat its form:
\begin{equation}\label{eq:ZIsing2}
Z[h,J] 
= \sum_{\{S\}}  \exp \! \Big( 
\sum_j\sum_\mu h_j^\mu S_{j\mu} {+}
\sum_{\langle i,j\rangle} \sum_{\mu,\nu} J_{ij}^{\mu\nu} S_{i\mu}S_{j\nu}
\Big). 
\end{equation}
After introducing replicas as in \Eq{eq:replicasofpartitionfn},
we take the average (denoted by an overline) over the quenched random variables $h$ and $J$ via e.g.
\begin{align}
&\overline{\exp\!\lf 
\sum_{\mu} \sum_a S^a_\mu ( h^{\mu}_{\Re} + i h^{\mu}_{\Im })     
+ 
 \sum_{\mu}   \sum_A  \bar S^A_\mu  (h^{\mu}_{\Re } -i h^{\mu}_{\Im } )  
\ri
} \nn\\
&=\exp \lf
\f{{h^2}}{\sqrt{N_{\rm f}}} \sum_{\mu,a,A}  S^a_\mu \bar S^A_\mu
\ri.
\end{align}
This motivates defining the overlap matrix
\be\label{Y-def-app}
Y^{aA} := \f{1}{\sqrt{N_{\rm f}}} \sum_\mu S^a_\mu \bar{S}^A_\mu.
\ee
Averaging the $J$ terms similarly gives
\be
\mathcal{S}_0 [Y]= 
- h^2 \sum_j \sum_{aA} Y_j^{aA} 
- J^2 \sum_{\langle jk\rangle} \sum_{aA} Y_j^{aA} Y_k^{aA}.
\ee
The multi-replica partition function still involves a sum over  spins,
\be
\mathcal{Z} = \sum_{\{S_{j\mu}^a, \bar{S}_{j\mu}^a\}} \exp ( - \mathcal{S}_0 ).
\ee
In the limit of large $N_{\rm f}$, we may exchange these discrete sums for an integral over $Y$, with an appropriate weight ${\rm e}^{-\mathcal W}$  absorbed into the potential. It is sufficient to consider a single position $j$,
\be
\f{1}{2^{2N\times N_{\rm f}}} 
\sum_{ \{S^a_\mu, \bar{S}^A_\mu \}} 
\longrightarrow 
\int \dd X  {\rm e}^{ - \mathcal W(Y) }. 
\ee
We have ${\mathcal W(Y) = \mathcal W(-Y)}$ by the symmetry ${(S,\bar S) \rightarrow (-S, \bar S)}$ of the flat measure on $S$ and $\bar S$.
When $N_{\rm f}\rightarrow \infty$, $Y$ becomes Gaussian, since it is a sum of a large number of variables. As a result,   $\mathcal W$ is of the form $ \mathcal W= \f{1}{2} \sum_{aA} (Y^{aA})^2 + \mathcal  O(Y^4)$.
Once we have computed $ \mathcal W$,  the full lattice action is given as ${\mathcal{S} = \mathcal{S}_0 +  \mathcal W}$.

There are probably more efficient ways to compute $\mathcal W$ but here is one way. (We   neglect multiplicative factors in the partition function.) 
First, introduce 
a representation of a $\delta$-function to enforce the definition \eqref{Y-def-app}:
\begin{widetext}
\ba
{\rm e}^{-\mathcal W(Y)} & \sim  \left< \int \lf \prod_{aA} \dd \lambda^{aA} \ri
\exp \lf
i \lambda^{aA} Y^{aA} - \f{i}{N_{\rm f}^{1/2}} \lambda^{aA} S_\mu^a \bar S_\mu^A
\ri\right>,
\end{align}
where the brackets denote the 
normalized sum ${2^{-2N\times N_{\rm f}}} \sum_{\{S^a_\mu, \bar{S}^a_\mu \}}$, i.e.\ the 
flat average over $S$ and $\bar S$. 
Let us formally expand in the second term above (and abbreviate the notation a little):
\begin{align}
{\rm e}^{-\mathcal W(Y)} & \sim  \int \dd \lambda\,
{\rm e}^{i\lambda  \cdot Y} 
 \bigg[
 1 - \f{1}{2 N_{\rm f}} \lambda^{aA} \lambda^{bB}  
\brackets{S^a_\mu S^b_\nu}  \brackets{\bar S^A_\mu \bar S^B_\nu}
\nn\\
& ~~~~~~~~~~~~~~~~~~~~~~~~~~~+
 \f{1}{4! N_{\rm f}^2}
 \lambda^{aA} \lambda^{bB} \lambda^{cC} \lambda^{dD} 
 \brackets{S^a_\mu S^b_\nu S^c_\lambda S^d_\kappa}   \brackets{\bar S^A_\mu \bar S^B_\nu \bar S^C_\lambda \bar S^D_\kappa}
 + \ldots \bigg]. 
\end{align}
Using $\brackets{S^a_\mu S^b_\nu}=\delta^{ab}\delta_{\mu\nu}$ we see that the second term in the brackets $\sim \lambda\cdot \lambda$ gives the first
term in the expansion of the expected Gaussian weight,
\ba
{\rm e}^{-\mathcal W(Y)} & \sim   \int \dd \lambda\,
{\rm e}^{i\lambda\cdot Y} 
 \lf
 1 - \f{1}{2 } \lambda^{aA} \lambda^{aA}  
 +
 \f{1}{4! N_{\rm f}^2}
 \lambda^{aA} \lambda^{bB} \lambda^{cC} \lambda^{dD} \,
 \brackets{S^a_\mu S^b_\nu S^c_\lambda S^d_\kappa}   \brackets{\bar S^A_\mu \bar S^B_\nu \bar S^C_\lambda \bar S^D_\kappa}
 + \ldots \ri.
\end{align}
Let us absorb the Gaussian part explicitly,
\ba\label{eq:WabsorbedGaussian}
{\rm e}^{-\mathcal W(Y)} & \sim   \int \dd \lambda\,
{\rm e}^{i\lambda \cdot Y - \f{1}{2} \lambda \cdot\lambda} 
 \lf
 1  +
 \f{1}{4! N_{\rm f}^2}
 \lambda^{aA} \lambda^{bB} \lambda^{cC} \lambda^{dD} 
 \brackets{S^a_\mu S^b_\nu S^c_\lambda S^d_\kappa}   \brackets{\bar S^A_\mu \bar S^B_\nu \bar S^C_\lambda \bar S^D_\kappa}
 - \f{1}{8} (\lambda\cdot \lambda)^2
 + \ldots \ri.
\end{align}
\end{widetext}
The needed averages are of the form
\be\label{eq:Sav}
 \brackets{S^a_\mu S^b_\nu S^c_\lambda S^d_\kappa}  
 = (\delta^{ab}_{\mu\nu}  \delta^{cd}_{\lambda\kappa}  + \text{2 terms}) - 2 \delta^{abcd}_{\mu\nu\lambda\kappa},
\ee
where the parentheses include the three ways of pairing the four spins.
 $\delta^{ab}_{\mu\nu} =\delta^{ab}\delta_{\mu\nu}$, and  $\delta^{abcd}_{\mu\nu\lambda\kappa}=\delta^{abcd}\delta_{\mu\nu\lambda\kappa}$ (and $\delta^{abcd}$ is 1 only if $a=b=c=d$).
We   refer to the  terms in brackets in (\ref{eq:Sav}) as  paired terms.

Inserting this formula for the averages into the quartic term in Eq.~(\ref{eq:WabsorbedGaussian}) gives several terms.
The leading-order terms are those with two free sums over flavor indices. These come from choosing a paired term from the average over $S$, and the corresponding paired term from the average over $\bar S$. There are three such terms. They give $\f{1}{8}(\lambda\cdot \lambda)^2$, cancelling the final term in 
Eq.~(\ref{eq:WabsorbedGaussian}).

All other terms give a single free flavor-index sum, so are of order $1/N_{\rm f}$ once the prefactor is taken into account.
Let us define 
\be\label{g-def}
g^{abcd} = (\delta^{ab}  \delta^{cd}  + \text{2 terms}) - 2 \delta^{abcd},
\ee
which is the same structure of contractions as in (\ref{eq:Sav}), but without the flavour indices. Next set
\be\label{f-def}
f^{abcd}_{ABCD} = \delta^{ab} \delta^{cd} \delta_{AB} \delta_{CD} + \text{2 terms},
\ee
which is a sum over the three pairing patterns, with the same pattern for both row (lower case)  and column (upper case) indices.
Finally, define
\be\label{H-def}
H^{abcd}_{ABCD} = g^{abcd} g_{ABCD} - f^{abcd}_{ABCD}.
\ee
Here, the final term is subtracting off one of the terms that come from expanding the two $g$  tensors, since this term is of  order $(N_{\rm f})^0$, and was shown above to cancel.
Therefore
\begin{widetext}
\ba
\nonumber
{\rm e}^{-\mathcal W(Y)} & \sim   \int \dd \lambda\,
{\rm e}^{i\lambda \cdot Y - \f{1}{2} \lambda\cdot \lambda} 
 \lf
 1  +
 \f{1}{4! N_{\rm f}} H^{abcd}_{ABCD} 
 \lambda^{aA} \lambda^{bB} \lambda^{cC} \lambda^{dD} 
 +  \mathcal  O(\lambda^6/N_{\rm f}^2) \ri \\
 & = \int \dd \lambda\,
\exp\! \lf {i\lambda \cdot Y - \f{1}{2} \lambda\cdot \lambda +  \f{1}{4! N_{\rm f}} H^{abcd}_{ABCD} 
 \lambda^{aA} \lambda^{bB} \lambda^{cC} \lambda^{dD} 
 + \mathcal  O({\lambda^6}/{ N_{\rm f}^2} )} \ri. 
 \label{eq:Ylambda}
\end{align}
 Each additional power of $\lambda^2$ in the exponent comes with an additional factor of $1/N_{\rm f}$.\footnote{In more detail: we can see this by noting that (\ref{eq:Ylambda}) amounts to  writing the measure for $Y$
(defined by the sum over $S$ and $\bar S$)
as the Fourier transform of the corresponding generating function for $Y$. 
By the relation between the generating function of a random variable and its cumulants, this means that the term of order $\lambda^k$ is proportional to the $k$th cumulant of $Y$. From 
 \Eq{Y-def-app}, where $Y$ is written as a sum of independent random variables, we may argue that the $k$th cumulant of $Y$ is of order $N_{
 \rm f}^{1-k/2}$,  giving the scaling asserted above.}

Shifting ${\lambda \to \lambda + i Y}$ gives 
\ba
{\rm e}^{-\mathcal W(Y)}   \sim  &\int \dd \lambda\, 
{\rm e}^{- \f{1}{2} \lambda\cdot \lambda - \f{1}{2} Y\cdot Y}  
\exp\!  \lf  \f{1}{4! N_{\rm f}}   H^{abcd}_{ABCD} 
 (\lambda+iY)^{aA}( \lambda+iY)^{bB} (\lambda+iY)^{cC} (\lambda+iY)^{dD} 
 + \ldots  \ri. 
\end{align}
At leading order in $1/N_{\rm f}$, it is sufficient to take the first term in a cumulant expansion, where we average over $\lambda$ with ${\langle \lambda^{aA}\lambda^{bB} \rangle_\lambda = \delta^{ab}\delta^{AB}}$. As a result, 
\ba
{\rm e}^{-\mathcal{W}(Y)}  & =
{\rm e}^{- \f{1}{2} Y\cdot Y} 
\exp \lf  \f{1}{4! N_{\rm f}} H^{abcd}_{ABCD} 
\Big< (\lambda+iY)^{aA}( \lambda+iY)^{bB} (\lambda+iY)^{cC} (\lambda+iY)^{dD}  \Big>_{\lambda} \ri. 
\end{align}
Adding $\mathcal S_0[ Y]$ and the contribution from $\mathcal W(Y_j)$, summed over all sites $j$, we obtain the action as
\ba\notag
\mathcal{S}[Y] &= \mathcal{S}_0[Y] + \sum_j\mathcal{W}(Y_j)\nn\\ \nn
& =\f{J^2}{2} \sum_{\<jk\>}\sum_{aA} (Y^{aA}_j- Y^{aA}_k)^2  - h^2 \sum_{j}\sum_{aA} Y_j^{aA} + 
\f{1-z J^2}{2}\sum_{j} \sum_{aA} (Y_j^{aA})^2 
\\
&  -  \f{1}{4! N_{\rm f}}
\sum_j \sum_{abcdABCD}  H^{abcd}_{ABCD}  \Big< (\lambda_j+iY_j)^{aA}( \lambda_j+iY_j)^{bB} (\lambda_j+iY_j)^{cC} (\lambda_j+iY_j)^{dD}  \Big>_\lambda ,
\label{eq:actionbeforeshift}
\end{align}
\end{widetext}
where $z=4$ is the coordination number of the   2d Ising  lattice. 

The above construction generalizes to other lattices: 
for example in $d=3$ we could take the tensors $T$ to live on the sites of the diamond lattice, in which case the spins $\bf S$ live on the sites of the pyrochlore lattice, with $z=6$. In general the spins $\bf S$ live on the ``medial lattice'', whose sites are centered on the bonds of the initial tensor network.
(Generically this lattice may have more than one site in the unit cell: 
in this case, taking the continuum limit involves isolating the locally ``uniform'' mode and integrating out the ``staggered'' modes with a unit cell. 
However, this may be done trivially as a result of the weakness of interactions at large $N_{\rm f}$.)

\subsection{Derivation of field theory for the $N\rightarrow 0$ limit}
\label{app:DerivationfieldtheoryforNto0}

In order to simplify the action, we proceed differently for the $N\rightarrow 0$ and $N\rightarrow 1$ limits. 
In the $N\rightarrow 0$ limit we must work with the unconstrained field $Y$, 
while for the $N\rightarrow 1$ limit it is possible to isolate a part of $Y$ transforming in a single irreducible representation of $G_N$ symmetry. 

We consider the $N\rightarrow 0$ case first. We are free to shift the field by a constant,
\ba\label{eq:fieldshift}
Y^{aA}_{j}& \mapsto Y^{aA}_j + u.
\end{align}
Since there is no $Y\rightarrow -Y$ symmetry, we must impose a convention for fixing $u$. 
A natural choice is to choose $u$ so that the coefficient of the linear term, $\sum_{aA} Y^{aA}$, vanishes.  This is analogous to shifting to the minimum of the potential in a more conventional Landau-Ginsburg theory.
Then the massless line, i.e.\ the critical line in the $(h,J)$ plane, is defined by the condition that the coefficient of the mass, $\sum_{aA} (Y^{aA})^2$, \textit{also} vanishes. 

In order to both determine the equation $J=J_c(h)$ for the critical line, and   to find the values of the other couplings on the critical line, 
it is sufficient to  impose two equations: the vanishing of the couplings for both   $\sum_{aA} Y^{aA}$ and
$\sum_{aA} (Y^{aA})^2$.  Solving these two equations  determines, first, the relation between $J$ and $h$ required for criticality, and, second,  the value for $u$ required to put the action in the standard form.  

The shift in Eq.~(\ref{eq:fieldshift}) gives
\begin{widetext}
\ba \label{eq:Sintermediate}
\mathcal{S}[Y] &=\f{J^2}{2} \sum_{\<jk\>}\sum_{aA} (Y^{aA}_j- Y^{aA}_k)^2   
- h^2 \sum_{j}\sum_{aA} Y_j^{aA} 
+ 
\f{1-z J^2}{2}\sum_{j} \sum_{aA} (Y_j^{aA}+u)^2 
+ \Delta {\cal S}, 
\end{align}
with 
\be\label{eq:DeltaS}
\Delta \mathcal{S} \equiv 
-  \f{1}{4! N_{\rm f}} \sum_j
\sum_{abcdABCD}  H^{abcd}_{ABCD}  \Big< (\lambda_j{+}iu {+}iY_j)^{aA}( \lambda_j{+}iu{+}iY_j)^{bB} (\lambda_j{+}iu{+}iY_j)^{cC} (\lambda_j{+}iu{+}iY_j)^{dD}  \Big>_{\!\lambda}.
\ee
\end{widetext}
The above term generates quadratic, cubic and quartic terms in $Y$ when it is expanded out.
For the tensor network with nonzero $h$ fields we can 
neglect the quartic terms as we are only interested in  the terms with the largest tree-level scaling dimensions. 
For the same reason, we also neglect quadratic and cubic terms that have ``too many'' index sums and are therefore irrelevant according to the dimensional analysis sketched  around \Eq{y-dim}.

We therefore want the action in the following form:
\begin{eqnarray}\nonumber
\label{eq:actiondesiredform}
 \mathcal{S}[Y] &=& 
\f{J^2}{2} \sum_{\<jk\>}\sum_{aA} (Y^{aA}_j- Y^{aA}_k)^2   
 \nn\\
&& \nonumber
+ r \sum_j\sum_{aA} Y_j^{aA} +  \f{m^2}{2}  \sum_j\sum_{aA} (Y_j^{aA})^2  \\
&& + \frac{\sigma}2 \sum_j \lf \sum_{aAB} Y_j^{aA} Y_j^{aB} + \sum_{aBA} Y_j^{aA} Y_j^{bA} \ri
\nn\\
&&+ g \sum_j \sum_{aA} (Y_j^{aA})^3 + \ldots
\end{eqnarray}
Before we impose a condition fixing $u$, these coefficients   depend on $u$. 

In order to expand out the action it is helpful to note the following identities, which we obtained both   graphically and with computer algebra: 
\ba
\label{rel1}
& \!\! H^{abcd}_{ABCD} \delta^{cd} \delta_{CD} = - 2 \delta_{abcd}\delta^{ABCD}\quad \text{(no sum)} ,\\
\label{rel2}
&\!\! \sum_{cdCD}H^{abcd}_{ABCD} {\delta^{cd} \delta_{CD}   } 
=
- 2 \delta^{ab} \delta_{AB} , \\
\label{rel3}
&\!\! \sum_{bcdBCD}H^{abcd}_{ABCD}{ \delta^{cd} \delta_{CD}} =
- 2 ,  \\
\label{rel4}
&\!\! \sum_{cdCD}H^{abcd}_{ABCD}  
=
2(N-2) (\delta^{ab} + \delta_{AB}) \nn\\
&\!\! \qquad\qquad\qquad  + 4(1-N) \delta^{ab} \delta_{AB} + 2, \\
\label{rel5}
&\!\!\sum_{bcdBCD}H^{abcd}_{ABCD}  
= 2 ( 3 N^2 - 6 N + 2).
\end{align}
(Note that  $N$ is the number of replicas, not to be confused with the number $N_{\rm f}$ of flavors.) In expanding out $\Delta \mathcal{S}$, it is   helpful to note that the tensor $H$ is invariant under permutations, so long as the upper- and lower-case indices are permuted together. 

The contribution to the linear term in $\Delta \mathcal{S}$ is
\begin{widetext}
\ba
\Delta \mathcal{S}_1[Y] & = -\frac1{3! N_{\rm f}} \sum_{\text{ all indices}} H^{abcd}_{ABCD}  Y_j^{aA} \left( u^3 -3 u \delta^{cd}\delta_{CD} \right) = \gamma \sum_{aA} Y_j^{aA},\\
\gamma  &  = - \f{1}{3N_{\rm f}} \left[
3 u + (3 N^2 - 6 N + 2) u^3
\right].
\end{align}
Next, consider the quadratic terms in $\Delta \mathcal{S}$:
\ba
\Delta \mathcal{S}_2[Y] & =   \f{6}{4! N_{\rm f}}  \sum_{\text{all indices}} H^{abcd}_{ABCD} Y_j^{aA} Y_j^{bB} 
\< (\lambda^{cC} + iu) (\lambda^{dD}+ i u)  \>_\lambda \nn \\ 
& =
  \f{1}{4 N_{\rm f}} \sum_{\text{all indices}}H^{abcd}_{ABCD} Y_j^{aA} Y_j^{bB} 
\lf 
{ \delta^{cd}\delta_{CD}} - u^2
\ri .
\end{align}
With the help of Eqs.~\eqref{rel2} and \eqref{rel3} this yields 
\bea\label{eq:expandquadratic}
\Delta \mathcal{S}_2 &= & \frac{\sigma}2 \sum_j \lf \sum_{aAB} Y_j^{aA} Y_j^{aB} + \sum_{aBA} Y_j^{aA} Y_j^{bA} \ri
+ \mu  \sum_j\sum_{aA} (Y_j^{aA})^2
- \f{u^2}{2 N_{\rm f}}\sum_j \sum_{abAB} Y_j^{aA} Y_j^{bB}
\\
\sigma 
& =& u^2 \f{2-N}{N_{\rm f}},
\qquad 
\mu= -   \f{1+2(1-N)u^2}{2 N_{\rm f}}. 
\eea
\end{widetext}
The final term in Eq.~(\ref{eq:expandquadratic}) will be dropped as it is irrelevant in the regime of interest.
Finally the  cubic term is given by the term linear in $u$ in $\Delta \mathcal{S}$ (\ref{eq:DeltaS}):
\ba 
\Delta \mathcal{S}_3 & = -
 \f{  u}{3! N_{\rm f}}
\sum_{\text{all indices}}
H^{abcd}_{ABCD} Y_j^{aA} Y_j^{bB} Y_j^{cC}\nn\\
&   = -\f{2 u}{3 N_{\rm f}}
\sum_{j}\sum_{aA}
  (Y_j^{aA})^3  + ...
  \label{eq:DeltaS3}
\end{align}
The dropped terms have  triple and quadruple replica sums. 

For future reference, let us also note the quartic term  obtained from Eqs.~(\ref{eq:DeltaS}) and (\ref{H-def}), although we will not use this for the generic theory:
\begin{eqnarray}
\label{DeltaS4}
\Delta {\cal S}_4 &=& -\frac1{6 N_{\rm f}} \sum_j \sum_{aA} (Y_j^{aA})^4 \nn\\
&&+  \frac1{4N_{\rm f} } \sum_j \left[\sum_{abA} (Y_j^{aA} Y_j^{bA})^2 + \sum_{aAB} (Y_j^{aA} Y_j^{aB})^2\right] \nn\\
&& - \frac1{4N_{\rm f}} \sum_j
\sum_{abAB} Y^{a A}_j Y_j^{a B}  Y_j^{bA} Y_j^{b B}.
\end{eqnarray}
Combining the terms we have obtained from the expansion of $\Delta \mathcal{S}$  with the other terms in Eq.~\eqref{eq:Sintermediate}, 
the coefficients in the action (\ref{eq:actiondesiredform}) are:
\ba\label{eq:rval}
r & = - h^2 + u (1-zJ^2) 
\nn\\
&- \f{1}{3N_{\rm f}} \left[
3 u + (3 N^2 - 6 N + 2) u^3
\right],
\\ \label{eq:msqval}
 m^2 & = (1-zJ^2) { \bf -} \f{1+2(1-N)u^2}{N_{\rm f}},
\\  \label{eq:sigval}
\sigma 
& = u^2 \f{2-N}{N_{\rm f}},\\ \label{eq:gval}
g & = -\f{2 u}{3 N_{\rm f}}.
\end{align}
Let us specify to the limit $N\rightarrow 0$
\ba
r & = - h^2 + u (1-zJ^2) 
- \f{3 u + 2 u^3}{3N_{\rm f}},
\\
m^2 & = (1-zJ^2) { \bf -} \f{1+2u^2}{N_{\rm f}}
\\ 
\sigma 
& = \f{2  u^2}{N_{\rm f}},\\
g & = -\f{2 u}{3 N_{\rm f}}.
\end{align}
Finally, we can   impose the two conditions $r=0$, $m^2=0$ that were discussed above. 
It is clear from the first of these conditions that $u$ must be much larger than one. Working to leading order in $u$,
\ba\label{eq:uJrelns}
u & = \lf \f{3 N_{\rm f} h^2}{4} \ri^{1/3} + \ldots,
\\
 J_c(h)^2 & =\f{1}{z} \left[   
1 - \lf \f{9h^4}{2N_{\rm f}} \ri^{1/3}
+ \ldots  \right].
\end{align}
 We see that at large $N_{\rm f}$ the critical coupling $J_c(h)$ becomes independent of $h$.

On the critical line, and finally rescaling the field   ${Y_j^{aA}\to   Y_j^{aA}/ J}$, the action takes the form
\begin{align}
\label{eq:actiondesiredform-crit-line}
\mathcal{S}[Y] = & 
\f{1}{2} \sum_{\<jk\>}\sum_{aA} (Y^{aA}_j- Y^{aA}_k)^2  
\nonumber\\
&+ \frac{\sigma}2 \sum_j \lf \sum_{aAB} Y_j^{aA} Y_j^{aB} + \sum_{aBA} Y_j^{aA} Y_j^{bA} \ri \nn\\
&+ g \sum_j \sum_{aA} (Y_j^{aA})^3 + \ldots
\end{align}
with the couplings $\sigma$, $g$ given by inserting Eq.~(\ref{eq:uJrelns}) into Eqs.~{(\ref{eq:sigval})-(\ref{eq:gval})}:
\ba
\sigma & = \frac1{J^2}\lf \f{9 h^4}{2 N_{\rm f}} \ri^{\!1/3},
& 
g & = - \frac1{J^3}\lf \f{2 h^2}{9 N_{\rm f}^2} \ri^{\!1/3}.
\end{align}
As discussed in Sec.~\ref{s:FTforNt00}, the coupling constant relevant for perturbation theory is not $g$  but the combination
\be
g \sigma =  -\f{h^2}{ N_{\rm f} J^5}.
\ee
  In Eq.~\eqref{eq:actiondesiredform-crit-line} the equation is written on the lattice. The continuum form follows immediately from a derivative expansion: for example,
\ba
\f{1}{2} \sum_{\<j,k\>} (Y_j^{aA}-Y_k^{aA})^2 
\simeq 
\f{1}{2}
\int \dd^2 x \, \big[\nabla Y^{aA}(x)\big]^2.
\end{align}
Keeping the lowest order in the derivative expansion is   justified here. 
The reason for this is standard. 
Higher terms in the derivative expansion are non-negligible for high-momentum modes (those with wavelengths comparable to the lattice spacing).
The weakness of the bare interactions means that these modes can be integrated out with only a negligible effect on the low-momentum modes of interest. 

We have discussed the case of the square lattice explicitly because of its simple geometry, but as noted above the expansion can be performed for any lattice without qualitative changes.

\subsection{Derivation of field theory for the $N\rightarrow 1$ limit}
\label{app:deriveNto1theory}

In order to address the $N\rightarrow 1$ limit, relevant to the measurement transition,  we separate the field $  X$ into   pieces that transform in distinct representations of the global $G_N$ symmetry group. 
Going back to Eq.~(\ref{eq:Sintermediate}), let us write
\be\label{eq:fielddecomposition2}
Y^{aA} :=  X^{aA} + L^a + R^A + \phi,
\ee
where $  X$ vanishes if any index is summed, and similarly for $L$ and $R$. 
This is a decomposition into four irreducible representations of ${S_N\times S_N}$ symmetry
(though $L$ and $R$ combine into a single irrep of the full $G_N$ symmetry of the model).\footnote{The global symmetry group is ${G_N=(S_N\times S_N)\rtimes \mathbb{Z}_2}$. The $\mathbb{Z}_2$ generator can be taken to be transposition of $X$. Since this operation exchanges $L$ and $R$,
these two fields combine into a single irrep of $G_N$. The fact that $L$ and $R$ are in the same irrep ensures that they have the same mass.}

Writing the action as 
\be
\mathcal{S}[X]=\sum_j \mathcal{V}[X_j] + \text{(derivative terms)},
\ee
we examine the form of $\mathcal{V}$ that results from the above substitution.
At linear order in the fields, we have
\be
\mathcal{V}_1[X] = N^2 r \phi,
\ee
where $r$ was given in Eq.~(\ref{eq:rval}).
When $N\rightarrow 1$, this is 
\be\label{eq:rvalNto1}
r = - h^2 + u(1-J^2 z) - \f{3u-u^3}{3N _{\rm f}}.
\ee
In a moment, we will fix the field shift $u$ so that this linear term vanishes. 

For the quadratic terms, it is useful to note that 
\ba
&\big( \sum_{aA} X^{aA} \big)^2 = N^4 \phi^2, 
\\
& \sum_{aA} (X^{aA})^2  = 
\sum_{aA} \lf   X^{aA}\ri^2
+ N^2 \phi \notag \\ 
&\quad\quad \quad \quad \quad + N \Big[ \sum_a (L^a)^2 + \sum_A (R^A)^2 \Big],
\\ \notag
& \sum_{aAB} X_j^{aA} X_j^{aB} + \sum_{aBA} X_j^{aA} X_j^{bA}
\\ & \quad \quad \quad
= 2 N^3 \phi^2 +
N^2 \Big[ \sum_a (L^a)^2 + \sum_A (R^A)^2 \Big].
\end{align} 
Using these formulas, and the results for the quadratic terms in Eq.~(\ref{eq:expandquadratic}),
we obtain the (bare) mass terms in the limit of $N\to 1$:
\ba \notag
\mathcal{V}_2[X]
& = 
\f{m^2}{2} \sum_{aA} (  X^{aA})^2
\\ \notag
& +
\f{m^2 + \sigma}{2} \lf\sum_a (L^a)^2  + \sum_A (R^A)^2 \ri  
\\
& + 
\f{m^2 + 2 \sigma - N_{\rm f}^{-1} u^2}{2} \phi^2.
\label{eq:massesforNto1case}
\end{align}
The coefficients  are given by setting ${N\to 1}$ in expressions (\ref{eq:msqval}),~(\ref{eq:sigval}),
\ba\label{eq:massNto1}
m^2 & = (1-zJ^2) - \f{1}{N _{\rm f}},
&
\sigma & = \f{u^2}{N _{\rm f}},
\end{align}
so that \Eq{eq:massesforNto1case} simplifies to
\ba \notag
\mathcal{V}_2[X]
& = 
\f{m^2}{2} \sum_{aA} (  X^{aA})^2
\\ \notag
& +
\f{m^2 + N _{\rm f}^{-1} u^2}{2} \lf\sum_a (L^a)^2  + \sum_A (R^A)^2 \ri  
\\
& + 
\f{m^2 + N _{\rm f}^{-1} u^2}{2} \phi^2.
\label{eq:massesforNto1case2}
\end{align}

The key point here is that at the critical point, where $  X$ becomes massless ($m^2=0$), the other fields have positive masses. 
Therefore we will be able to integrate them out of the critical theory.

Let us mention as an aside that the procedure we are   following for   ${N\to 1}$   cannot be applied to the ${N\to 0}$ case. If the constants in Eq.~(\ref{eq:massesforNto1case}) are rederived for the ${N\to 0}$ case, it is found that all three masses vanish simultaneously. This is why, in the previous section, it was necessary to retain all of the degrees of freedom in $X$ for the critical theory.

Let us consider the theory close to the line where the bare mass $m^2$ vanishes.
We fix the value of $u$ by requiring that $r$ in Eq.~(\ref{eq:rvalNto1}) vanishes. This equation may be written as
\be
r = - h^2 + m^2 u + \f{u^3}{3 N _{\rm f}}.
\ee
If $m^2$ is small
(smaller than order $N _{\rm f}^{-1/3}$)
we must take
\be
u = \lf 3 h^2 N _{\rm f}\ri^{1/3}
\ee
to leading order in $N _{\rm f}$.

Next, we may consider the cubic terms, which come from
inserting the decomposition of $X$ into the first line of Eq.~\eqref{eq:DeltaS3}.
First, there is a term that involves only  $  X$:
\be\label{eq:Nto1cubiccoefficient}
\mathcal{V}_3[X]= - \f{2 u}{3 N _{\rm f}} \sum_{aA} \lf   X^{aA} \ri^3 + \ldots
\ee
The other terms involve 
either the massive fields ($L$, $R$, $\phi$) alone, or couplings between $  X$ and the massive fields.
By counting powers of $N _{\rm f}$,\footnote{To do this, it is convenient to rescale the massive fields so that their masses are of order 1.}   we may check that integrating out the massive fields only contributes a subleading correction to the cubic coupling of $X$. 
 Integrating out the massive fields also renormalizes the mass of $X$, but only by an amount which is much smaller than the difference in bare mass between the fields.  This, together with the finite renormalization of the mass in the resulting theory for $X$ alone, just contributes to a small shift to the position of the critical line, 
which is given to leading order in $N _{\rm f}$ by
\be
J^2_c = \f{1}{z}.
\ee 
Again, this critical line is, at leading order in $N _{\rm f}$, a horizontal line in the $(h,J)$ plane.

We now have an action for $  X$ alone. Absorbing again a factor of $J$ into each field by setting $X\to X/J$, this becomes
\ba \label{eq:SwidetildeXlattice}
\mathcal{S}[  X] &=\f{1}{2} \sum_{\<jk\>}\sum_{aA} (  X^{aA}_j-   X^{aA}_k)^2  
+ \f{\tilde m^2}{2} \sum_{aA} (  X^{aA})^2 \nn\\
&+ \f{g}{3!}\sum_{aA}   (X^{aA})^3,
\end{align}
where
\ba
 \tilde m^2 &=   1- z J^2 + \cdots ,
 &
g &= - \frac1{J^3}\lf \f{192 h^2}{N _{\rm f}^2} \ri^{1/3}.
\end{align}
As discussed at the end of the previous subsection, this may be directly converted into a continuum action by replacing lattice differences with derivatives.

\section{Details of relation between square tensor network for $N_{\rm f}=1$ and nonunitary kicked Ising model}
\label{app:detailsTNtoQC}

When we view the diagonal direction in Fig.~\ref{f:TN} as time, we have a quantum circuit, with a brickwork structure. For  ${N_{\rm f}=1}$, this circuit acts on a one-dimensional chain of qubits.
Each tensor  $T$ defines the matrix elements of a two-site gate $V$ via:
\be
T_{{S}_1,{S}_2,{S}_3,{S}_4}
= 
\bra{S_4, S_3} V \ket{S_1, S_2},
\ee
where $\ket{\pm 1, \pm 1}$ are basis states, 
and each $V$ is independently random. 
The gate $V$ above may  be written in terms of Pauli $X$ and $Z$ operators for the two spins it acts on. Up to an unimportant constant of proportionality (the following kinds of rewriting are standard,  see e.g.\ Refs.~\cite{GopalakrishnanLamacraft2019,BertiniKosProsen2018}):

\be\label{eq:Vgate}
V \propto 
\exp\lf {J_3 Z\otimes Z} \ri 
\left[ 
W_1\otimes W_2
\right]
 \exp\lf {J_1 Z \otimes Z}\ri,
\ee
where 
\ba
W_1 & = \exp({\tilde h_4 Z})\exp({f_1 X})\exp({ \tilde h_1 Z }),
\\
W_2 & = \exp({\tilde h_3 Z_2}) \exp({ f_2 X_2}) \exp({\tilde h_2 Z_2}),
\end{align}
with ${\tanh f_1 = e^{-2 J_4}}$,  ${\tanh f_2 = e^{-2 J_2}}$. 

Now when we consider the entire circuit, the fact that  operators of the form $e^{J Z\otimes Z}$ in (\ref{eq:Vgate}) commute means that 
we can redraw the circuit as in Fig.~\ref{f:TNtoQC}. The horizontal bars represent a layer of commuting gates applied at the same time. The boxes are   the ``$W$'' gates in \Eq{eq:Vgate}. 
In the discussion around Fig.~\ref{f:TNtoQC} these single-site gates written as $e^{\vec g.\sigma}$.
This is always possible,~e.g.
\be 
e^{\vec g_1. \vec \sigma} =
\exp({\tilde h_4 Z})\exp({f_1 X})\exp({ \tilde h_1 Z })
\ee
(here we are mixing two notations for Pauli operators: ${\sigma^x = X}$ etc.). This is a standard decomposition of an $\mathrm{SL}(2,\mathbb{C})$ matrix. The probability distribution for $g$ is nontrivial and is inherited from the Gaussian distributions of the original variables.

The resulting circuit acts on a chain of spin-1/2 degrees of freedom $\vec \sigma_{r}$ ($r\in \mathbb{Z}$). 
Neglecting  boundary conditions, it is a product of single-time-step time-evolution operators of the schematic form
\begin{eqnarray}
    K(t) &=&  \exp \lf \sum_{r} \vec g(r,t) . \vec \sigma_{r} \ri  \nn \\
   && \times \exp \lf  \sum_{r} J(r,t)  \sigma^z_{r}  \sigma^z_{r+1}    \ri.
\end{eqnarray}
It would be possible to generalize this rewriting to large $N_{\rm f}$, and it would also be possible to take a continuous-time limit by appropriate scalings of the couplings.


\ifx\doi\undefined
\providecommand{\doi}[2]{\href{http://dx.doi.org/#1}{#2}}
\else
\renewcommand{\doi}[2]{\href{http://dx.doi.org/#1}{#2}}
\fi
\providecommand{\link}[2]{\href{#1}{#2}}
\providecommand{\arxiv}[1]{\href{http://arxiv.org/abs/#1}{#1}}
\providecommand{\hal}[1]{\href{https://hal.archives-ouvertes.fr/hal-#1}{hal-#1}}
\providecommand{\mrnumber}[1]{\href{https://mathscinet.ams.org/mathscinet/search/publdoc.html?pg1=MR&s1=#1&loc=fromreflist}{MR#1}}

\tableofcontents

\end{document}